\documentclass[onecolumn,preprintnumbers]{revtex4}
\usepackage{amsmath}
\usepackage{amssymb}
\usepackage{latexsym}
\usepackage{epsfig}
\usepackage{color}
\usepackage{hyperref}

\begin{document}

\title{Dark energy models from a parametrization of $H$: A comprehensive
analysis and observational constraints}
\author{S. K. J. Pacif}
\email{shibesh.math@gmail.com}
\affiliation{Department of Mathematics, School of Advanced Sciences, Vellore Institute of
Technology, Vellore 632014, Tamil Nadu, India}
\date{\today }

\begin{abstract}
The presented paper is a comprehensive analysis of two dark energy (DE)
cosmological models wherein exact solutions of the Einstein field equations
(EFEs) are obtained in a model-independent way (or by cosmological
parametrization). A simple parametrization of Hubble parameter ($H$) is
considered for the purpose in the flat Friedmann-Lemaitre-Robertson-Walker
(FLRW) background. The parametrization of $H$ covers some known models for
some specific values of the model parameters involved. Two models are of
special interest which show the behavior of cosmological phase transition
from deceleration in the past to acceleration at late-times. The model parameters are
constrained with $57$ points of Hubble datasets together with the $%
580$ points of Union $2.1$ compilation supernovae datasets and baryonic
acoustic oscillation (BAO) datasets. With the constrained values of the
model parameters, both the models are analyzed and compared with the
standard $\Lambda $CDM model and showing nice fit to the datasets. Two
different candidates of DE is considered, cosmological constant $%
\Lambda $ and a general scalar field $\phi $ and their dynamics are
discussed on the geometrical base built. The geometrical and physical
interpretations of the two models in consideration are discussed in details
and the evolution of various cosmological parameters are shown graphically.
The age of the Universe in both models are also calculated. Various cosmological parametrization schemes used
in the past few decades to find exact solutions of the EFEs are also
summarized at the end which can serve as a unified reference for the readers.
\end{abstract}

\maketitle

\section{Introduction}

Late-time cosmic acceleration is an essential constituent of precision
cosmology at present. The idea of cosmic acceleration was first evidenced by
the observations of high redshift supernova of type Ia \cite{perlmutter,
riess}. The idea of cosmic acceleration was later accepted quickly by the
scientific community largely because of the independent observations with
different methodology adopted by the supernova search teams lead by
Perlmutter and Riess and also the CMB and the large scale structure data
were providing substantial evidence for a cosmological constant, indirectly 
\cite{bernadis, hanany, netterfield}. Later on some robust analysis and
precise observations strengthen the idea of cosmic acceleration and a flat
Universe consistent with $\Omega _{\Lambda }=1-\Omega _{m}=0.75$ \cite%
{mould, spergel, komastu, essence}. What causing the accelerating expansion
is still a mystery and we are mostly in dark in this context. However, the
theoretical predictions and advanced surveys in observational point of view
indicating the presence of a weird form of energy in the Universe with high
negative pressure with increasing density. The mysterious energy is named as 
\textit{dark energy} \cite{turner} as it's nature, characteristics is
speculative only without any laboratory tests. Also, the candidature of dark
energy is a debatable topic at present cosmological studies. Moreover, the
age crisis in the standard model need cosmic acceleration \cite{globular}.
Although, the modification of gravity theory at infra red scale attracted
attention to explain the late-time acceleration without invoking any extra
source term \cite{capozilo, dvali}, but the theory of dark energy became
quite popular \cite{copeland-sami, bamba-de}.

Very recently, gravitational wave detection and the picture of black hole
shadow strenghten the Einstein's general theory of relativity and any
modifications in the Einstein's theory (specifically to the geometry part)
is not worth appreciated. However, Einstein himself was not convinced with
the matter distribution in the Universe i.e. the right hand side of his
field equations (representing matter sector) is considered to made up of low
grade wood while the geometry part is of solid marble (representing the
space-time). Any extra source term such as Einstein's cosmological constant
(representing energy density of vacuum) could be added into the energy
momentum tensor\ and serve as a candidate of dark energy. The most favoured
candidate of dark energy is the well known cosmological constant $\Lambda $.
Also, $\Lambda $CDM models have the best fit with many observational
datasets. However, with this significant $\Lambda $, due to its non
dynamical and the long standing fine tuning problem, researchers thought
beyond it for a better candidate of dark energy. So, scalar field models
were discussed after the cosmological constant for which $\Lambda $ could
also be generated from particle creation effect \cite{sahni}. The
dynamically evolving scalar field models have been utilized for the purpose
are quintessence \cite{zlatev,brax,barreiro}, K-essence \cite%
{mukhanov1,chiba,steinhardt1}, phantom \cite{caldwell} and tachyonic field 
\cite{sen1, garousi1, bergshoeff}. The exotic fluid is also serve the
purpose to explain the cosmic acceleration phenomenology that considered an
equation of state producing large negative pressure e.g. Chapligyn gas
equation of state \cite{gorini}, Polytropic gas equation of state \cite%
{chavanis} etc.

Soon after the formulation of EFEs, theoreticians worked on finding exact
solutions. The first exact solution of the EFEs is the Schwarzschild
exterior solution \cite{schwar} wherein the prefect fluid equation of state
was considered as a suplementary condition. Despite of the high non
linearity of the EFEs, various exact solutions are obtained for static and
spherically symmetric metrics. Einstein's static solution \cite{static},
de-Sitter solution \cite{de-Sitter}, Tolman's solutions \cite{Tolman},
Adler's solutions \cite{Adlar}, Buchdahl's solution \cite{Buchd}, Vaidya and
Tikekar solution \cite{Vaidya}, Durgapal's solutions \cite{Durgapal},
Knutsen's solutions \cite{Knutsen} and many more well-known solutions of
EFEs are obtained which are summarized in the literature \cite{Kramer} and
also discussed in \cite{Negi}. Milne's model \cite{milne}, steady state
model \cite{Steady} are some different models proposed. All those
phenomenological cosmological models explain the Universe theoretically very
well. However, observations play a major role in modern cosmology which
validate or discard a model. Now, numerical computations are also playing
big role in modern cosmology and estimating cosmological parameters and also
parameters of a model. In this study, an important discussion is given on a
technique of finding exact solution of EFEs known as model independent way
study. Moreover, two models are discussed and analyzed comprehensively with
current trendz in theoretical comology.

The paper is organized as follows. The first section is an introduction to
present cosmological scenario. The second section describes the Einstein's
field equations in general relativity in presence of dark energy. The third
section is a motivation to the idea of model independent way or the
cosmological parametrization study to obtain exact solutions to EFEs. A
simple parametrization of Hubble parameter is considered in the light of
cosmographical study in the fourth section. In the fifth section,
observational constraints have been found for the model parameters involved
in the functional form of $H$ for the two models obtained. The sixth section
is devoted to the geometrical dynamics and analysis of some important
cosmological parameters describing the geometrical behavior of the Universe
for both the models. In the seventh section, two candidates of dark energy
is explored, cosmological constant and a general scalar field and the
physical parameters such as energy density, density parameter, potential of
scalar field and equation of state parameter are discussed for both the
models under considerations. In the eighth section, the age of the Universe
for the obtained models are calculated. The final section summarizes the
physical insights of the results obtained and concluded. A brief summary of
the various parametrization schemes used in the past few decades are given
in the appendix.

\section{EFEs in presence of Dark Energy}

The nature of dark energy and its candidature is a mystery and it is a
matter of speculation to express it as a source term into the Einstein field
equations. However, DE is speculated to be homogeneous permeating all over
the space for which the energy momentum tensor can be represented in the
form of a perfect fluid as%
\begin{equation}
T_{ij}^{DE}=(\rho _{DE}+p_{DE})U_{i}U_{j}+p_{DE}~g_{ij},  \label{01}
\end{equation}%
with its equation of state in the form $p_{DE}=\omega _{DE}\rho _{DE}$,
where $\omega _{DE}$ is the equation of state (EoS) parameter and is a
function of time in general satisfying the inequality $\omega _{DE}<0$.
There is hot debate going on for a suitable value of $\omega _{DE}$ and the
analysis of some observational data shows that its value lies in the range $%
-1.61<\omega _{DE}<-0.78$ \cite{spergel, knop, tegmark}. But recent analysis
of datasets provide more tighter constraints on $\omega _{DE}$ \cite{w-de1,
w-de2}. The different values of $\omega _{DE}$\ in certain ranges gives rise
to different candidates and can broadly be be classified as follows. For

$\blacktriangleright $ \ \ $\omega _{DE}=-1$, the case is for the
cosmological constant;

$\blacktriangleright $ \ \ $\omega _{DE}=constant\neq -1$, the case is for
cosmic strings, domain walls, etc.;

$\blacktriangleright $ \ \ $\omega _{DE}\neq constant$, the cases for scalar
fields (quintessence, k-essence etc.), braneworlds, Dirac-Born-Infeld(DBI)
action, Chaplygin gas etc.;

$\blacktriangleright $ \ \ $\omega _{DE}<-1$, the case is for phantom models.

For a broader list of dark energy models see \cite{copeland-sami, bamba-de}
(and refs. therein). There are interesting cases in each of them with some
problems though. For example, Cosmological constant $\Lambda $ is the most
consistent model for dark energy explaining observations but is plagued with
fine tuning problem. Similarly, the phantom models are interesting where the
weak energy condition ($\rho >0,~\rho +p>0$) is violated with the feature of
finite time singularity \cite{caldwell}.

In general relativity, dark energy can be introduced by supplementing the
energy momentum tensor $T_{ij}^{DE}$ into the Einstein field equations $%
G_{ij}=-8\pi GT_{ij}$ together with the matter source $T_{ij}^{M}$ as a
perfect fluid,

\begin{equation}
T_{ij}^{tot}=T_{ij}^{m}+T_{ij}^{de}=(\rho
_{tot}+p_{tot})U_{i}U_{j}+p_{tot}~g_{ij},  \label{02}
\end{equation}%
with $\rho _{tot}=\sum \rho +\rho _{de}$ and $p_{tot}=\sum p+p_{de}$
denoting the total energy densities and total pressure due to all types of
matter (baryonic matter, dark matter and radiation) and dark energy
respectively. $U_{i}$ is the usual four velocity vector and $g_{ij}$ is the
metric tensor. Now, the modified Einstein Field Equations for a flat FLRW
metric 
\begin{equation}
ds^{2}=-dt^{2}+a^{2}(t)\left[ dr^{2}+r^{2}(d\theta ^{2}+\sin ^{2}\theta
d\phi ^{2})\right] \text{, }  \label{2}
\end{equation}%
where $a(t)$ is the scale factor of the Universe, can be written as 
\begin{equation}
M_{pl}^{-2}\rho _{tot}=3\left( \frac{\dot{a}}{a}\right) ^{2}=3H^{2},
\label{03}
\end{equation}

\begin{equation}
M_{pl}^{-2}p_{tot}=-2\frac{\ddot{a}}{a}-\left( \frac{\dot{a}}{a}\right)
^{2}=(2q-1)H^{2}.  \label{04}
\end{equation}

The conservation of energy-momentum (or from Eqs. (\ref{03}) and (\ref{04})
yields 
\begin{equation}
\dot{\rho}_{tot}+3(p_{tot}+\rho _{tot})\frac{\dot{a}}{a}=0.  \label{5}
\end{equation}

The continuity equation (\ref{5}) play significant role in the evolution as
it deals with the matter and its interaction. In current cosmology, two
kinds of dark energy models generally discussed; interacting models of dark
energy (considering the interaction between cold dark matter and dark
energy) \cite{zimdahl, bertolemi, banerjee} and non-interacting models of
dark energy where all the matters allowed to evolve separately \cite{ellis,
sahni-2, saini, simon}. Up to date, there are no known interaction other
than gravity between the matter and dark energy. The present study refers to
non interacting models only. The system of equations are non linear ordinary
differential equations and is difficult to find exact solutions. There are
tremendous efforts to find both the exact and numerical solutions to EFEs in
the past. In the next section, the solution techniques of the above system
of equations will be discussed elaborately.

\section{Cosmological parametrization\qquad}

The above system of equations (\ref{03}), (\ref{04}) and (\ref{5}) possesses
only two independent equations with five unknowns $a$, $\rho $, $p$, $\rho
_{de}$, $p_{de}$ (or $\omega _{de}$). Due to the homogeneous distribution of
matter in the Universe at large scale, it is customary to consider the
barotropic equation of state $p=\omega \rho $, $\omega \in \left[ 0,1\right] 
$. The equation of state describes different types of matter source in the
Universe depending upon the discrete or dynamical values of equation of
state parameter $\omega $ that includes baryonic matter ($\omega =0$), dark
matter ($\omega =0$), radiation ($\omega =1/3$), stiff matter ($\omega =1$),
etc. This additional equation provides the third constraint equation.
Another constraint equation can be the consideration of equation of state of
dark energy ($\omega _{de}=$ constant or a function of time $t$ or function
of scale factor $a$ or function of redshift $z$) - best known as
parametrization of dark energy equation of state. These four equations can
explain the cosmological dynamics of the Universe where all the geometrical
parameters (Hubble parameter $H$, deceleration parameter $q$, jerk parameter 
$j$, etc.) or physical parameters (densities $\rho $, $\rho _{de}$,
pressures $p$, $p_{de}$, EoS parameter $\omega _{de}$, density parameter $%
\Omega _{i}$, etc.) are expressed as functions of either scale factor $a$ or
the redshift $z$ ($=\frac{a_{0}}{a}-1$, $a_{0}$ being the present value of
scale factor generally normalized to $a_{0}=1$). But, there is still one
more equation short to close the system for the complete determination of
the system; the time evolution of scale factor $a$ is yet to be determined.
In literature, there are various schemes of parametrization of the scale
factor and it's higher order derivatives ($H$, $q$, $j$ etc.) providing the
complete solution of the EFEs i.e. the explicit forms of cosmological
parameters as a function of cosmic time $t$.

In fact, a critical analysis of the solution techniques of EFEs in general
relativity theory or in modified theories has two aspects; one is the
parametrization of geometrical parameters $a$, $H$, $q$, $j$ giving the time
dependent functions of all the cosmological parameters; another is the
parametrization of the physical parameters $\rho $, $p$, $\rho _{de}$, $%
p_{de}$ (or $\omega _{de}$) giving the scale factor dependence or redshift
dependence of all the cosmological parameters. See the appendix for a broad
list of different schemes of parametrization of geometrical parameters and
physical parameters and also some phenomenological ansatzs used in the past
few decades to find the exact solutions of Einstein field equations. If we,
observe carefully, we can say that the first kind of parametrization schemes
(of geometrical parameters) are considered to find exact solutions that
discusses the expansion dynamics of the Universe and provides the time
evolution of the physical parameters $\rho $, $p$, $\rho _{de}$, $p_{de}$
(or $\omega _{de}$). This method is generally known as \textit{model
independent way} study of cosmological models or the cosmological
parametrization \cite{chuna, edvard, pacif2016}. The method do not affect
the background theory anyway and provide solutions to the EFEs explicitly
and also has an advantage of reconstructing the cosmic history of the
Universe explaining some phenomena of the Universe. Also, this method
provides the simplest way to resolve some of the problems of standard model
e.g. the initial singularity problem, cosmological constant problem, etc.
and also explain the late-time acceleration conundrum, theoretically. While
the second kind of parametrization (of physical parameters) are generally
considered to discuss all the physical aspects (thermodynamics, structure
formation, nucleosynthesis etc.) of the Universe. However, both the schemes
of parametrization are adhoc choices or some phenomenological ansatzs (e.g. $%
\Lambda $-varying cosmologies). All parametrization schemes (see appendix-1)
contain some arbitrary constants, referred to as \textit{model parameters}
which are constrained through any observational datasets.

The purpose here is to obtain an exact solution of the Einstein field
equations in standard general relativity theory with a simple
parametrization of the Hubble parameter $H$ and discuss the reconstructed
cosmic evolution.

\section{Parametrization of $H$ \& the models}

The cosmographic analysis provide clues to study the evolution of the
observable Universe in a model independent way in terms of the kinematic
variables \cite{capozilo-lakoz}. Moreover, analysis of cosmographic
parameters helps in studying the dark energy without any assumption of any
particular cosmological model except only the cosmological principle. In the
standard approximation the scale factor can be expanded in Taylors series
around the present time $t_{0}$ (which is the current age of the Universe
also) and is the simple strategy adopted in cosmographical analysis. Here
and afterwards a suffix $0$ denotes the value of the parameter at present
time $t_{0}$. The Taylor's series expansion can be written as:

\begin{eqnarray}
a^{(n)} &=&1+H_{0}(t-t_{0})-\frac{1}{2!}q_{0}H_{0}^{2}(t-t_{0})^{2}+\frac{1}{%
3!}j_{0}H_{0}^{3}(t-t_{0})^{3}  \notag \\
&&+\frac{1}{4!}s_{0}H_{0}^{4}(t-t_{0})^{4}+\frac{1}{5!}%
l_{0}H_{0}^{5}(t-t_{0})^{5}+.....  \label{Taylor}
\end{eqnarray}%
where $H(t)=\frac{1}{a}\frac{da}{dt}$ is the Hubble parameter measuring
velocity, $q(t)=-\frac{1}{a}\frac{d^{2}a}{dt^{2}}\left[ \frac{1}{a}\frac{da}{%
dt}\right] ^{-2}$ is the deceleration parameter measuring acceleration, $%
j(t)=\frac{1}{a}\frac{d^{3}a}{dt^{3}}\left[ \frac{1}{a}\frac{da}{dt}\right]
^{-3}$ jerk parameter measuring jerk, $s(t)=\frac{1}{a}\frac{d^{4}a}{dt^{4}}%
\left[ \frac{1}{a}\frac{da}{dt}\right] ^{-4}$ is the snap parameter and $%
l(t)=\frac{1}{a}\frac{d^{5}a}{dt^{5}}\left[ \frac{1}{a}\frac{da}{dt}\right]
^{-5}$ is the lerk parameter. All of these parameters play significant roles
in the cosmographic analysis of the Universe (specifically the $H$, $q$ and $%
j$) and distinguish various dark energy models.

Motivated by the above discussion, in this paper, a simple parametrization
of the Hubble parameter ($H$) is considered as an explicit function of
cosmic time `$t$' in the form \cite{pacif2016}%
\begin{equation}
H(t)=\frac{k_{2}t^{m}}{\left( t^{n}+k_{1}\right) ^{p}}  \label{mainansatz}
\end{equation}%
where $k_{1},$ $k_{2}\neq 0,$ $m,$ $n,$ $p$ are real constants (or model
parameters). $k_{1}$ and $k_{2}$ both have the dimensions of time. Some
specific values of the parameters $m,$ $n,$ $p$ suggest some distinguished
models which are elaborated by Pacif et al \cite{pacif2016}. It is easy to
see that, the single parametrization (\ref{mainansatz}) generalizes several
known models e.g. $\Lambda $CDM model, power law model, hybrid expansion
model, bouncing model, linearly varying deceleration parameter model and
some more. Out of the twelve models deduced for some integral or non
integral values of \thinspace $m$, $n$, $p$ in the functional form of HP in (%
\ref{mainansatz}), two models (with $m=-1$, $p=1$, $n=1$ and with $m=-1$, $%
p=1$, $n=2$) show the possibility of describing the phenomena of
cosmological phase transition for negative $k_{1}$ \& $k_{2}$ and is
described as in the following Table-1.

\begin{center}
\begin{tabular}{|l|c|c|c|}
\hline
\multicolumn{4}{|c|}{Table-1: The models} \\ \hline
Models & $H(t)$ & $a(t)$ & $q(t)$ \\ \hline
M1 & \multicolumn{1}{|l|}{$\frac{k_{2}}{t\left( k_{1}-t\right) }$} & 
\multicolumn{1}{|l|}{$\beta \left( \frac{t}{k_{1}-t}\right) ^{\frac{k_{2}}{%
k_{1}}}$} & \multicolumn{1}{|l|}{$-1+\frac{k_{1}}{k_{2}}-\frac{2}{k_{2}}t$}
\\ \hline
M2 & \multicolumn{1}{|l|}{$\frac{k_{2}}{t\left( k_{1}-t^{2}\right) }$} & 
\multicolumn{1}{|l|}{$\beta \left( \frac{t^{2}}{k_{1}-t^{2}}\right) ^{\frac{%
k_{2}}{2k_{1}}}$} & \multicolumn{1}{|l|}{$-1+\frac{k_{1}}{k_{2}}-\frac{3}{%
k_{2}}t^{2}$} \\ \hline
\end{tabular}
\end{center}

where, $\beta $ is an integrating constant which also play an important role
in the evolution. Pacif et al. obtained solutions for these two models in a
scalar field background and also found the observational constraints on
model parameters with $28$ points of $H(z)$ datasets. The present paper is
an extension of the same study for these two models M1 and M2 wherein much
deeper analysis have been done.

\begin{figure}[tbph]
\label{aM2.pdf}
\par
\begin{center}
$%
\begin{array}{c@{\hspace{.1in}}c}
\includegraphics[width=2.7 in, height=2.3 in]{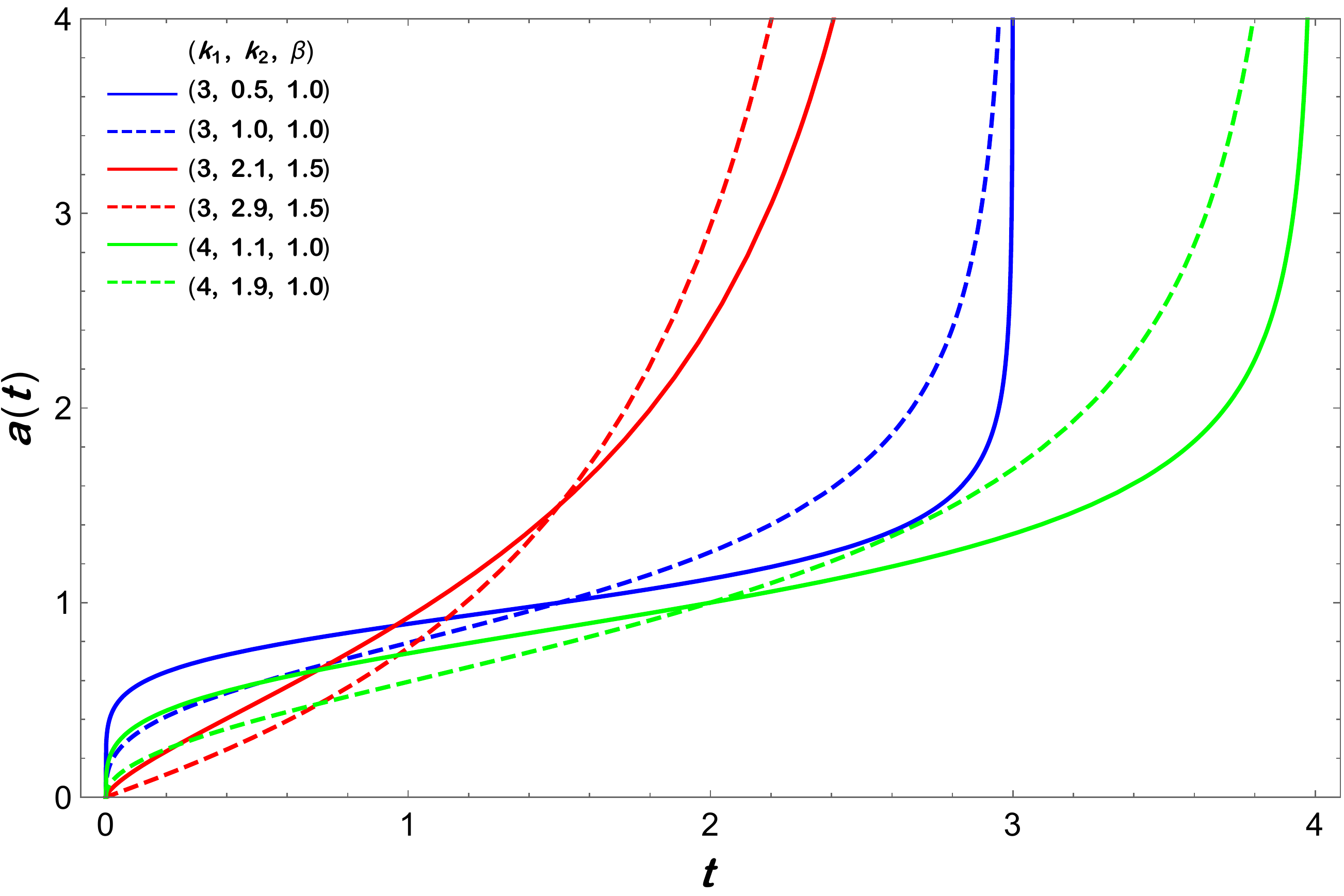} & %
\includegraphics[width=2.7 in, height=2.3 in]{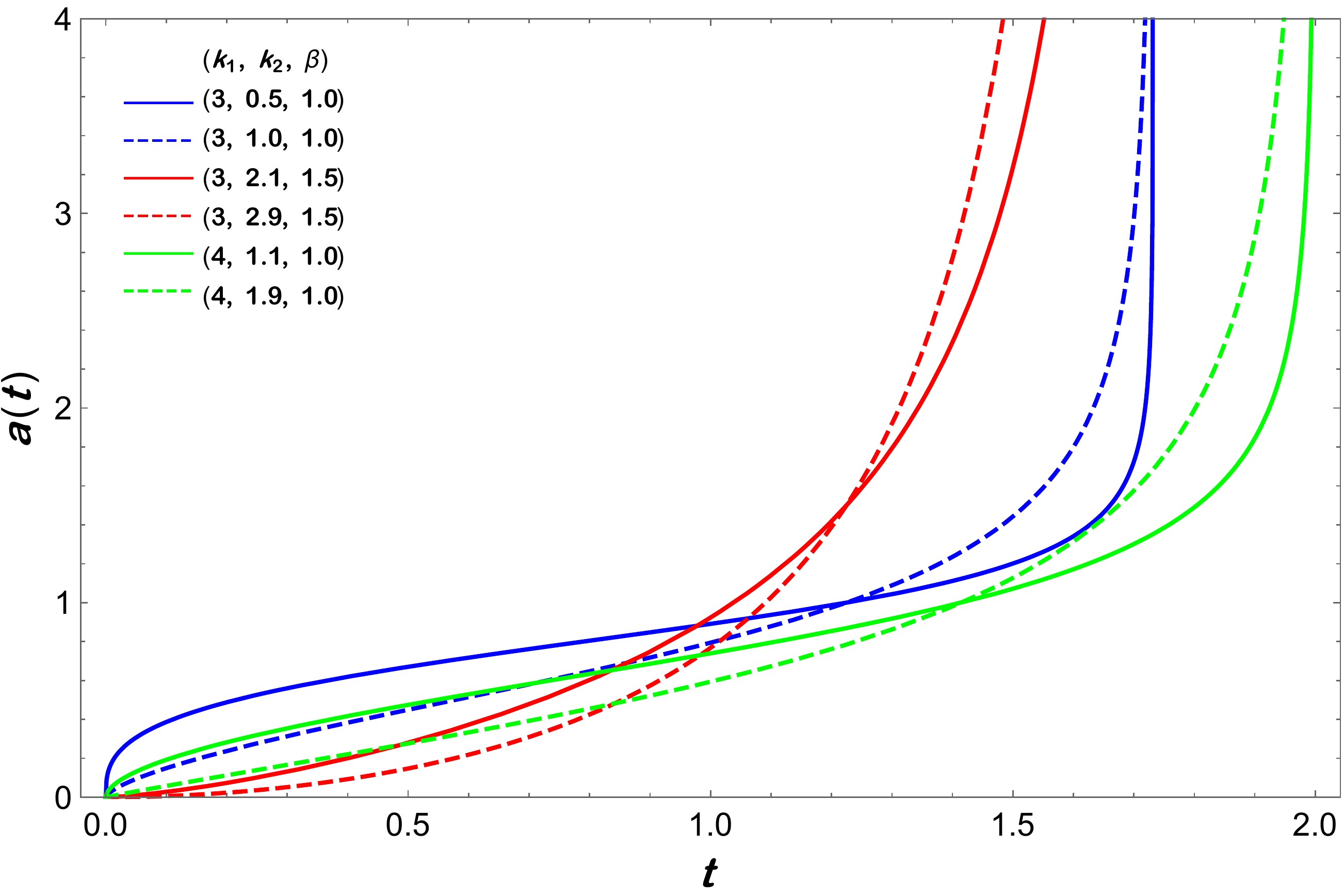} \\ 
\mbox (a) & \mbox (b)%
\end{array}
$%
\end{center}
\caption{ Figures (a) and (b) respectively show rough sketches of the
evolotion of the scale factor w.r.t. cosmic time `$t$' for both models M1
and M2 with some arbitrary values of the model parameters $k_{1}$, $k_{2}$, $%
\protect\beta $.}
\label{SF}
\end{figure}

\begin{figure}[tbph]
\label{HM2.pdf}
\par
\begin{center}
$%
\begin{array}{c@{\hspace{.1in}}c}
\includegraphics[width=2.7 in, height=2.3 in]{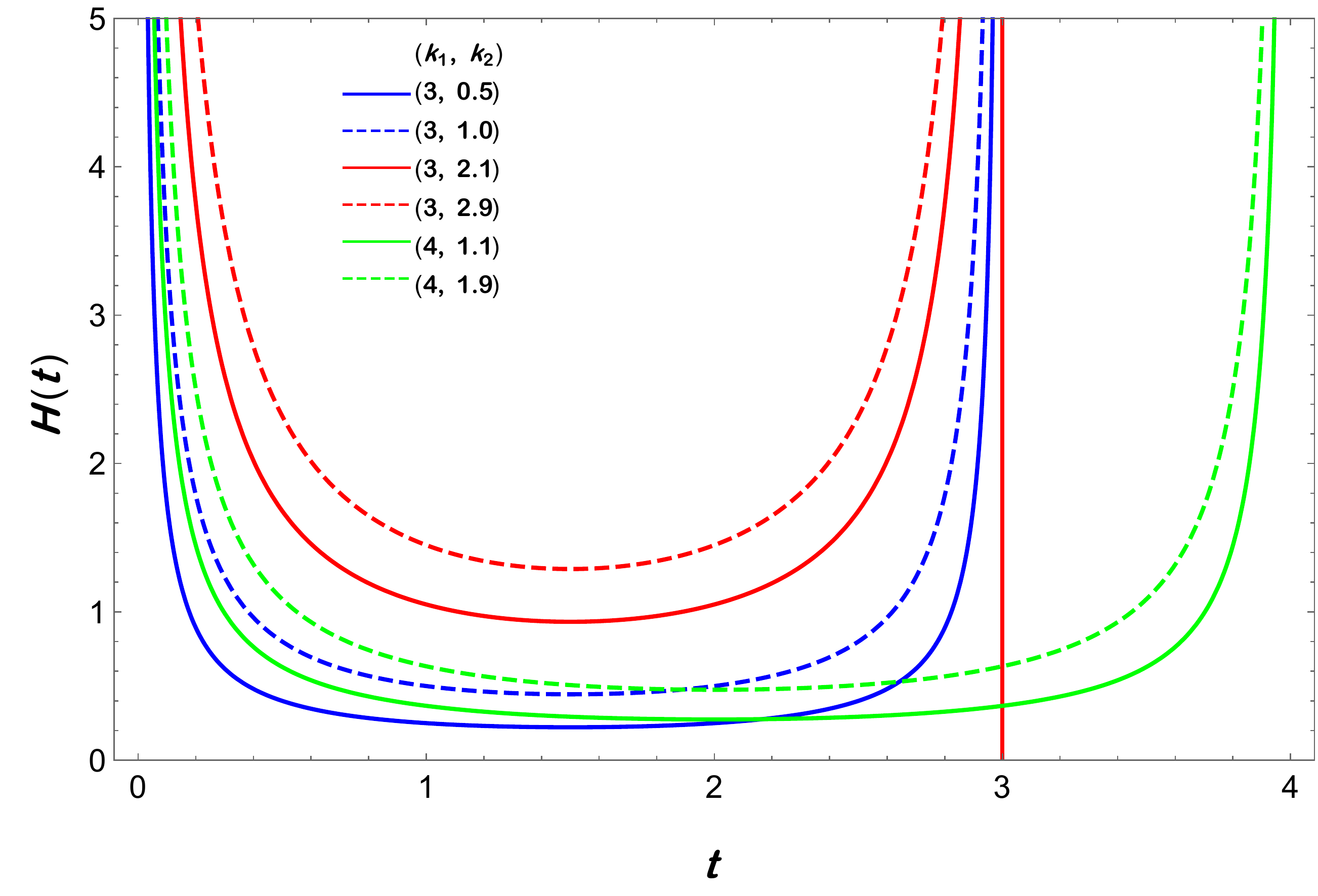} & %
\includegraphics[width=2.7 in, height=2.3 in]{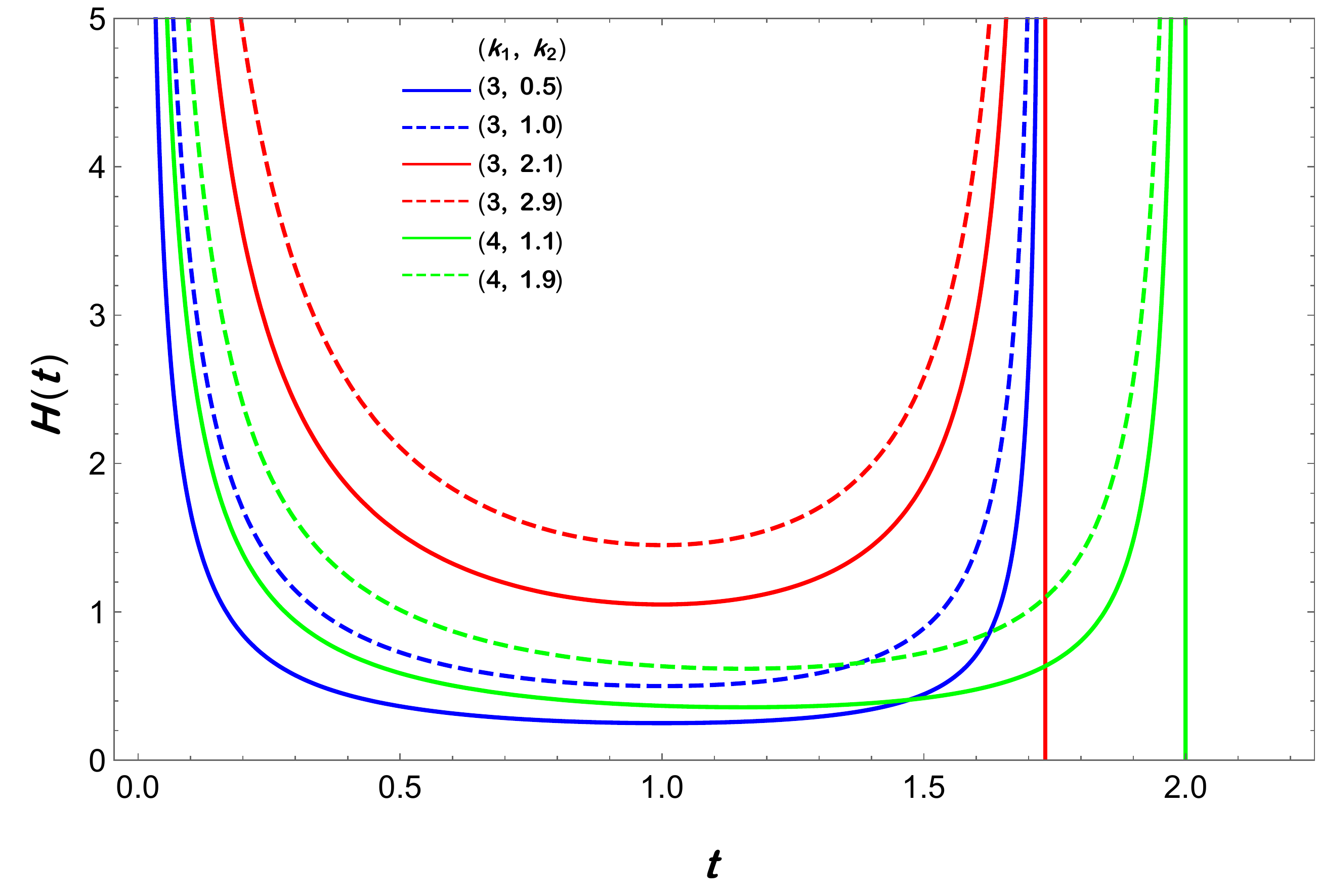} \\ 
\mbox (a) & \mbox (b)%
\end{array}
$%
\end{center}
\caption{ Figures (a) and (b) respectively show rough sketches of the
evolotion of the Hubble parameter w.r.t. cosmic time `$t$' for both models
M1 and M2 with some arbitrary values of the model parameters $k_{1}$, $k_{2}$%
, $\protect\beta $.}
\label{Ht}
\end{figure}

One can see, for both the models M1 and M2, the Hubble parameter and scale
factor both diverge in finite time and show a big rip singularity in near
future at $t=t_{s}=k_{1}$ for model M1 and at $t=t_{s}=\sqrt{k_{1}}$ for
model M2. The phase transition occurs at time $t_{tr}=\frac{k_{1}-k_{2}}{2}$
for model M1 and at time $t_{tr}=\sqrt{\frac{k_{1}-k_{2}}{3}}$ for model M2
and suggest that $k_{1}$ must be greater than $k_{2}$. With some suitable
choice of model parameters $k_{1}$, $k_{2}$ and $\beta $, rough sketches for
the time-evolution of scale factor (SF) and the Hubble parameter (HP) are
made and are shown graphically in the figures FIG. \ref{SF} and FIG. \ref{Ht}
respectively showing that $a(t)$ diverges in finite time and the $H(t)$
becomes asymptotic showing big rip in near future.

\qquad In order to check the consistency of the theoretical models obtained
here with the observations, some available datasets are used in the next
section. The model parameters are constrained through these datasets.

\section{Observational constraints}

\qquad Three datasets are considered here for our analysis namely Hubble
datasets ($Hz$), Type Ia supernovae datasets ($SN$) and Baryon Acoustic
Oscillations datasets ($BAO$). The detailed datasets and the method used are
explained below.

In the study of late-time Universe and the observational studies, it is
convenient to express all the cosmological parameters as functions of
redshift $z$. As the cosmological parameters here are functions of cosmic
time $t$, the time-redshift relationship must be established. The $t$-$z$
relations are obtained as: 
\begin{equation}
t(z)=k_{1}\left[ 1+\left\{ \beta \left( 1+z\right) \right\} ^{\frac{k_{1}}{%
k_{2}}}\right] ^{-1},  \label{tzM1}
\end{equation}%
for model M1 and%
\begin{equation}
t(z)=\sqrt{k_{1}}\left[ 1+\left\{ \beta \left( 1+z\right) \right\} ^{2\frac{%
k_{1}}{k_{2}}}\right] ^{-\frac{1}{2}}  \label{tzM2}
\end{equation}%
for model M2. The above expressions (\ref{tzM1}) and (\ref{tzM2}) contain
three parameters $\beta $, $k_{1}$ and $k_{2}$ but actually two model
parameters are sufficient to describe these models by taking $\frac{k_{1}}{%
k_{2}}=\alpha $ which is also beneficial for further analysis and numerical
computations for which the expressions for the Hubble parameter for both the
models M1 and M2 are written in terms of redshift $z$ as follows:

\begin{equation}
H(z)=H_{0}\left( 1+\beta ^{\alpha }\right) ^{-2}(1+z)^{-\alpha }\left[
1+\left\{ \beta \left( 1+z\right) \right\} ^{\alpha }\right] ^{2},
\label{HM1}
\end{equation}%
for model M1 and

\begin{equation}
H(z)=H_{0}\left( 1+\beta ^{2\alpha }\right) ^{-\frac{3}{2}}(1+z)^{-2\alpha }%
\left[ 1+\left\{ \beta \left( 1+z\right) \right\} ^{2\alpha }\right] ^{\frac{%
3}{2}}  \label{HM2}
\end{equation}%
for model M2. The different datasets are described below.

\subsection{H(z) datasets}

\qquad It is well known that the Hubble parameter ($H=\frac{\dot{a}}{a}$)
directly probes the expansion history of the Universe where $\dot{a}$ is the
rate of change of the scale factor $a$ of the Universe. Hubble parameter is
also related to the differential redshift as, $H(z)$ $=-\frac{1}{1+z}\frac{dz%
}{dt}$, where $dz$ is obtained from the spectroscopic surveys and so a
measurement of $dt$ provides the Hubble parameter which will be independent
of the model. In fact, two methods are generally used to measure the Hubble
parameter values $H(z)$ at certain redshift and are extraction of $H(z)$
from line-of-sight BAO data and differential age method \cite{H1}-\cite{H19}
estimating $H(z)$. Here, in this paper, an updated list of $57$ data points
are used as listed in Table-2 out of which $31$ data points measured with DA
method and $26$ data points are obtained with BAO and other methods in the
redshift range $0.07\leqslant z\leqslant 2.42$ \cite{sharov}. Moreover, the
value of $H_{0}$ is taken as prior for our analysis as $H_{0}=67.8$ $%
Km/s/Mpc $ \cite{H0-value}. The chi square function to determine the mean
values of the model parameters $\alpha $ \& $\beta $ (which is equivalent to
the maximum likelihood analysis) is,

\begin{equation}
\chi _{H}^{2}(\alpha ,\beta )=\sum\limits_{i=1}^{28}\frac{%
[H_{th}(z_{i},\alpha ,\beta )-H_{obs}(z_{i})]^{2}}{\sigma _{H(z_{i})}^{2}},
\label{chihz}
\end{equation}%
where, $H_{th}$ is the theoretical, $H_{obs}$ is the observed value and $%
\sigma _{H(z_{i})}$ is the standard error in the observed value of the
Hubble parameter $H$. The $57$ points of Hubble parameter values $H(z)$ with
errors $\sigma _{H}$ from differential age ($31$ points) method and BAO and
other ($26$ points) methods are tabulated in Table-2 with references.

\begin{center}
\begin{tabular}{|c|c|c|c||c|c|c|c||c|c|c|c||c|c|c|c|}
\hline
\multicolumn{16}{|c|}{Table-2: 57 points of $H(z)$ datasets} \\ \hline
\multicolumn{8}{|c||}{31 points from DA method} & \multicolumn{8}{|c|}{26
points from BAO \& other method} \\ \hline
$z$ & $H(z)$ & $\sigma _{H}$ & Ref. & $z$ & $H(z)$ & $\sigma _{H}$ & Ref. & $%
z$ & $H(z)$ & $\sigma _{H}$ & Ref. & $z$ & $H(z)$ & $\sigma _{H}$ & Ref. \\ 
\hline
$0.070$ & $69$ & $19.6$ & \cite{H1} & $0.4783$ & $80$ & $99$ & \cite{H5} & $%
0.24$ & $79.69$ & $2.99$ & \cite{H8} & $0.52$ & $94.35$ & $2.64$ & \cite{H10}
\\ \hline
$0.90$ & $69$ & $12$ & \cite{H2} & $0.480$ & $97$ & $62$ & \cite{H1} & $0.30$
& $81.7$ & $6.22$ & \cite{H9} & $0.56$ & $93.34$ & $2.3$ & \cite{H10} \\ 
\hline
$0.120$ & $68.6$ & $26.2$ & \cite{H1} & $0.593$ & $104$ & $13$ & \cite{H3} & 
$0.31$ & $78.18$ & $4.74$ & \cite{H10} & $0.57$ & $87.6$ & $7.8$ & \cite{H14}
\\ \hline
$0.170$ & $83$ & $8$ & \cite{H2} & $0.6797$ & $92$ & $8$ & \cite{H3} & $0.34$
& $83.8$ & $3.66$ & \cite{H8} & $0.57$ & $96.8$ & $3.4$ & \cite{H15} \\ 
\hline
$0.1791$ & $75$ & $4$ & \cite{H3} & $0.7812$ & $105$ & $12$ & \cite{H3} & $%
0.35$ & $82.7$ & $9.1$ & \cite{H11} & $0.59$ & $98.48$ & $3.18$ & \cite{H10}
\\ \hline
$0.1993$ & $75$ & $5$ & \cite{H3} & $0.8754$ & $125$ & $17$ & \cite{H3} & $%
0.36$ & $79.94$ & $3.38$ & \cite{H10} & $0.60$ & $87.9$ & $6.1$ & \cite{H13}
\\ \hline
$0.200$ & $72.9$ & $29.6$ & \cite{H4} & $0.880$ & $90$ & $40$ & \cite{H1} & $%
0.38$ & $81.5$ & $1.9$ & \cite{H12} & $0.61$ & $97.3$ & $2.1$ & \cite{H12}
\\ \hline
$0.270$ & $77$ & $14$ & \cite{H2} & $0.900$ & $117$ & $23$ & \cite{H2} & $%
0.40$ & $82.04$ & $2.03$ & \cite{H10} & $0.64$ & $98.82$ & $2.98$ & \cite%
{H10} \\ \hline
$0.280$ & $88.8$ & $36.6$ & \cite{H4} & $1.037$ & $154$ & $20$ & \cite{H3} & 
$0.43$ & $86.45$ & $3.97$ & \cite{H8} & $0.73$ & $97.3$ & $7.0$ & \cite{H13}
\\ \hline
$0.3519$ & $83$ & $14$ & \cite{H3} & $1.300$ & $168$ & $17$ & \cite{H2} & $%
0.44$ & $82.6$ & $7.8$ & \cite{H13} & $2.30$ & $224$ & $8.6$ & \cite{H16} \\ 
\hline
$0.3802$ & $83$ & $13.5$ & \cite{H5} & $1.363$ & $160$ & $33.6$ & \cite{H7}
& $0.44$ & $84.81$ & $1.83$ & \cite{H10} & $2.33$ & $224$ & $8$ & \cite{H17}
\\ \hline
$0.400$ & $95$ & $17$ & \cite{H2} & $1.430$ & $177$ & $18$ & \cite{H2} & $%
0.48$ & $87.79$ & $2.03$ & \cite{H10} & $2.34$ & $222$ & $8.5$ & \cite{H18}
\\ \hline
$0.4004$ & $77$ & $10.2$ & \cite{H5} & $1.530$ & $140$ & $14$ & \cite{H2} & $%
0.51$ & $90.4$ & $1.9$ & \cite{H12} & $2.36$ & $226$ & $9.3$ & \cite{H19} \\ 
\hline
$0.4247$ & $87.1$ & $11.2$ & \cite{H5} & $1.750$ & $202$ & $40$ & \cite{H2}
&  &  &  &  &  &  &  &  \\ \hline
$0.4497$ & $92.8$ & $12.9$ & \cite{H5} & $1.965$ & $186.5$ & $50.4$ & \cite%
{H7} &  &  &  &  &  &  &  &  \\ \hline
$0.470$ & $89$ & $34$ & \cite{H6} &  &  &  &  &  &  &  &  &  &  &  &  \\ 
\hline
\end{tabular}
\end{center}

\subsection{SN Ia datasets}

\qquad The first indication for the accelerating expansion of the Universe
is due to observations of supernovae of type \textit{Ia}. Since then,
several new SN \textit{Ia} datasets have been published. In this analysis,
the Union $2.1$ compilation supernovae datasets is considered containing $%
580 $ points from \cite{SNeIa} which provides the estimated values of the
distance moduli $\mu _{i}=\mu _{i}^{obs}$ for a particular redshift $z_{i}$
in the interval $0<z_{i}\leq 1.41$. The model parameters of the models are
to be fitted with, comparing the observed $\mu _{i}^{obs}$ value to the
theoretical $\mu _{i}^{th}$ value of the distance moduli which are the
logarithms $\mu _{i}^{th}=\mu (D_{L})=m-M=5\log _{10}(D_{L})+\mu _{0}$,
where $m$ and $M$ are the apparent and absolute magnitudes and $\mu
_{0}=5\log \left( H_{0}^{-1}/Mpc\right) +25$ is the nuisance parameter that
has been marginalized. The luminosity distance is defined by, 
\begin{eqnarray*}
D_{l}(z) &=&\frac{c(1+z)}{H_{0}}S_{k}\left( H_{0}\int_{0}^{z}\frac{1}{%
H(z^{\ast })}dz^{\ast }\right) , \\
\text{where }S_{k}(x) &=&\left\{ 
\begin{array}{c}
\sinh (x\sqrt{\Omega _{k}})/\Omega _{k}\text{, }\Omega _{k}>0 \\ 
x\text{, \ \ \ \ \ \ \ \ \ \ \ \ \ \ \ \ \ \ \ \ \ \ \ }\Omega _{k}=0 \\ 
\sin x\sqrt{\left\vert \Omega _{k}\right\vert })/\left\vert \Omega
_{k}\right\vert \text{, }\Omega _{k}<0%
\end{array}%
\right. .
\end{eqnarray*}%
Here, $\Omega _{k}=0$ (flat space-time). For our cosmological models M1 and
M2 with theoretical value $H(z)$ which are depending on the model parameters 
$\alpha $ \& $\beta $, the distance $D_{L}(z)$ is calculated and the
corresponding chi square function measuring differences between the SN 
\textit{Ia} observational data and values predicted by the models is given
by,

\begin{equation}
\chi _{SN}^{2}(\mu _{0},\alpha ,\beta )=\sum\limits_{i=1}^{580}\frac{[\mu
^{th}(\mu _{0},z_{i},\alpha ,\beta )-\mu ^{obs}(z_{i})]^{2}}{\sigma _{\mu
(z_{i})}^{2}},  \label{chisn}
\end{equation}%
$\sigma _{\mu (z_{i})}^{2}$ is the standard error in the observed value.
Following \cite{perivolar} after marginalizing $\mu _{0}$, the chi square
function is written as,

$\qquad \qquad \qquad \qquad \qquad \qquad \qquad \qquad \chi
_{SN}^{2}(\alpha ,\beta )=A(\alpha ,\beta )-[B(\alpha ,\beta )]^{2}/C(\alpha
,\beta )$

where

$\qquad \qquad \qquad \qquad \qquad \qquad \qquad \qquad A(\alpha ,\beta
)=\sum\limits_{i=1}^{580}\frac{[\mu ^{th}(\mu _{0}=0,z_{i},\alpha ,\beta
)-\mu ^{obs}(z_{i})]^{2}}{\sigma _{\mu (z_{i})}^{2}},$

$\qquad \qquad \qquad \qquad \qquad \qquad \qquad \qquad B(\alpha ,\beta
)=\sum\limits_{i=1}^{580}\frac{[\mu ^{th}(\mu _{0}=0,z_{i},\alpha ,\beta
)-\mu ^{obs}(z_{i})]^{2}}{\sigma _{\mu (z_{i})}^{2}},$

$\qquad \qquad \qquad \qquad \qquad \qquad \qquad \qquad C(\alpha ,\beta
)=\sum\limits_{i=1}^{580}\frac{1}{\sigma _{\mu (z_{i})}^{2}}$.

\subsection{BAO datasets}

\qquad Baryonic acoustic oscillations is an analysis dealing with the early
Universe. It is known that the early Universe filled with baryons, photons
and dark matter. Moreover, baryons and photons together act as single fluid
(coupled tightly through the Thompson scattering) and can not collapse under
gravity rather oscillate due to the large pressure of photons. These
oscillations are termed a Baryonic acoustic oscillations (BAO). The
characteristic scale of BAO is governed by the sound horizon $r_{s}$ at the
photon decoupling epoch $z_{\ast }$ is given as, 
\begin{equation*}
r_{s}(z_{\ast })=\frac{c}{\sqrt{3}}\int_{0}^{\frac{1}{1+z_{\ast }}}\frac{da}{%
a^{2}H(a)\sqrt{1+(3\Omega _{0b}/4\Omega _{0\gamma })a}},
\end{equation*}%
where $\Omega _{0b}$ stands for the baryon density and $\Omega _{0\gamma }$
stands for the photon density at present time.

The BAO sound horizon scale is also used to derive the angular diameter
distance $D_{A}$ and the Hubble expansion rate $H$ as a function of $z$. If $%
\triangle \theta $ be the measured angular separation of the BAO feature in
the 2 point correlation function of the galaxy distribution on the sky and
the $\triangle z$ be the measured redshift separation of the BAO feature in
the 2 point correlation function along the line of sight then,

$\triangle \theta =\frac{r_{s}}{d_{A}(z)}$ where $d_{A}(z)=\int_{0}^{z}\frac{%
dz^{\prime }}{H(z^{\prime })}$ and $\triangle z=H(z)r_{s}$.

In this work, BAO datasets of $d_{A}(z_{\ast })/D_{V}(z_{BAO})$ from the
references \cite{BAO1, BAO2, BAO3, BAO4, BAO5, BAO6} is considered where the
photon decoupling redshift is $z_{\ast }\approx 1091$ and $d_{A}(z)$ is the
co-moving angular diameter distance and $D_{V}(z)=\left(
d_{A}(z)^{2}z/H(z)\right) ^{1/3}$ is the dilation scale. The data used for
this analysis is given in the Table-3

\begin{center}
\begin{tabular}{|c|c|c|c|c|c|c|}
\hline
\multicolumn{7}{|c|}{Table-3: Values of $d_{A}(z_{\ast })/D_{V}(z_{BAO})$
for distinct values of $z_{BAO}$} \\ \hline
$z_{BAO}$ & $0.106$ & $0.2$ & $0.35$ & $0.44$ & $0.6$ & $0.73$ \\ \hline
$\frac{d_{A}(z_{\ast })}{D_{V}(z_{BAO})}$ & $30.95\pm 1.46$ & $17.55\pm 0.60$
& $10.11\pm 0.37$ & $8.44\pm 0.67$ & $6.69\pm 0.33$ & $5.45\pm 0.31$ \\ 
\hline
\end{tabular}
\end{center}

\qquad The chi square function for BAO is given by \cite{BAO6} 
\begin{equation}
\chi _{BAO}^{2}=X^{T}C^{-1}X\,,  \label{chibao}
\end{equation}%
where 
\begin{equation*}
X=\left( 
\begin{array}{c}
\frac{d_{A}(z_{\star })}{D_{V}(0.106)}-30.95 \\ 
\frac{d_{A}(z_{\star })}{D_{V}(0.2)}-17.55 \\ 
\frac{d_{A}(z_{\star })}{D_{V}(0.35)}-10.11 \\ 
\frac{d_{A}(z_{\star })}{D_{V}(0.44)}-8.44 \\ 
\frac{d_{A}(z_{\star })}{D_{V}(0.6)}-6.69 \\ 
\frac{d_{A}(z_{\star })}{D_{V}(0.73)}-5.45%
\end{array}%
\right) \,,
\end{equation*}%
and $C^{-1}$ is the inverse covariance matrix defined in \cite{BAO6}. 
\begin{equation*}
C^{-1}=\left( 
\begin{array}{cccccc}
0.48435 & -0.101383 & -0.164945 & -0.0305703 & -0.097874 & -0.106738 \\ 
-0.101383 & 3.2882 & -2.45497 & -0.0787898 & -0.252254 & -0.2751 \\ 
-0.164945 & -2.454987 & 9.55916 & -0.128187 & -0.410404 & -0.447574 \\ 
-0.0305703 & -0.0787898 & -0.128187 & 2.78728 & -2.75632 & 1.16437 \\ 
-0.097874 & -0.252254 & -0.410404 & -2.75632 & 14.9245 & -7.32441 \\ 
-0.106738 & -0.2751 & -0.447574 & 1.16437 & -7.32441 & 14.5022%
\end{array}%
\right) \,
\end{equation*}

With the above samples of Hubble ($Hz$), supernovae of type \textit{Ia} ($SN$%
) and baryon acoustic oscillations ($BAO$) datasets, the chi square
functions (\ref{chihz}), (\ref{chisn}) and (\ref{chibao}) are minimized to
get the average values of the model parameters $\alpha $ \& $\beta $. The
maximum likelihood contours for the model parameters $\alpha $ \& $\beta $
are shown in the following figures FIG. \ref{Hz}, FIG. \ref{Hz-Sn}, FIG. \ref%
{Sn-Bao}, and FIG. \ref{Hz-Sn-Bao} for independent $Hz$ datasets and
combined $Hz+SN$, $SN+BAO$ and $Hz+SN+BAO$ datasets respectively with $1$-$%
\sigma $, $2$-$\sigma $ and $3$-$\sigma $ error contours in the $\alpha $-$%
\beta $ plane.

\begin{figure}[tbph]
\label{cont-hz-M2-Red.pdf}
\par
\begin{center}
$%
\begin{array}{c@{\hspace{.1in}}c}
\includegraphics[width=2.7 in, height=2.3 in]{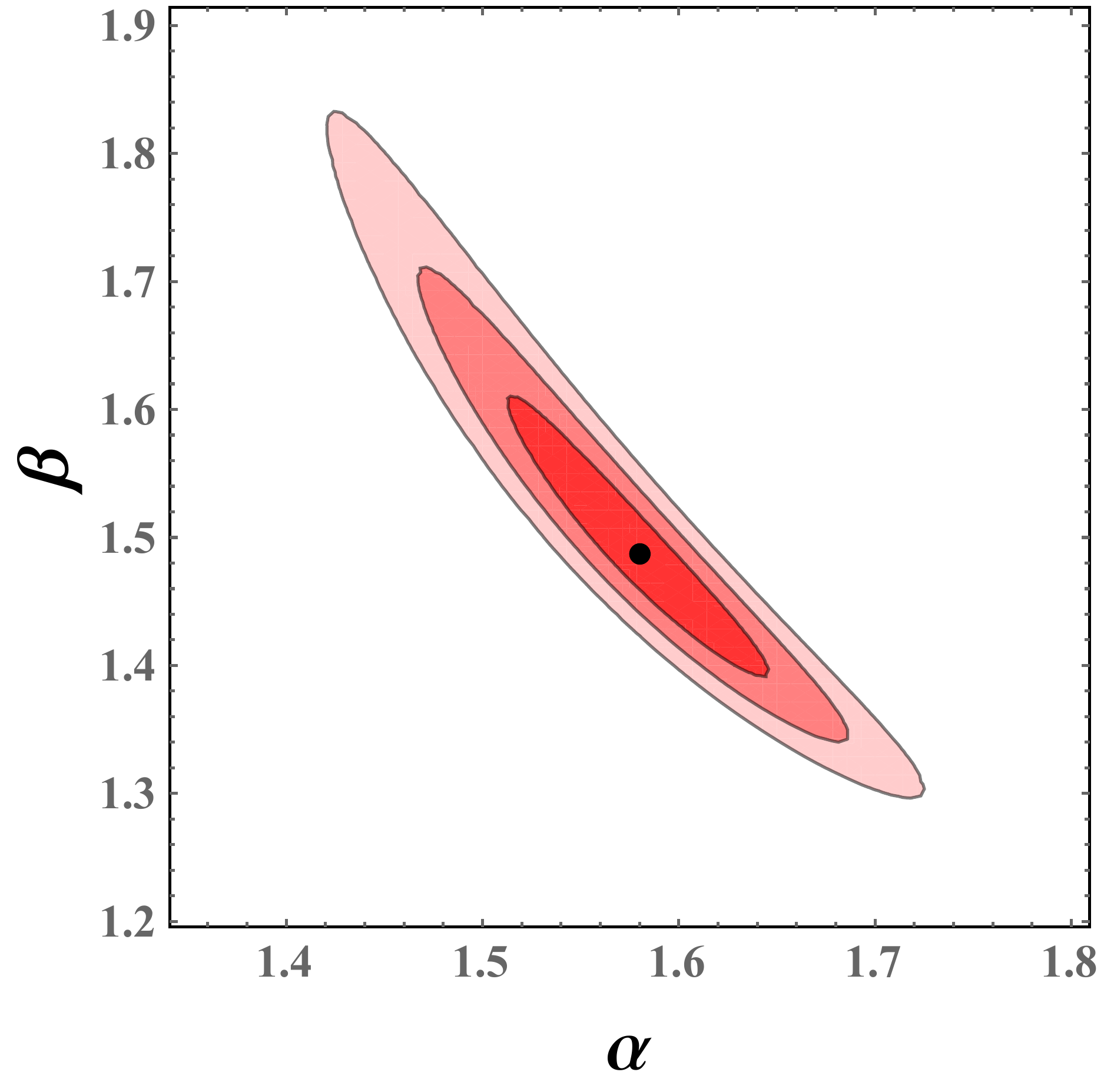} & %
\includegraphics[width=2.7 in, height=2.3 in]{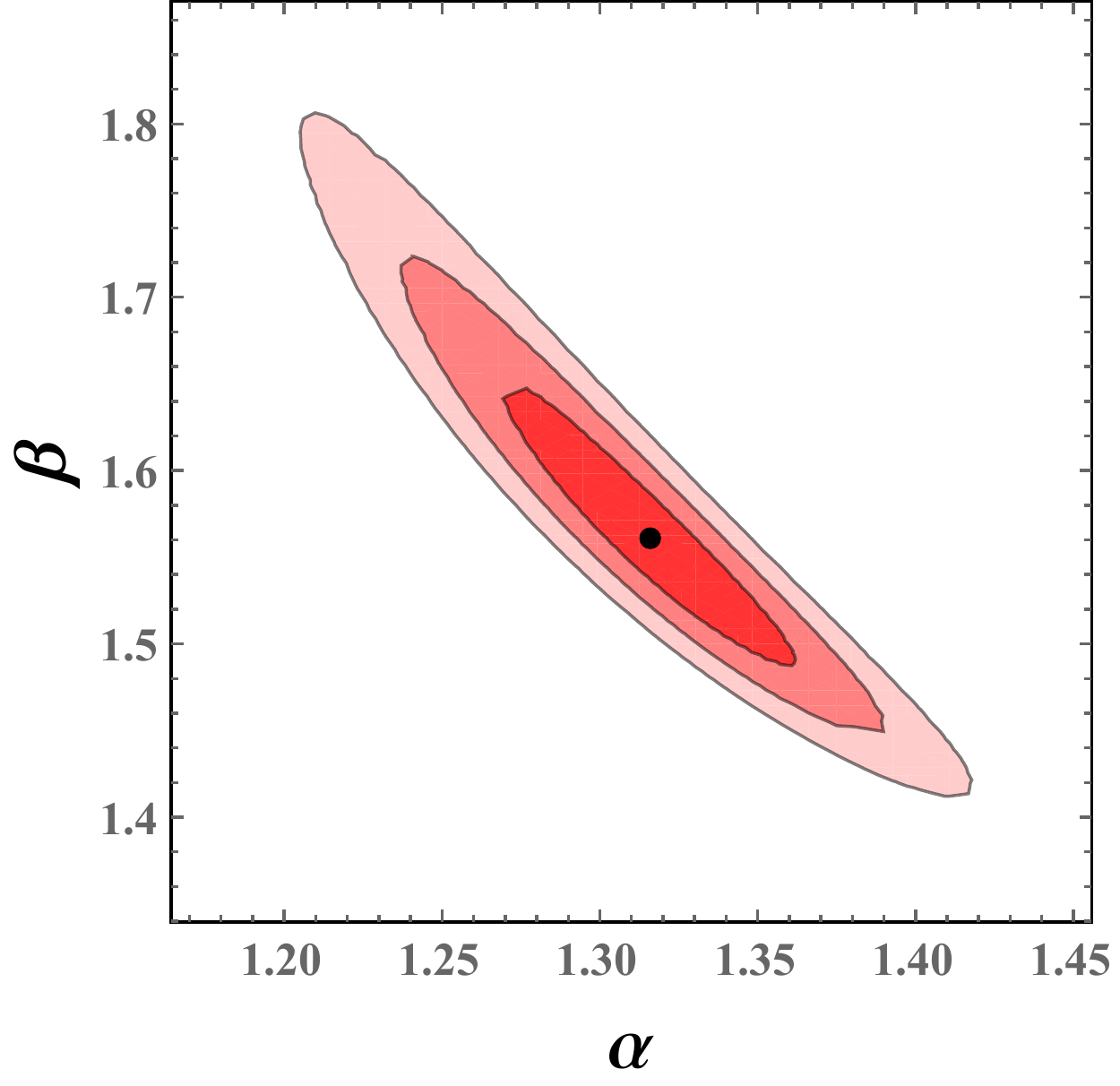} \\ 
\mbox (a) & \mbox (b)%
\end{array}
$%
\end{center}
\caption{ Figures (a) and (b) are contour plots for Hubble datasets ($Hz$)
for models M1 and M2 respectvely.}
\label{Hz}
\end{figure}

\begin{figure}[tbp]
\label{cont-hz+sn-M2-Yellow.pdf}
\par
\begin{center}
$%
\begin{array}{c@{\hspace{.1in}}c}
\includegraphics[width=2.7 in, height=2.3 in]{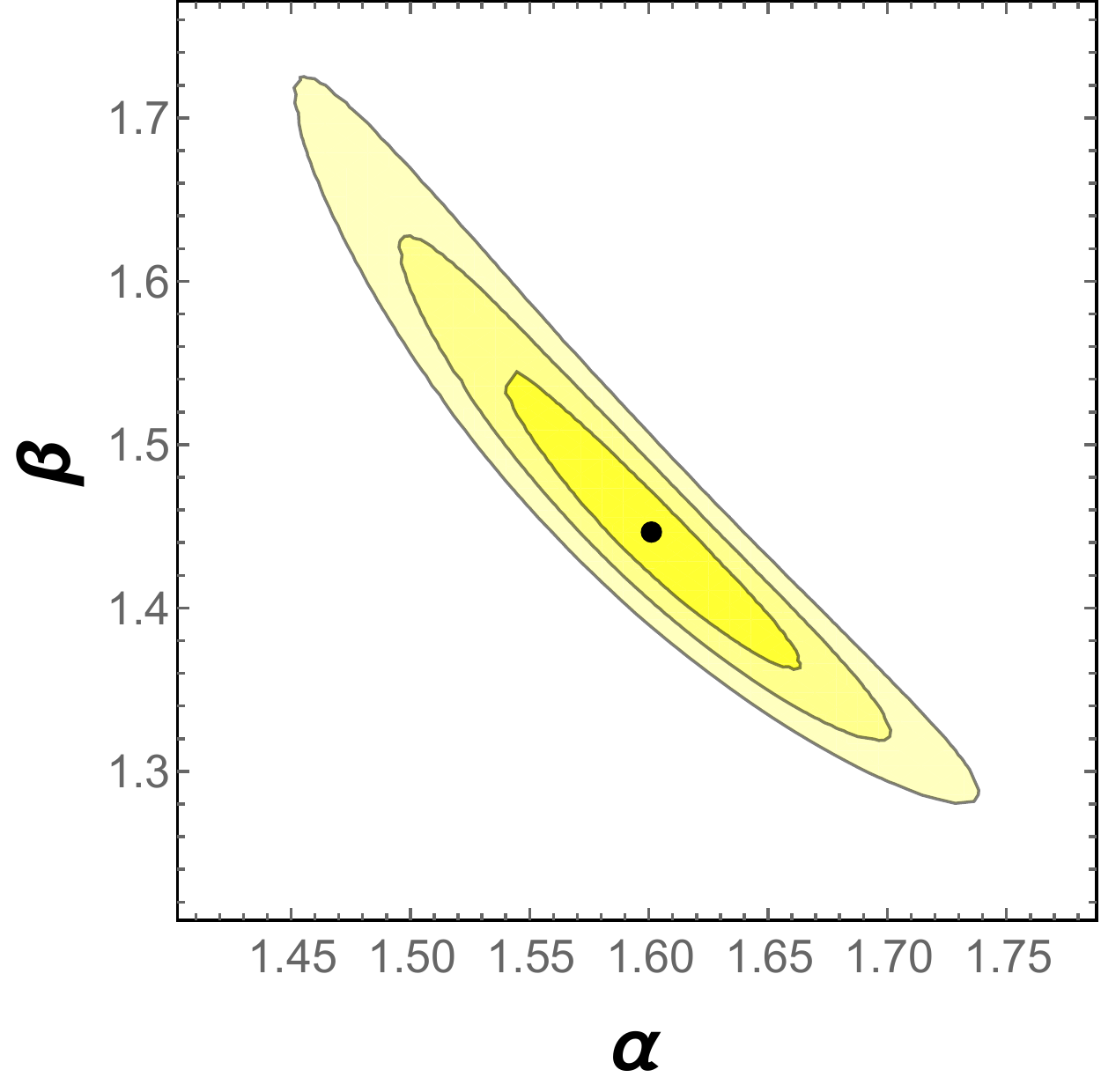} & %
\includegraphics[width=2.7 in, height=2.3 in]{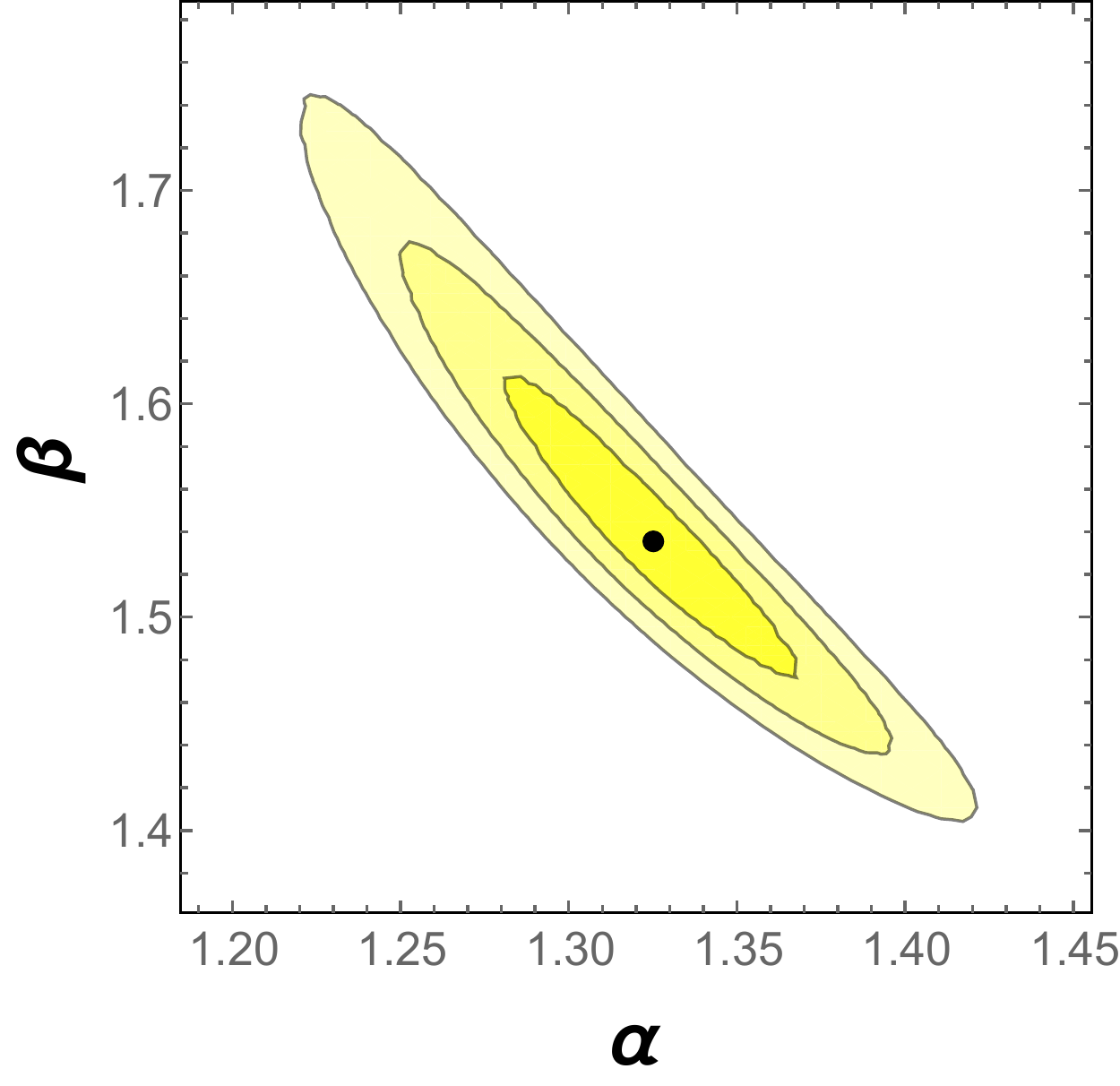} \\ 
\mbox (a) & \mbox (b)%
\end{array}
$%
\end{center}
\caption{ Figures (a) and (b) are contour plots for combined $Hz+SN$
datasets for models M1 and M2 respectvely.}
\label{Hz-Sn}
\end{figure}

\begin{figure}[tbp]
\label{cont-sn+bao-M2-Green.pdf}
\par
\begin{center}
$%
\begin{array}{c@{\hspace{.1in}}c}
\includegraphics[width=2.7 in, height=2.3 in]{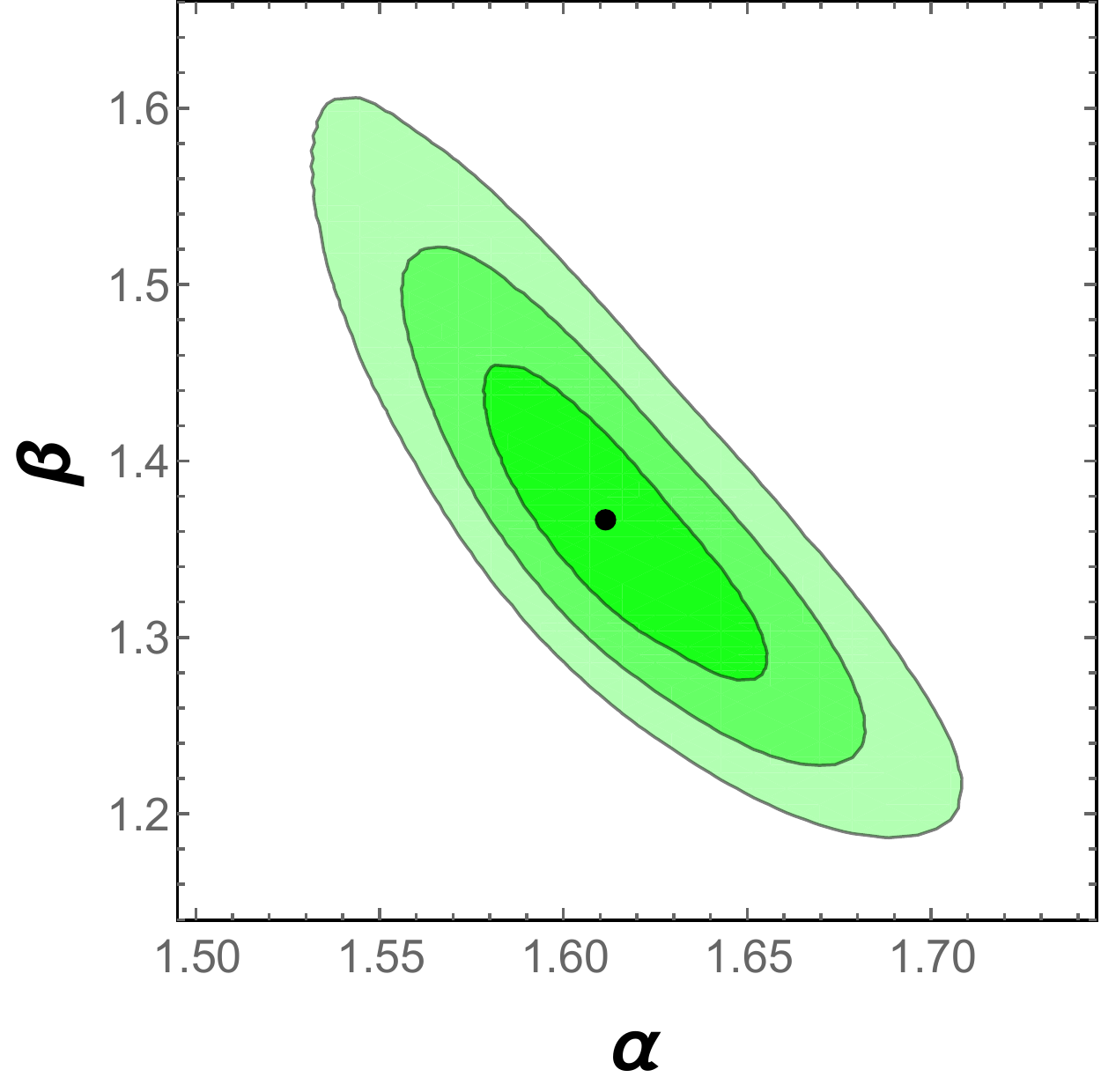} & %
\includegraphics[width=2.7 in, height=2.3 in]{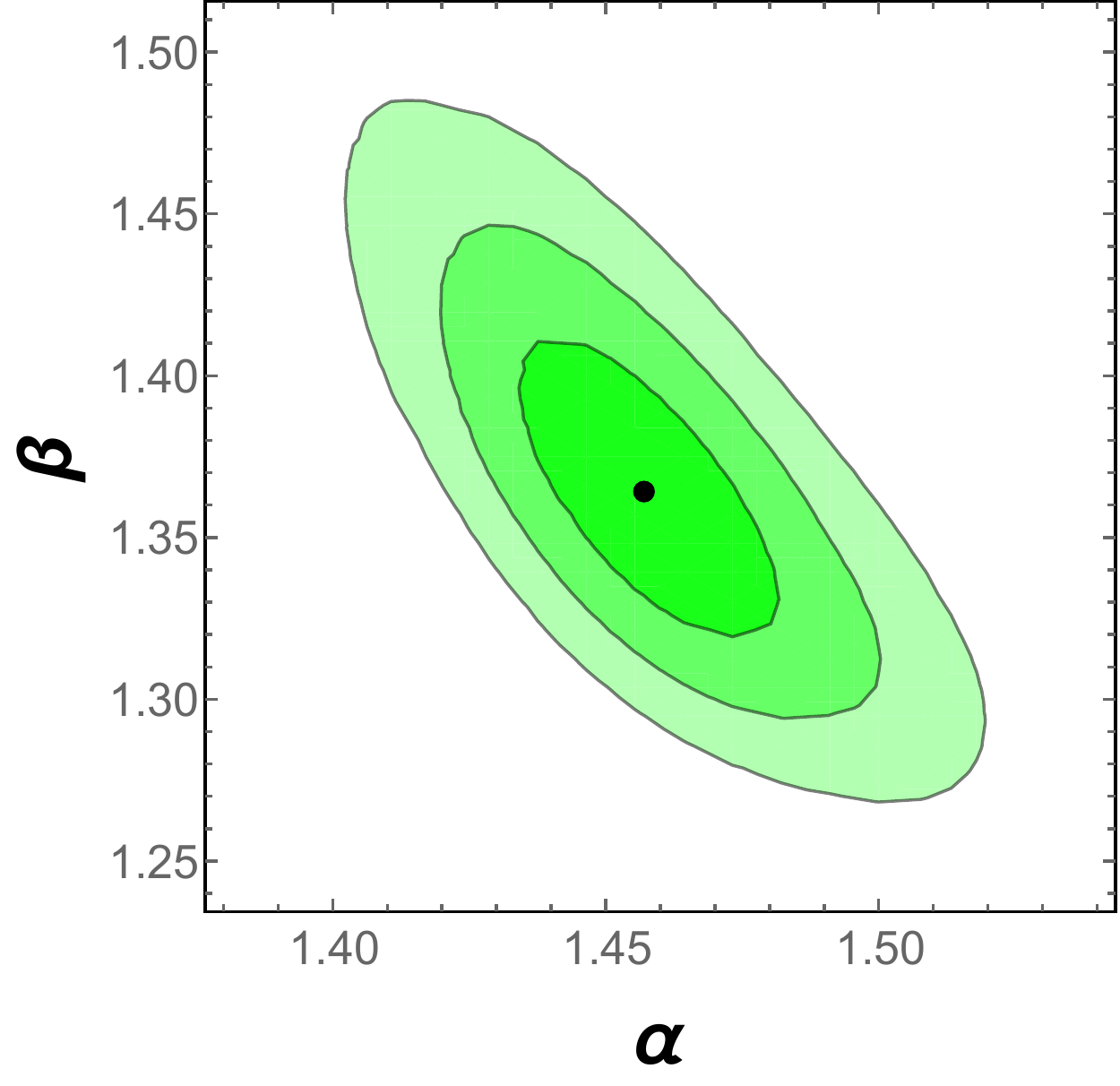} \\ 
\mbox (a) & \mbox (b)%
\end{array}
$%
\end{center}
\caption{ Figures (a) and (b) are contour plots for combibed $SN+BAO$
datasets for models M1 and M2 respectvely.}
\label{Sn-Bao}
\end{figure}

\begin{figure}[tbp]
\label{cont-hz+sn+bao-M2-Blue.pdf}
\par
\begin{center}
$%
\begin{array}{c@{\hspace{.1in}}c}
\includegraphics[width=2.7 in, height=2.3 in]{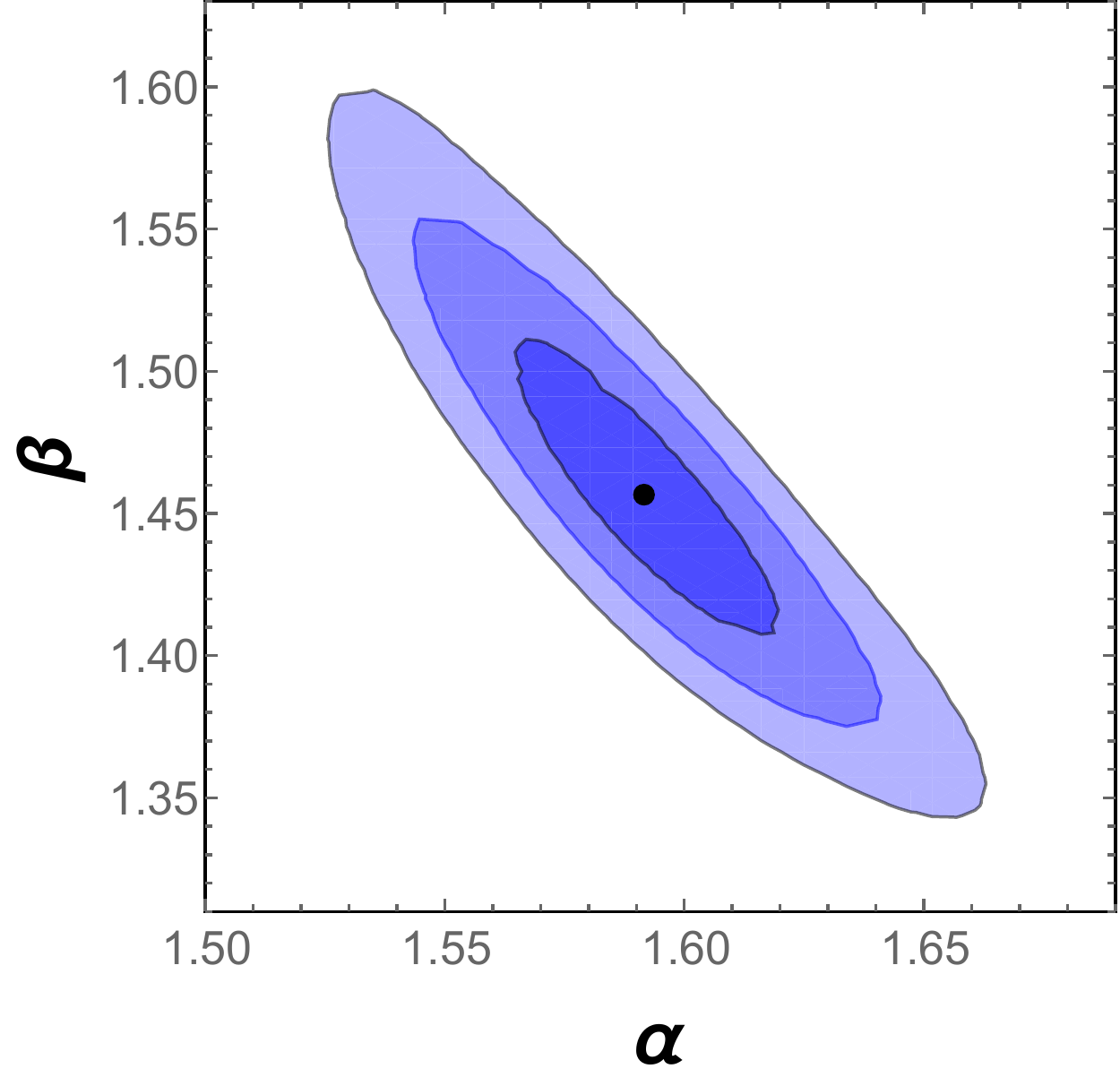} & %
\includegraphics[width=2.7 in, height=2.3 in]{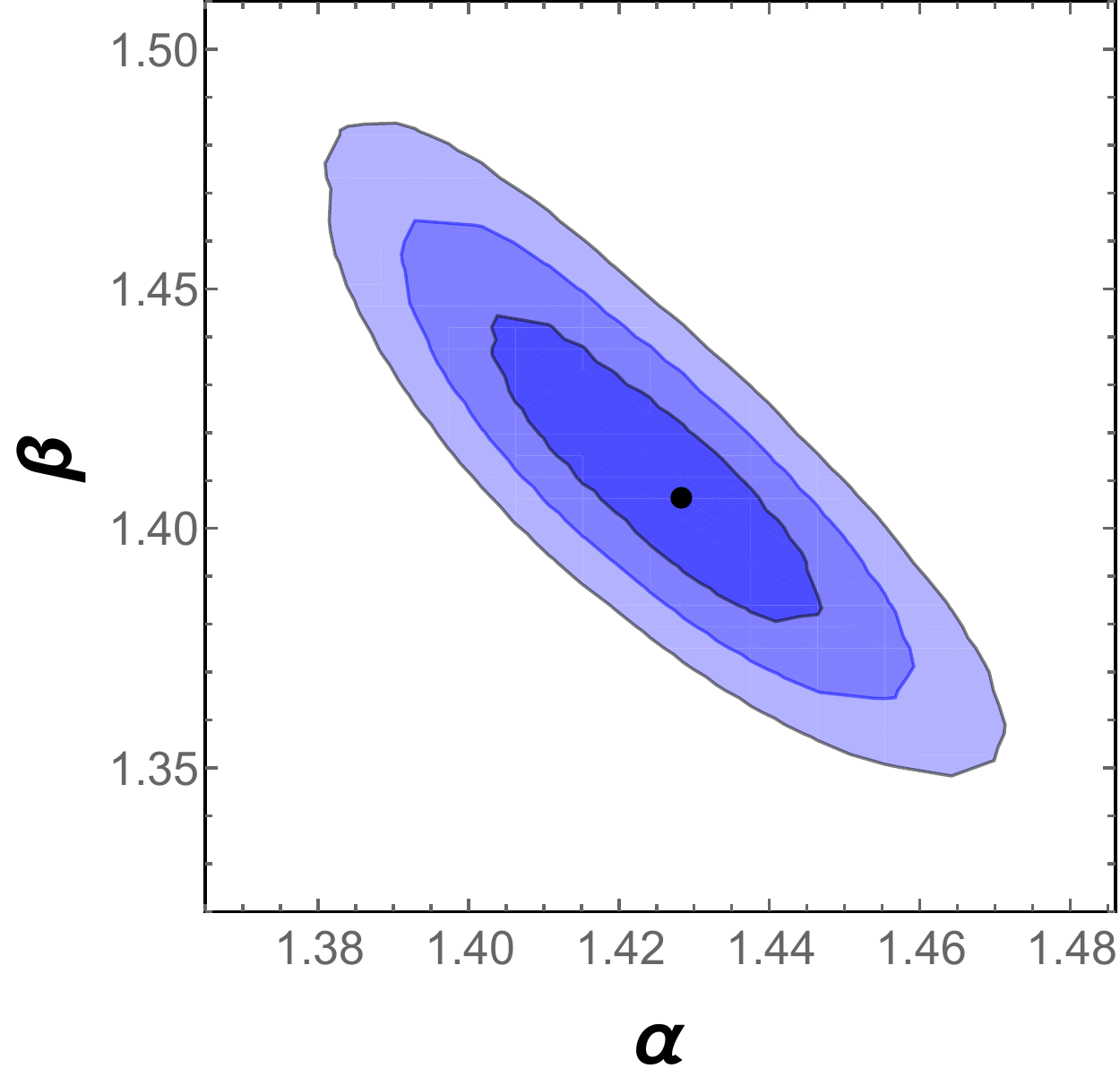} \\ 
\mbox (a) & \mbox (b)%
\end{array}
$%
\end{center}
\caption{ Figures (a) and (b) are contour plots for combined $Hz+SN+BAO$
datasets for models M1 and M2 respectvely.}
\label{Hz-Sn-Bao}
\end{figure}

The average mean values (constrained values) of the model parameters and the
minimum chi square values are tabulated in Table-4 for independent $Hz$
datasets and combined $Hz+SN$, $SN+BAO$ and $Hz+SN+BAO$ datasets.

\begin{center}
\begin{tabular}{|l|c|c|c|c|c|}
\hline
\multicolumn{6}{|c|}{Table-4: Constrained values of model parameters and chi
square values} \\ \hline
Datasets & Models & $\alpha $ & $\beta $ & $\chi _{\min }^{2}$ & $\chi
^{2}/dof$ \\ \hline
$H(z)$ & 
\begin{tabular}{|c|}
\hline
M1 \\ \hline
M2 \\ \hline
\end{tabular}
& 
\begin{tabular}{|c|}
\hline
$1.58064$ \\ \hline
$1.31611$ \\ \hline
\end{tabular}
& 
\begin{tabular}{|c|}
\hline
$1.48729$ \\ \hline
$1.56124$ \\ \hline
\end{tabular}
& 
\begin{tabular}{|c|}
\hline
$31.329529$ \\ \hline
$29.972660$ \\ \hline
\end{tabular}
& 
\begin{tabular}{|c|}
\hline
$0.56962$ \\ \hline
$0.54495$ \\ \hline
\end{tabular}
\\ \hline
$H(z)+SN$ & 
\begin{tabular}{|c|}
\hline
M1 \\ \hline
M2 \\ \hline
\end{tabular}
& 
\begin{tabular}{|c|}
\hline
$1.60094$ \\ \hline
$1.32551$ \\ \hline
\end{tabular}
& 
\begin{tabular}{|c|}
\hline
$1.44572$ \\ \hline
$1.53587$ \\ \hline
\end{tabular}
& 
\begin{tabular}{|c|}
\hline
$596.49325$ \\ \hline
$595.02853$ \\ \hline
\end{tabular}
& 
\begin{tabular}{|c|}
\hline
$0.93935$ \\ \hline
$0.93705$ \\ \hline
\end{tabular}
\\ \hline
$SN+BAO$ & 
\begin{tabular}{|c|}
\hline
M1 \\ \hline
M2 \\ \hline
\end{tabular}
& 
\begin{tabular}{|c|}
\hline
$1.61116$ \\ \hline
$1.45677$ \\ \hline
\end{tabular}
& 
\begin{tabular}{|c|}
\hline
$1.36647$ \\ \hline
$1.36451$ \\ \hline
\end{tabular}
& 
\begin{tabular}{|c|}
\hline
$564.45777$ \\ \hline
$566.44641$ \\ \hline
\end{tabular}
& 
\begin{tabular}{|c|}
\hline
$0.96653$ \\ \hline
$0.96994$ \\ \hline
\end{tabular}
\\ \hline
$H(z)+SN+BAO$ & 
\begin{tabular}{|c|}
\hline
M1 \\ \hline
M2 \\ \hline
\end{tabular}
& 
\begin{tabular}{|c|}
\hline
$1.59173$ \\ \hline
$1.42829$ \\ \hline
\end{tabular}
& 
\begin{tabular}{|c|}
\hline
$1.45678$ \\ \hline
$1.40637$ \\ \hline
\end{tabular}
& 
\begin{tabular}{|c|}
\hline
$599.07805$ \\ \hline
$614.40132$ \\ \hline
\end{tabular}
& 
\begin{tabular}{|c|}
\hline
$0.93459$ \\ \hline
$0.95850$ \\ \hline
\end{tabular}
\\ \hline
\end{tabular}
\end{center}

The error bar plots of the $57$ points of $Hz$ and $580$ points of Union $%
2.1 $ compilation datasets are plotted and shown in the following figures
FIG. \ref{error-Hz} and FIG. \ref{error-Sn}, using the constrained values of
the model parameters as in Table-4 for both the models M1 and M2.

\begin{figure}[tbph]
\label{Error-HP-M2.pdf}
\par
\begin{center}
$%
\begin{array}{c@{\hspace{.1in}}c}
\includegraphics[width=2.7 in, height=2.3
in]{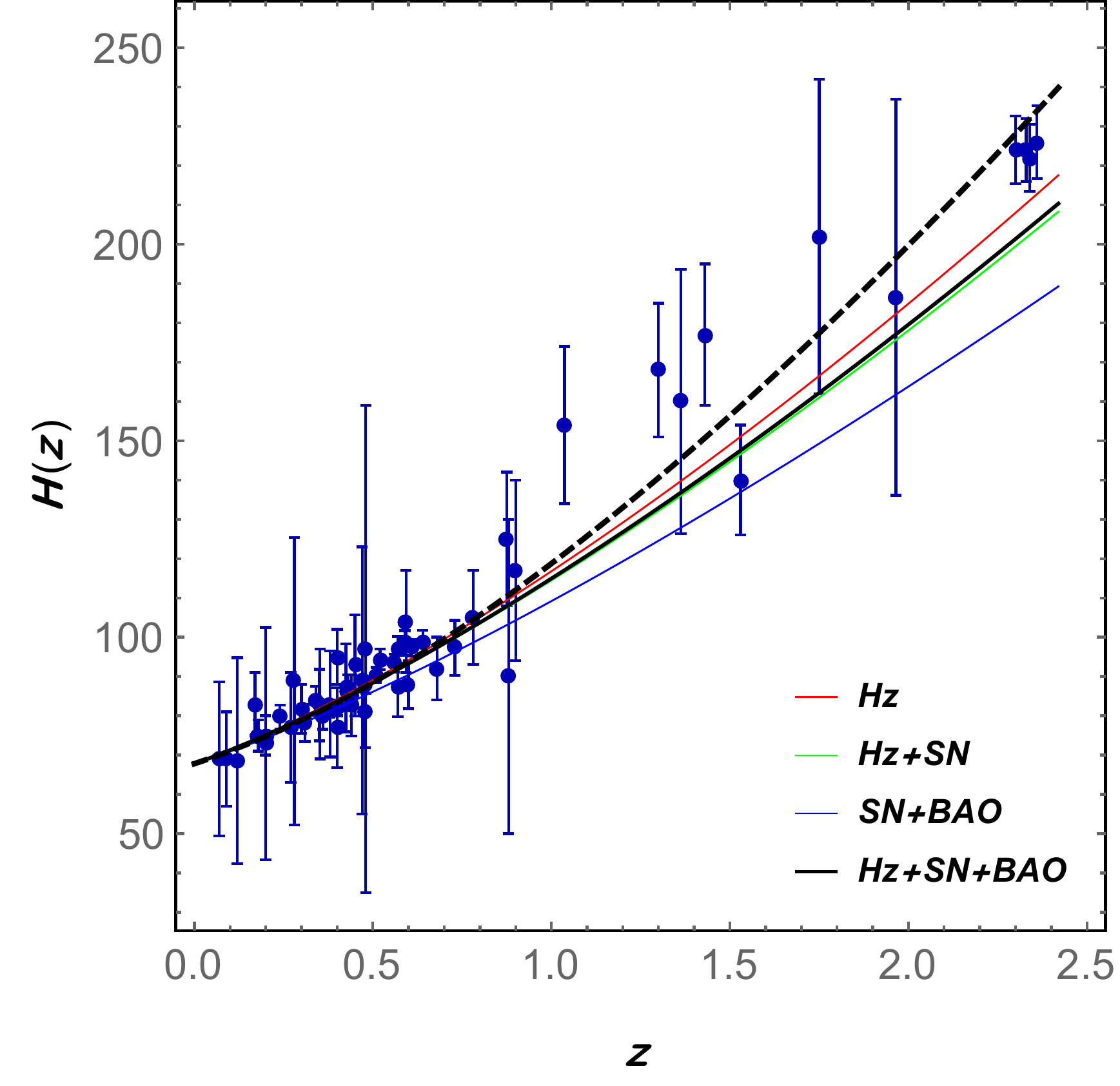} & 
\includegraphics[width=2.7 in,
height=2.3 in]{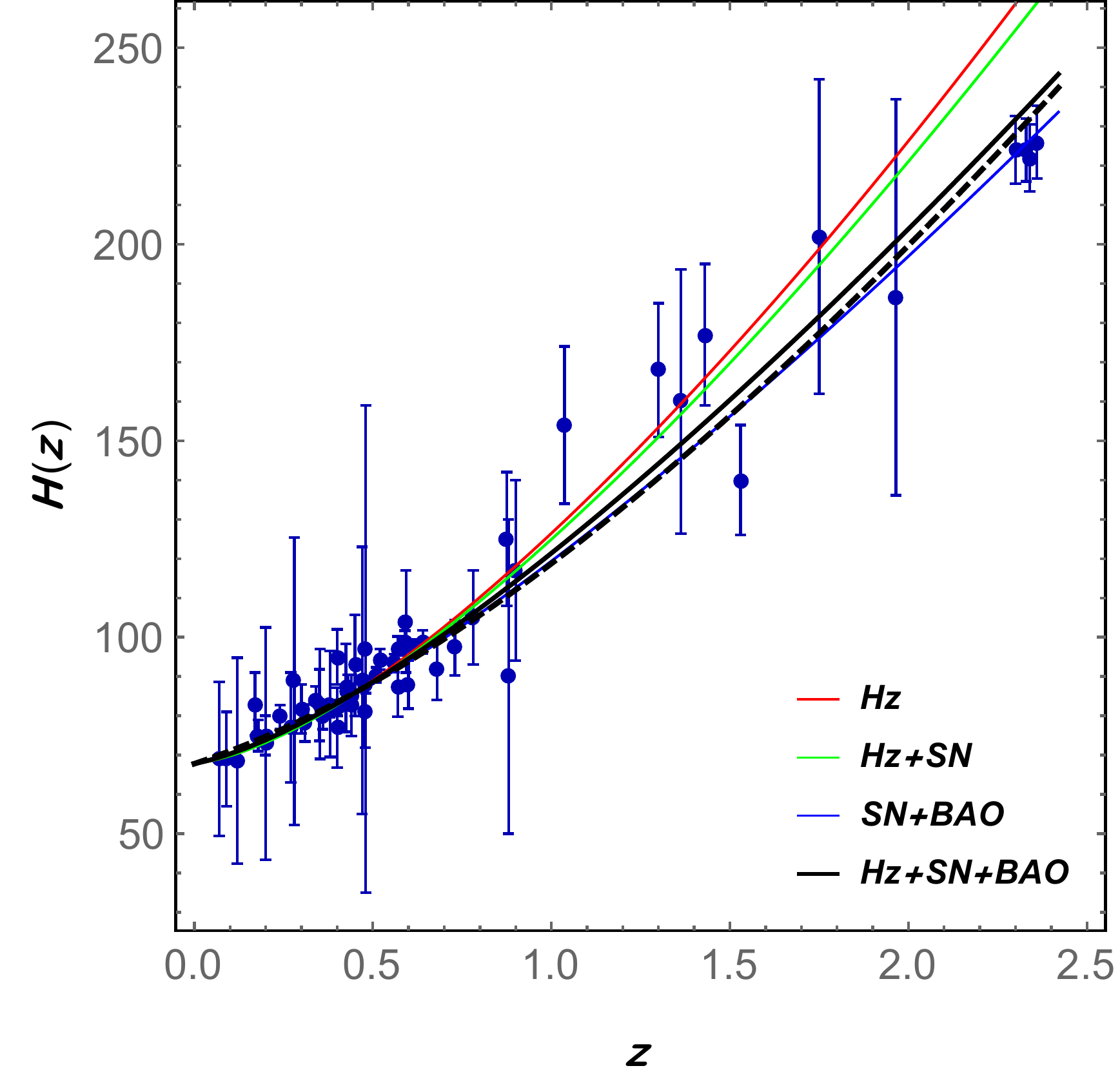} \\ 
\mbox (a) & \mbox (b)%
\end{array}
$%
\end{center}
\caption{ Figures (a) and (b) are the error bar plots for $57$ data points
from Hubble datasets together with the models M1 and M2 shown in solid red
lines respectively. The dashed lines in both the figures are $\Lambda $CDM
model shown for comparision.}
\label{error-Hz}
\end{figure}

\begin{figure}[tbph]
\label{Error-SN-M2.pdf}
\par
\begin{center}
$%
\begin{array}{c@{\hspace{.1in}}c}
\includegraphics[width=2.7 in, height=2.3
		in]{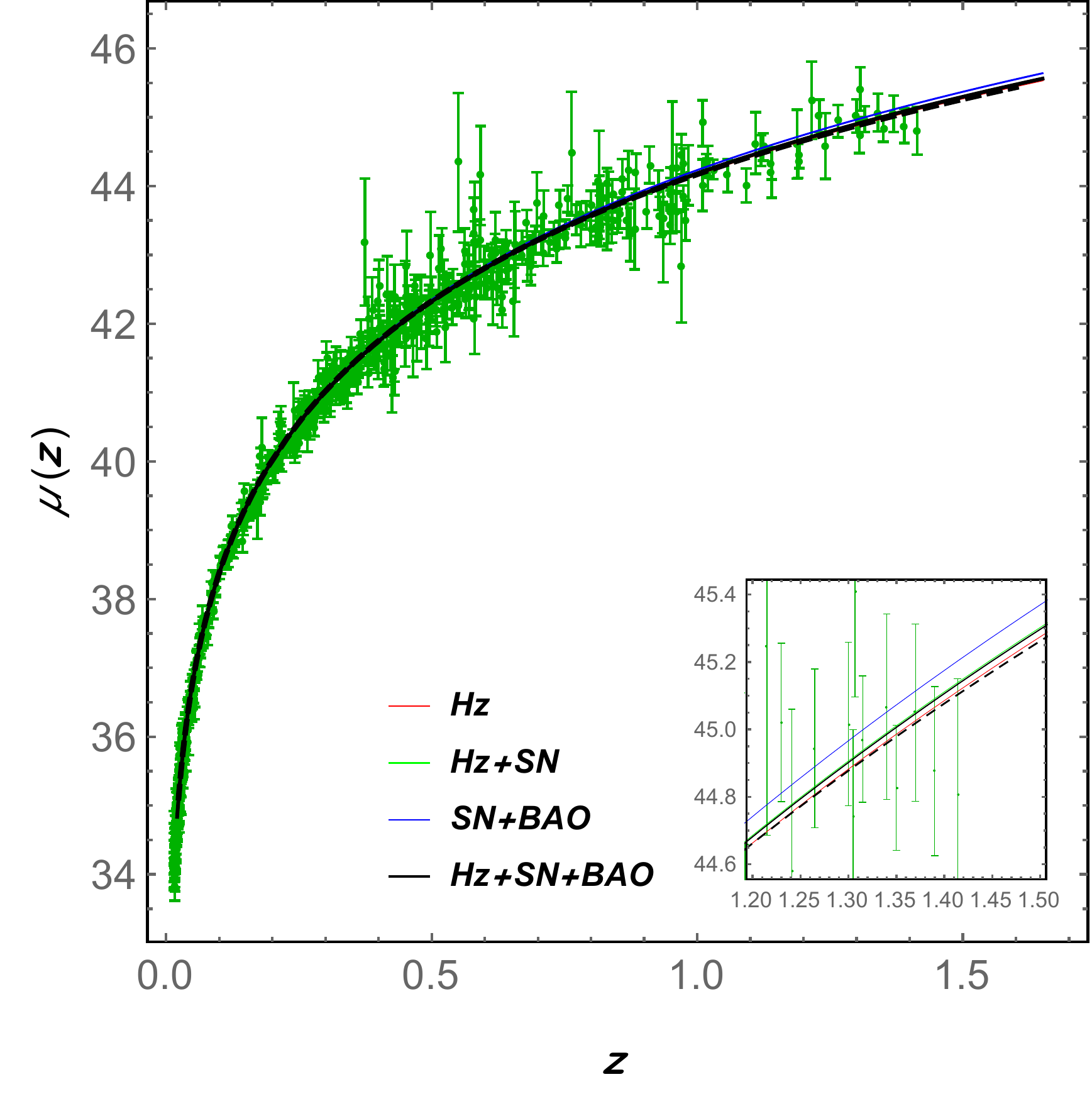} & 
\includegraphics[width=2.7 in,
		height=2.3 in]{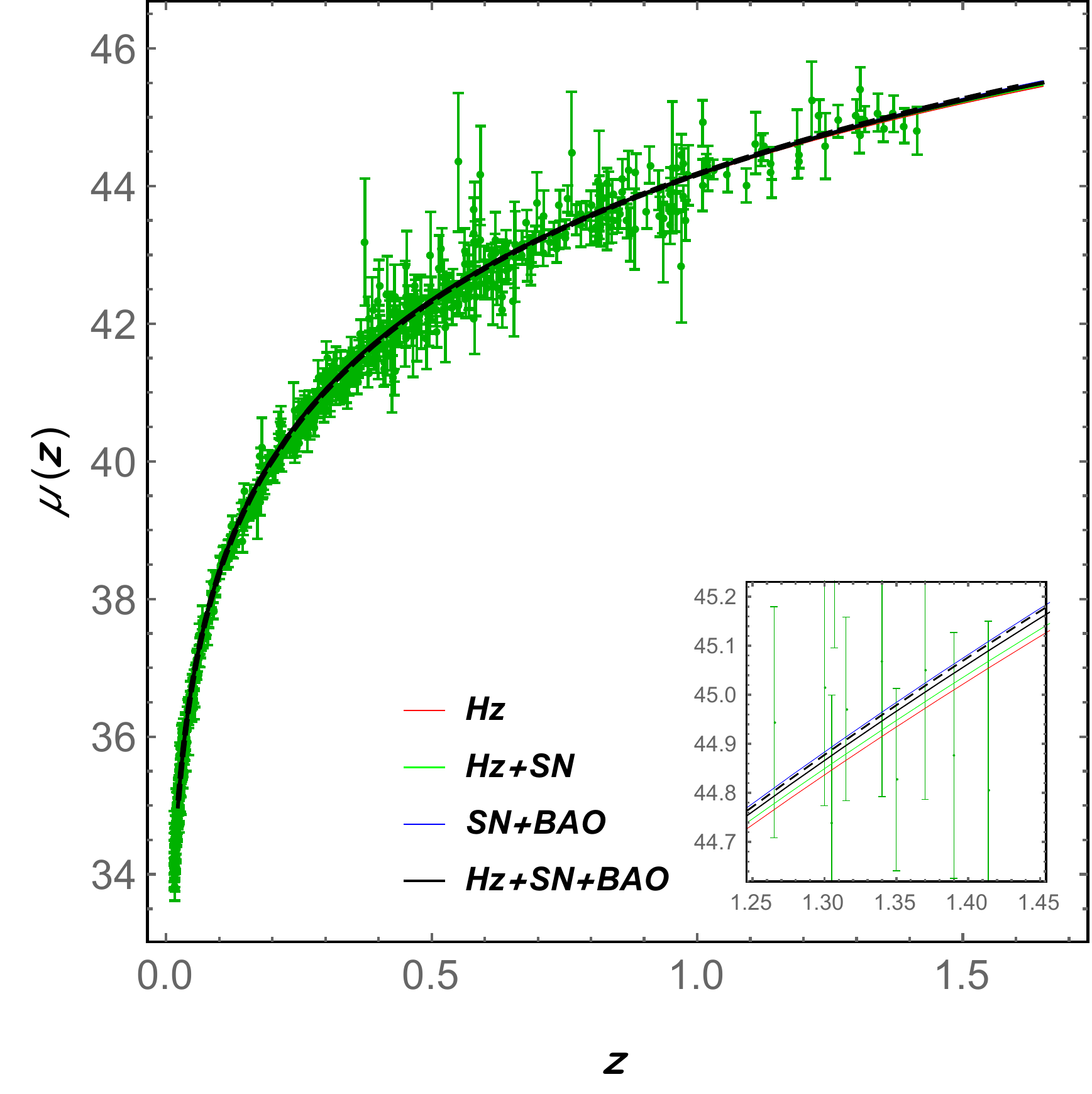} \\ 
\mbox (a) & \mbox (b)%
\end{array}
$%
\end{center}
\caption{ Figures (a) and (b) are the error bar plots for Union $2.1$
compilation supernovae datasets together with the models M1 and M2 shown in
solid red lines respectively. The dashed lines in both the figures are $%
\Lambda $CDM model shown for comparision.}
\label{error-Sn}
\end{figure}

\section{Geometrical Dynamics of the models}

\subsection{Deceleration parameter \& Phase transition}

The expressions for the deceleration parameter can be written in terms of
redshift $z$ as: 
\begin{equation}
q(z)=-1+\alpha -2\alpha \left[ 1+\left\{ \beta \left( 1+z\right) \right\}
^{\alpha }\right] ^{-1}  \label{qM1}
\end{equation}%
for model M1 and 
\begin{equation}
q(z)=-1+\alpha -3\alpha \left[ 1+\left\{ \beta \left( 1+z\right) \right\}
^{2\alpha }\right] ^{-1}  \label{qM2}
\end{equation}%
for model M2. The behavior of $q$ is shown in the following FIG. \ref{qz}
and the important values assumed by the deceleration parameter $q$ in the
course of evolution are tabulated in the following Table-5 for different
sets of $\alpha $ \& $\beta $ as obtained.

\begin{figure}[tbp]
\label{DP-M2.pdf}
\par
\begin{center}
$%
\begin{array}{c@{\hspace{.1in}}c}
\includegraphics[width=2.7 in, height=2.3 in]{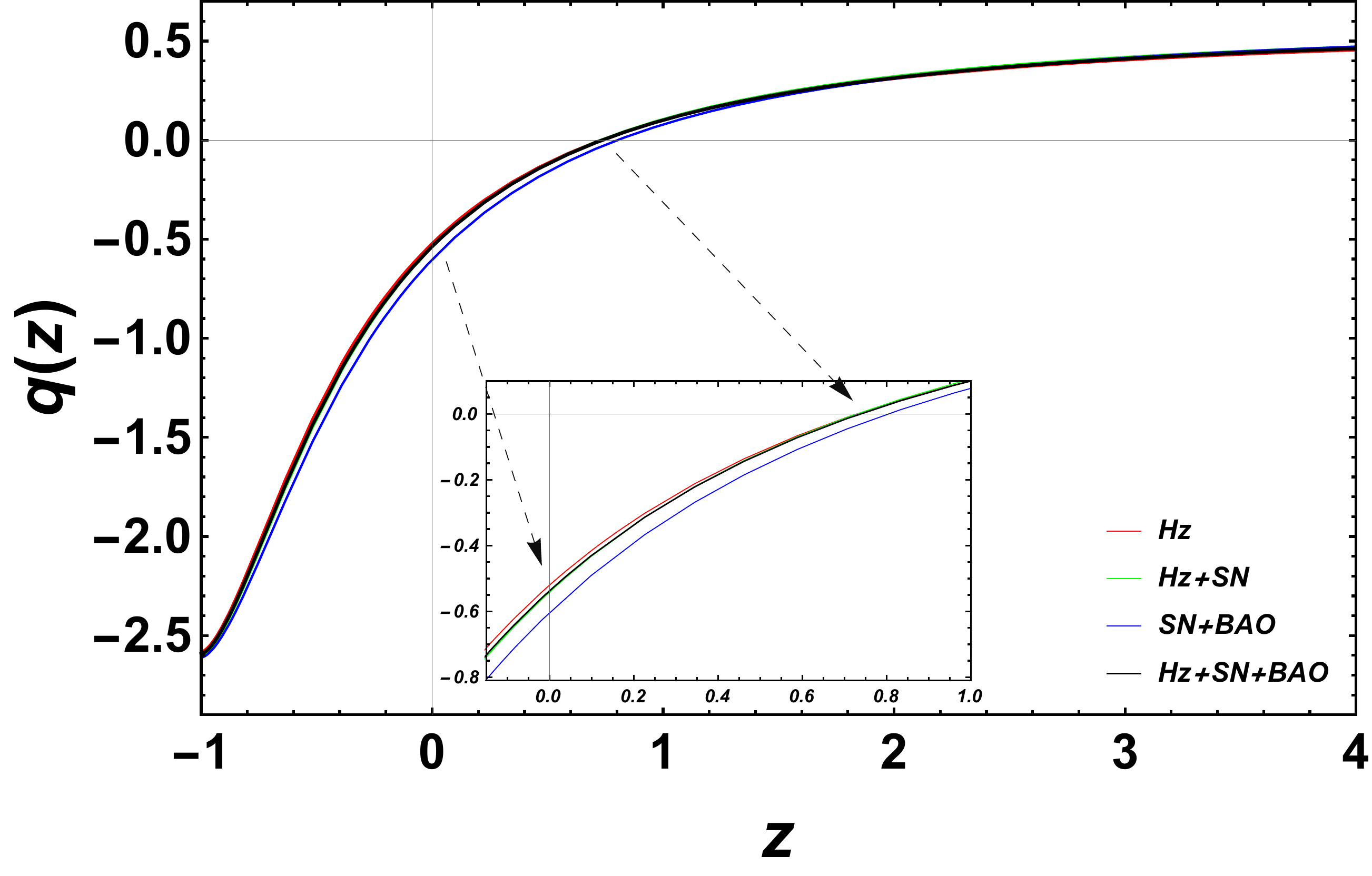} & %
\includegraphics[width=2.7 in, height=2.3 in]{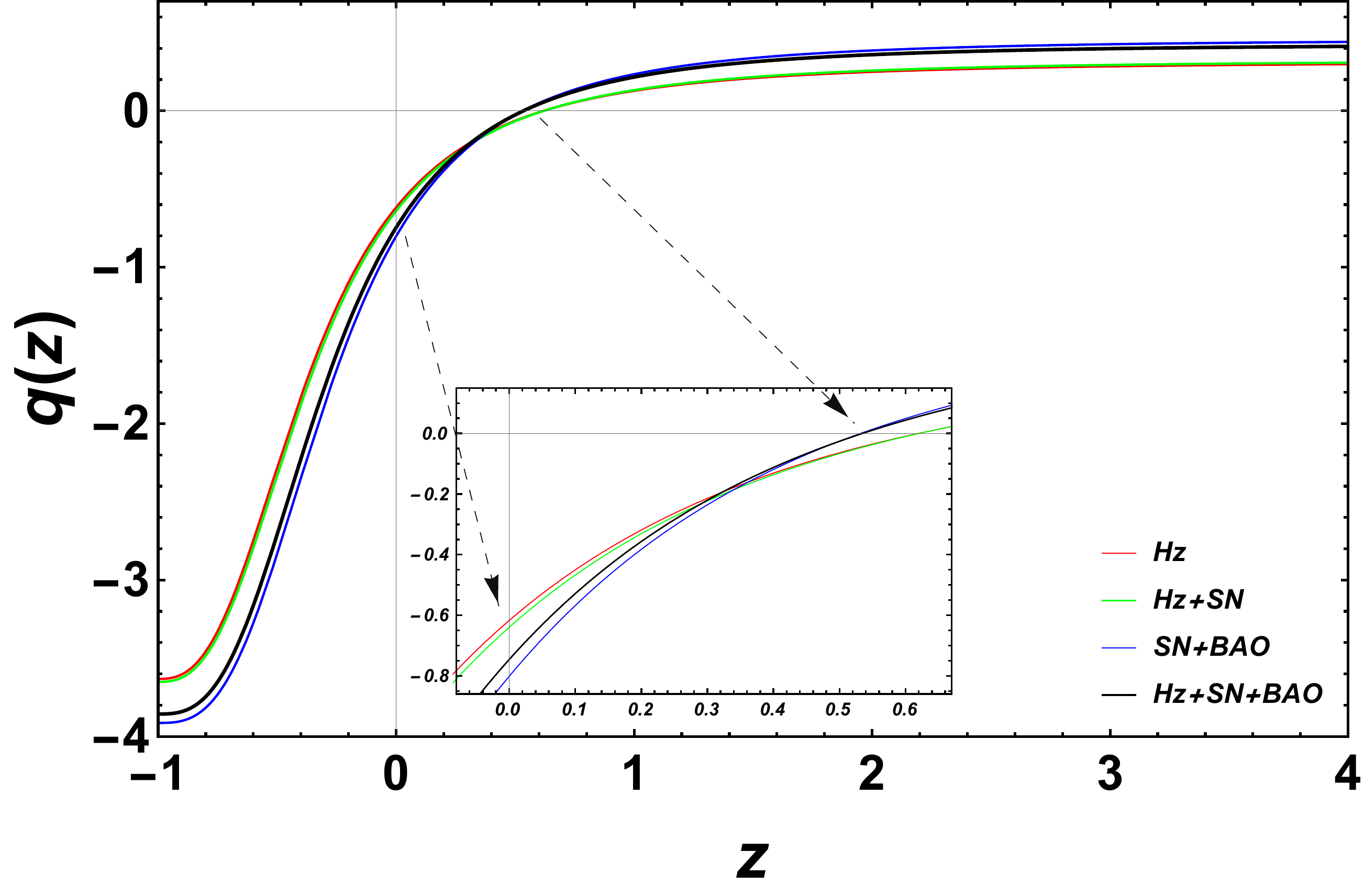} \\ 
\mbox (a) & \mbox (b)%
\end{array}
$%
\end{center}
\caption{ Figures (a) and (b) show the evolution of deceleration parameter
from past $(z=4)$ to far future with a phase transition for models M1 and M2
respectively.}
\label{qz}
\end{figure}

The initial value of the deceleration parameter ($q_{i}$ as $%
z\longrightarrow \infty $), the present value of the deceleration parameter (%
$q_{0}$ as $z\longrightarrow 0$) and the far future value of the
deceleration parameter ($q_{f}$ as $z\longrightarrow -1$) are calculated
together with the phase transition redshift ($z_{tr}$ for which $q=0$, the
redshift at which the Universe transited from decelerating expansion to
accelerating one) using the constrained (numerical) values of the model
parameters $\alpha $ \& $\beta $ (see Table-4) for both the models M1 and
model M2 in the following Table-5 and Table-6 respectively.

\begin{center}
\begin{tabular}{|c|c|c|c|c|c|}
\hline
\multicolumn{6}{|c|}{Table-5: Values of $q$ at different epochs \& phase
transition redshift for model M1} \\ \hline
redshift & formula & $Hz$ & $Hz+SN$ & $SN+BAO$ & $Hz+SN+BAO$ \\ \hline
\multicolumn{1}{|l|}{$z\longrightarrow \infty $ ($q_{i}$)} & 
\multicolumn{1}{|l|}{$q_{i}=-1+\alpha $} & $0.58064$ & $0.60094$ & $0.61116$
& $0.59173$ \\ \hline
\multicolumn{1}{|l|}{$z\longrightarrow 0$ ($q_{0}$)} & \multicolumn{1}{|l|}{$%
q_{0}=-1+\alpha -\frac{2\alpha }{1+\beta ^{\alpha }}$} & $-0.51977$ & $%
-0.54087$ & $-0.60308$ & $-0.53714$ \\ \hline
\multicolumn{1}{|l|}{$z\longrightarrow -1$ ($q_{f}$)} & \multicolumn{1}{|l|}{%
$q_{f}=-1-\alpha $} & $-2.58064$ & $-2.60094$ & $-2.61116$ & $-2.59173$ \\ 
\hline
\multicolumn{1}{|l|}{$z_{tr}$ ($q=0$)} & $z_{tr}=-1+\frac{1}{\beta }\left( 
\frac{\alpha +1}{\alpha -1}\right) ^{\frac{1}{\alpha }}$ & $0.72763$ & $%
0.72730$ & $0.80236$ & $0.73622$ \\ \hline
\end{tabular}

\begin{tabular}{|c|c|c|c|c|c|}
\hline
\multicolumn{6}{|c|}{Table-6: Values of $q$ at different epochs \& phase
transition redshift for model M2} \\ \hline
redshift & formula & $Hz$ & $Hz+SN$ & $SN+BAO$ & $Hz+SN+BAO$ \\ \hline
\multicolumn{1}{|l|}{$z\longrightarrow \infty $ ($q_{i}$)} & 
\multicolumn{1}{|l|}{$q_{i}=-1+\alpha $} & $0.31611$ & $0.32551$ & $0.45677$
& $0.42829$ \\ \hline
\multicolumn{1}{|l|}{$z\longrightarrow 0$ ($q_{0}$)} & \multicolumn{1}{|l|}{$%
q_{0}=-1+\alpha -\frac{3\alpha }{1+\beta ^{2\alpha }}$} & $-0.61721$ & $%
-0.63987$ & $-0.80152$ & $-0.74601$ \\ \hline
\multicolumn{1}{|l|}{$z\longrightarrow -1$ ($q_{f}$)} & \multicolumn{1}{|l|}{%
$q_{f}=-1-2\alpha $} & $-3.63222$ & $-3.65102$ & $-3.91354$ & $-3.85658$ \\ 
\hline
\multicolumn{1}{|l|}{$z_{tr}$ ($q=0$)} & $z_{tr}=-1+\frac{1}{\beta }\left( 
\frac{2\alpha +1}{\alpha -1}\right) ^{\frac{1}{2\alpha }}$ & $0.61941$ & $%
0.62055$ & $0.53182$ & $0.53471$ \\ \hline
\end{tabular}
\end{center}

From the above Table-5 and Table-6, it is observed that in both the models
M1 and M2, the Universe begins smoothly with deceleration (means no
inflationary phase in these models) and transit to accelerating phase at
around $z_{tr}\approx 0.72$ in model M1 for all numerical constrained values
of model parameters $\alpha $ \& $\beta $ while in model M2, the phase
transition occurs at $z_{tr}\approx 0.62$ $Hz$ and $Hz+SN$ constrained
values of $\alpha $ \& $\beta $ and $z_{tr}\approx 0.53$ $SN+BAO$ and $%
Hz+SN+BAO$ constrained values of $\alpha $ \& $\beta $. The present values
of the deceleration parameter $q_{0}$ found in both models can be seen in
the above tables consistent with predicted values. In the future, the
Universe enter into super acceleration phase ($q<-1$) for both the models M1
and M2 and finally attains maximum values $q_{f}<-2.58$ in model M1 and $%
q_{f}<-3.61$ in model M2 for all the values of $\alpha $ \& $\beta $.

\subsection{Statefinder diagnostics}

Statefinder diagnostics \cite{stfnd1, stfnd2, stfnd3, stfnd4} is a technique
generally used to distinguish various dark energy models and compare their
behavior using the higher order derivatives of the scale factor. The
parameters are $s$ \& $r$ and calculated using the relations:%
\begin{equation}
r=\frac{\dddot{a}}{aH^{3}}\text{, \ \ }s=\frac{r-1}{3(q-\frac{1}{2})}\text{.}
\label{statefinder}
\end{equation}%
The statefinder diagnostics pairs are constructed as $\{s,r\}$ and $\{q,r\}$
wherein different trajectories in the $s$-$r$ and $q$-$r$ planes are plotted
to see the temporal evolutions of different dark energy models. The fixed
points in this contexts are generally considered as $\{s,r\}=\{0,1\}$ for $%
\Lambda $CDM model and $\{s,r\}=\{1,1\}$ for SCDM (standard cold dark
matter) model in FLRW background and the departures of any dark energy model
from these fixed points are analyzed. The other diagnostic pair is $\left\{
q,r\right\} $ and the fixed points considered are $\left\{ q,r\right\}
=\left\{ -1,1\right\} $ for $\Lambda $CDM model and $\left\{ q,r\right\}
=\left\{ 0.5,1\right\} $ for SCDM model. The statefinder parameters for the
considered model M1 are calculated as,

\begin{equation}
r(z)=1+\alpha (2\alpha -3)+\frac{6\alpha }{1+\left\{ \beta (1+z)\right\}
^{\alpha }}\left[ 1-\alpha +\frac{\alpha }{1+\left\{ \beta (1+z)\right\}
^{\alpha }}\right]  \label{rM1}
\end{equation}

\begin{equation}
s(z)=\frac{2\alpha }{3}-\frac{\alpha }{1+\left\{ \beta (1+z)\right\}
^{\alpha }}+\frac{\alpha (3+2\alpha )}{3\left[ -3-2\alpha +(2\alpha
-3)\left\{ \beta (1+z)\right\} ^{\alpha }\right] }  \label{sM1}
\end{equation}%
and for model M2

\begin{equation}
r(z)=1-3\alpha +2\alpha ^{2}+\frac{12\alpha ^{2}}{\left[ 1+\left\{ \beta
(1+z)\right\} ^{2\alpha }\right] ^{2}}+\frac{3\alpha (3-2\alpha )}{1+\left\{
\beta (1+z)\right\} ^{2\alpha }}  \label{rM2}
\end{equation}

\begin{equation}
s(z)=\frac{2}{3}\left[ \alpha -\frac{2\alpha }{1+\left\{ \beta (1+z)\right\}
^{2\alpha }}+\frac{\alpha (3+4\alpha )}{\left[ -3-4\alpha +(2\alpha
-3)\left\{ \beta (1+z)\right\} ^{2\alpha }\right] }\right]  \label{sM2}
\end{equation}

\begin{figure}[tbp]
\label{srM2.pdf}
\par
\begin{center}
$%
\begin{array}{c@{\hspace{.1in}}c}
\includegraphics[width=2.7 in, height=2.3 in]{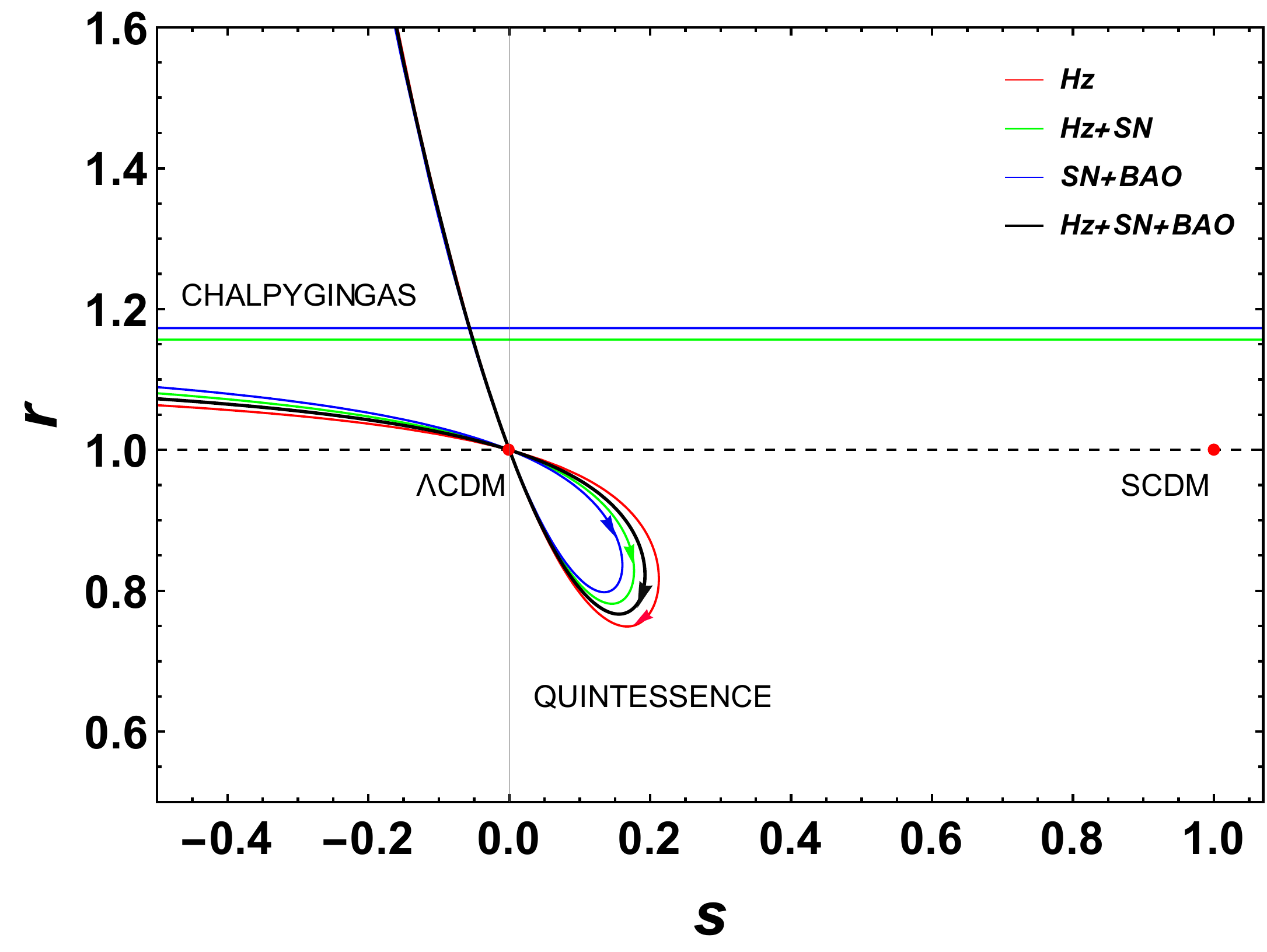} & %
\includegraphics[width=2.7 in, height=2.3 in]{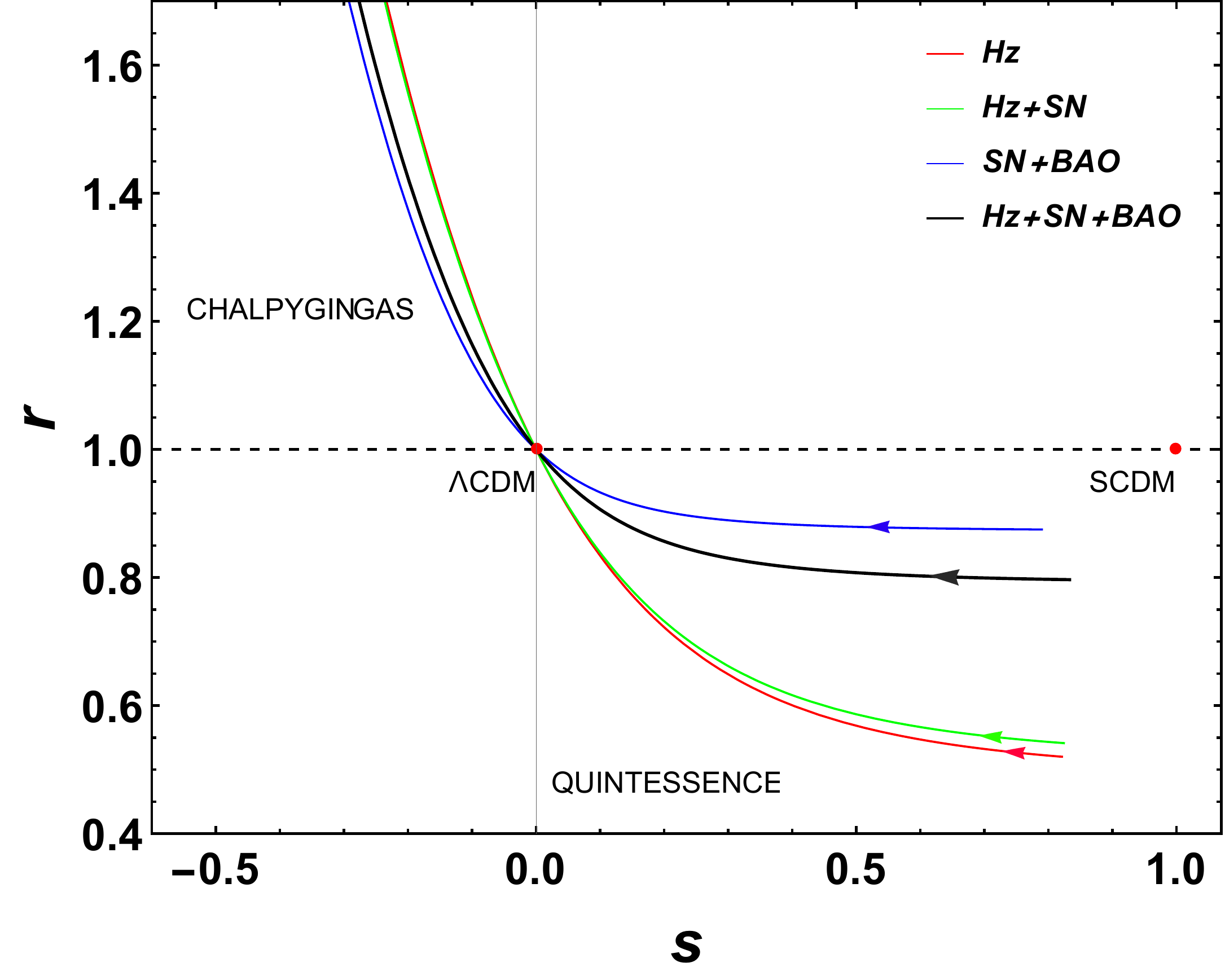} \\ 
\mbox (a) & \mbox (b)%
\end{array}
$%
\end{center}
\caption{ Figures (a) and (b) are the $s$-$r$ plots for model M1 and M2
respectively showing the different trajectories of the models.}
\label{s-r}
\end{figure}

\begin{figure}[tbp]
\label{qrM2.pdf}
\par
\begin{center}
$%
\begin{array}{c@{\hspace{.1in}}c}
\includegraphics[width=2.7 in, height=2.3 in]{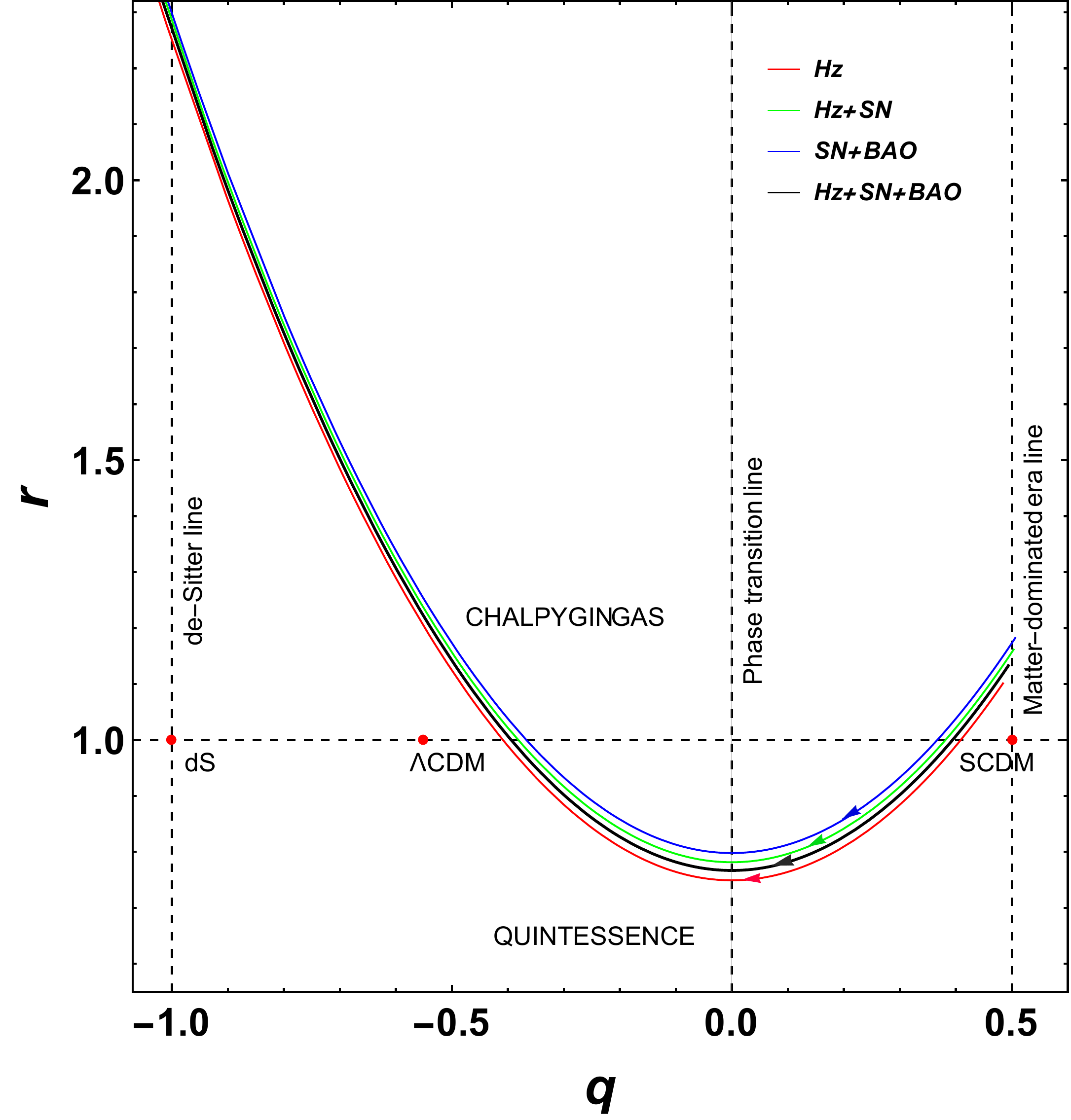} & %
\includegraphics[width=2.7 in, height=2.3 in]{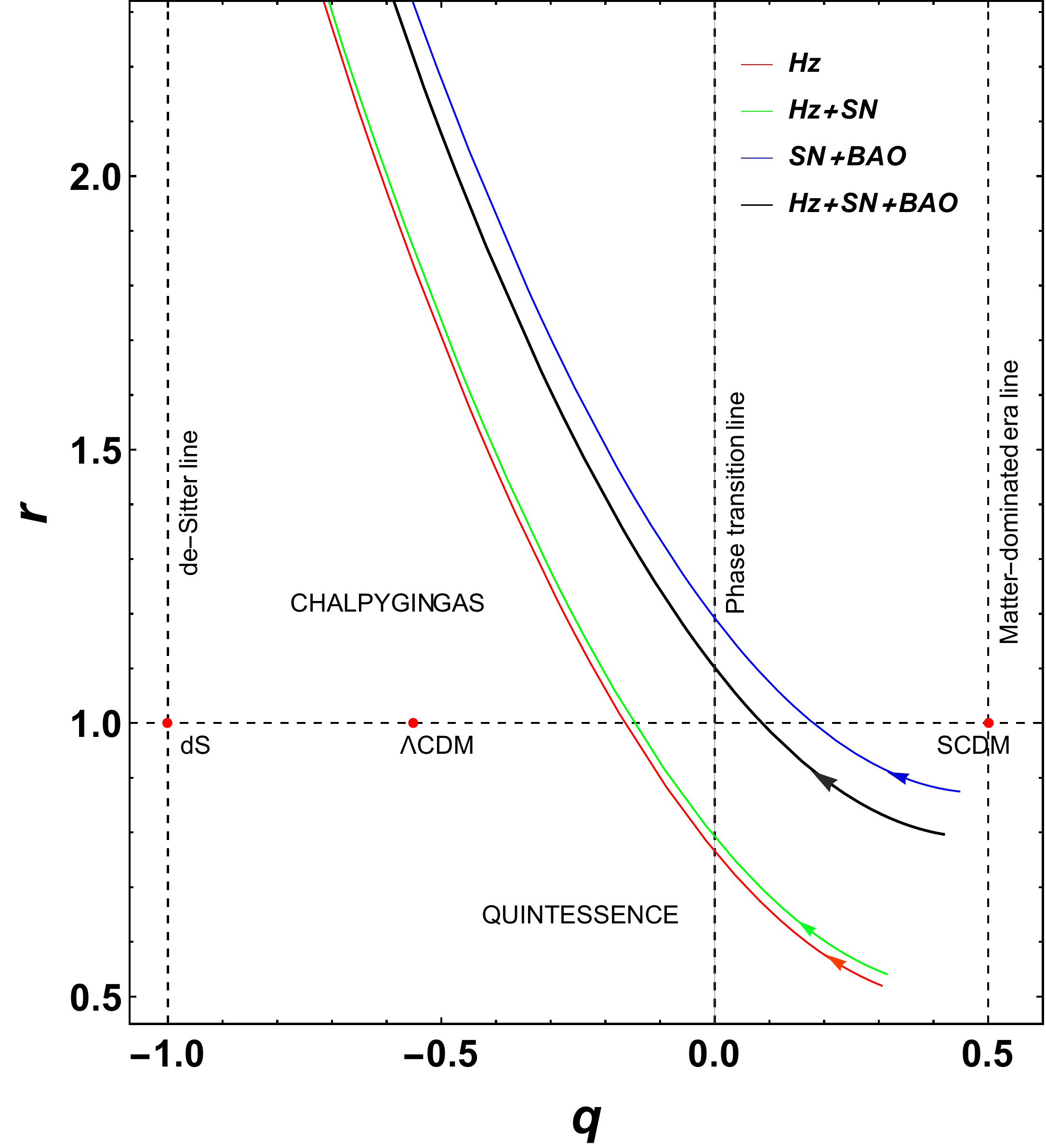} \\ 
\mbox (a) & \mbox (b)%
\end{array}
$%
\end{center}
\caption{ Figures (a) and (b) are $q$-$r$ plots for models M1 and M2
respectively showing different trajectories of the models.}
\label{q-r}
\end{figure}

In the above figure FIG. \ref{s-r}, one can see the diverge evolutions of
the model M1 and model M2 in the $s$-$r$ and $q$-$r$ planes. Both the models
showing distinctive features as compared to the other standard models. One
can observe that at early times, model M1 presumes values in the range $r>1$
and $s<0$ representing Chaplygin gas type DE model and evolutes to
quintessence region and again reverts to Chapligyn gas region at late times
by crossing the intermediate $\Lambda $CDM fixed point $\left\{ 0,1\right\} $
during evolution. But, the model M2 is different and evolutes from
quintessence region in the past and goes to Chaplygin gas region
intermediating the $\Lambda $CDM fixed point $\left\{ 0,1\right\} $ during
it's evolution for all cases. The figure FIG. \ref{q-r} depicts the temporal
evolution of the models M1 and M2 in the $\{q,r\}$ plane providing
additional information about the models M1 and M2 wherein the dashed lines
describe the evolution of the $\Lambda $CDM model below which quintessence
region and the upper one is Chaplygin gas region are shown. The evolution of
model M1 and M2 are clearly observed. Both the models M1 and M2 deviates
from de Sitter point ($-1,1$).

\subsection{$Om$ diagnostic}

\qquad $Om$ diagnostic is another tool introduced in \cite{omdig1, omdig2,
omdig3, omdig4}, using the Hubble parameter and serving the purpose of
providing a null test of the $\Lambda $CDM model. Like, statefinder
diagnostic, $Om$ diagnostic is also an effective method to discriminate
various DE models from $\Lambda $CDM model according to the slope variation
of $Om(z)$. Positive slope of diagnostic implies a Quintessence nature ($%
\omega >-1$), Negative slope of diagnostic implies a Phantom nature ($\omega
<-1$) and Constant slope with respect to redshift tells the nature of dark
energy coincide with that of the cosmological constant ($\omega =-1$).

The $Om(z)$ for a flat Universe is defined as: 
\begin{equation}
Om\left( z\right) =\frac{\left( \frac{H(z)}{H_{0}}\right) ^{2}-1}{\left(
1+z\right) ^{3}-1}\text{,{}}  \label{eq:om}
\end{equation}%
which can also be represented as $Om\left( z\right) =\Omega _{m0}+\left(
1-\Omega _{m0}\right) \frac{(1+z)^{3(1+\omega )}-1}{(1+z)^{3}-1}$. For a
constant EoS parameter $\omega $ imply $Om\left( z\right) =\Omega _{m0}$ and
different values of $Om\left( z\right) $ suggest whether the model is a $%
\Lambda $CDM model or quintessence or phantom models. For the models of
consideration here, the expressions for $Om(z)$ for models M1 and M2 are
obtained as, 
\begin{equation}
Om\left( z\right) =\frac{\frac{\left[ 1+\left\{ \beta \left( 1+z\right)
\right\} ^{\alpha }\right] ^{4}}{\left( 1+\beta ^{\alpha }\right)
^{4}(1+z)^{2\alpha }}-1}{\left( 1+z\right) ^{3}-1}\text{{}}  \label{om1}
\end{equation}%
\begin{equation}
Om\left( z\right) =\frac{\frac{\left[ 1+\left\{ \beta \left( 1+z\right)
\right\} ^{2\alpha }\right] ^{3}}{\left( 1+\beta ^{2\alpha }\right)
^{3}(1+z)^{4\alpha }}-1}{\left( 1+z\right) ^{3}-1}\text{{}}  \label{om2}
\end{equation}

The slope variation of $Om(z)$ vs. $z$ are shown in the following figure
FIG. \ref{om} for models M1 and M2. For both the models M1 and M2 and for
all values of $\alpha $ \& $\beta $, the $Om\left( z\right) $ values is less
than $\Omega _{m0}$ in the redshift range $z>0$ showing the models are in
quintessence region in the past and for the redshift range $z<0$, $Om\left(
z\right) $ values decreases sharply and becomes negative implying the both
the models enter into phantom region.

\begin{figure}[tbp]
\label{OmM2.pdf}
\par
\begin{center}
$%
\begin{array}{c@{\hspace{.1in}}c}
\includegraphics[width=2.7 in, height=2.2 in]{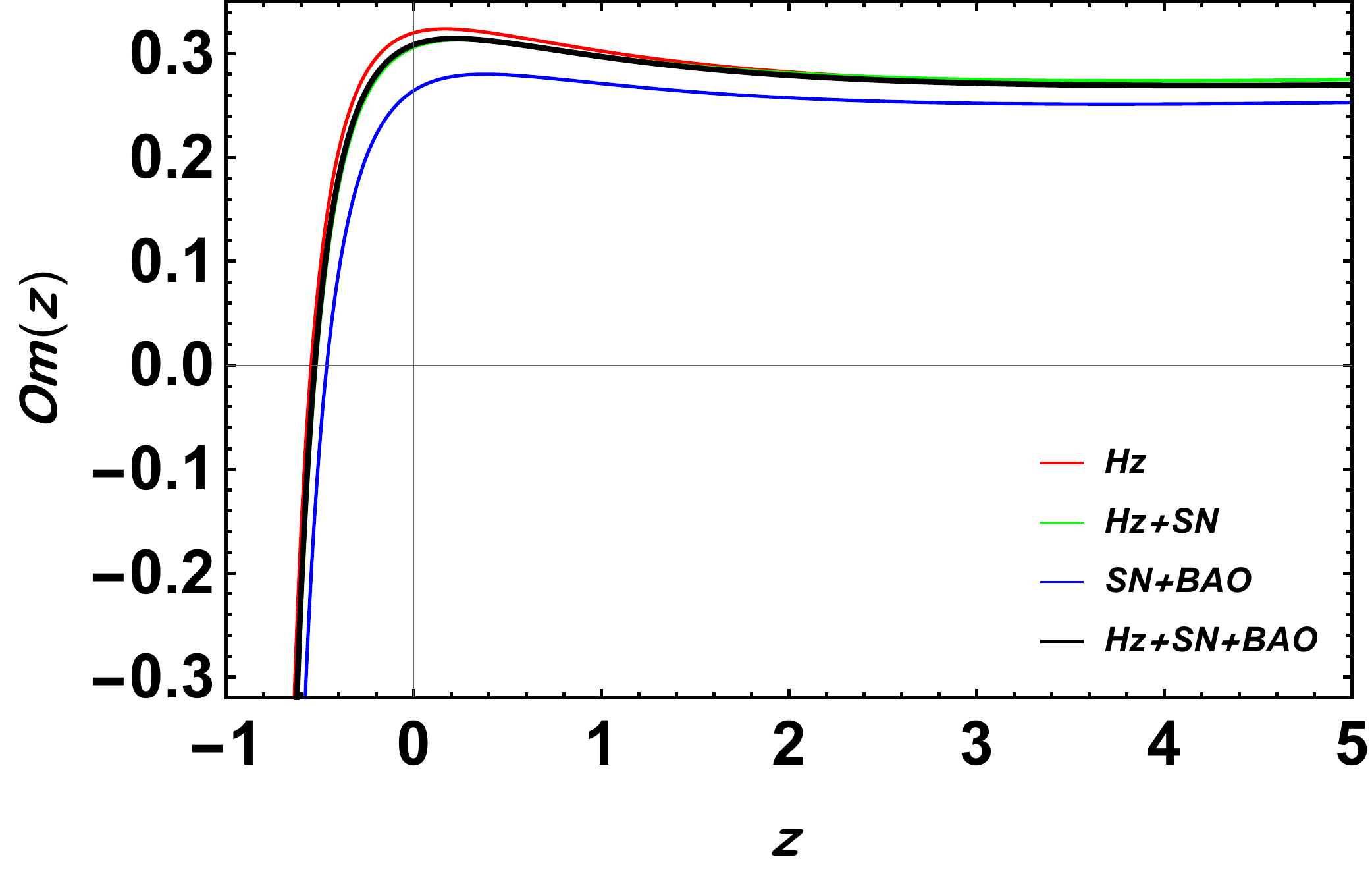} & %
\includegraphics[width=2.7 in, height=2.2 in]{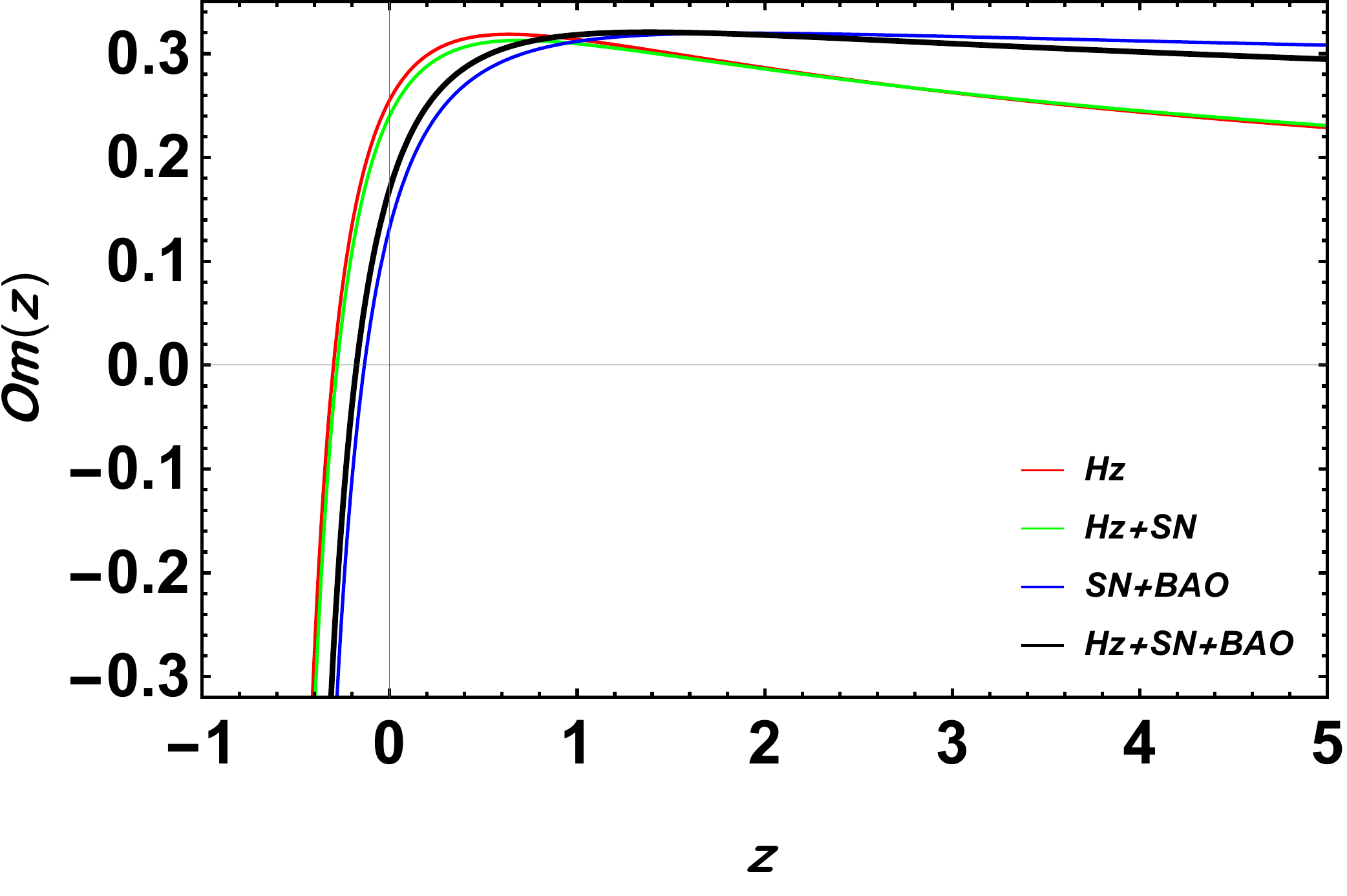} \\ 
\mbox (a) & \mbox (b)%
\end{array}
$%
\end{center}
\caption{ Figures (a) and (b) show the plots for $Om(z)$ vs. $z$ for models
M1 and M2 respectively.}
\label{om}
\end{figure}

\subsection{Jerk, Snap and Lerk parameters}

\begin{figure}[tbp]
\label{jerk-M2.pdf}
\par
\begin{center}
$%
\begin{array}{c@{\hspace{.1in}}c}
\includegraphics[width=2.7 in, height=2.2 in]{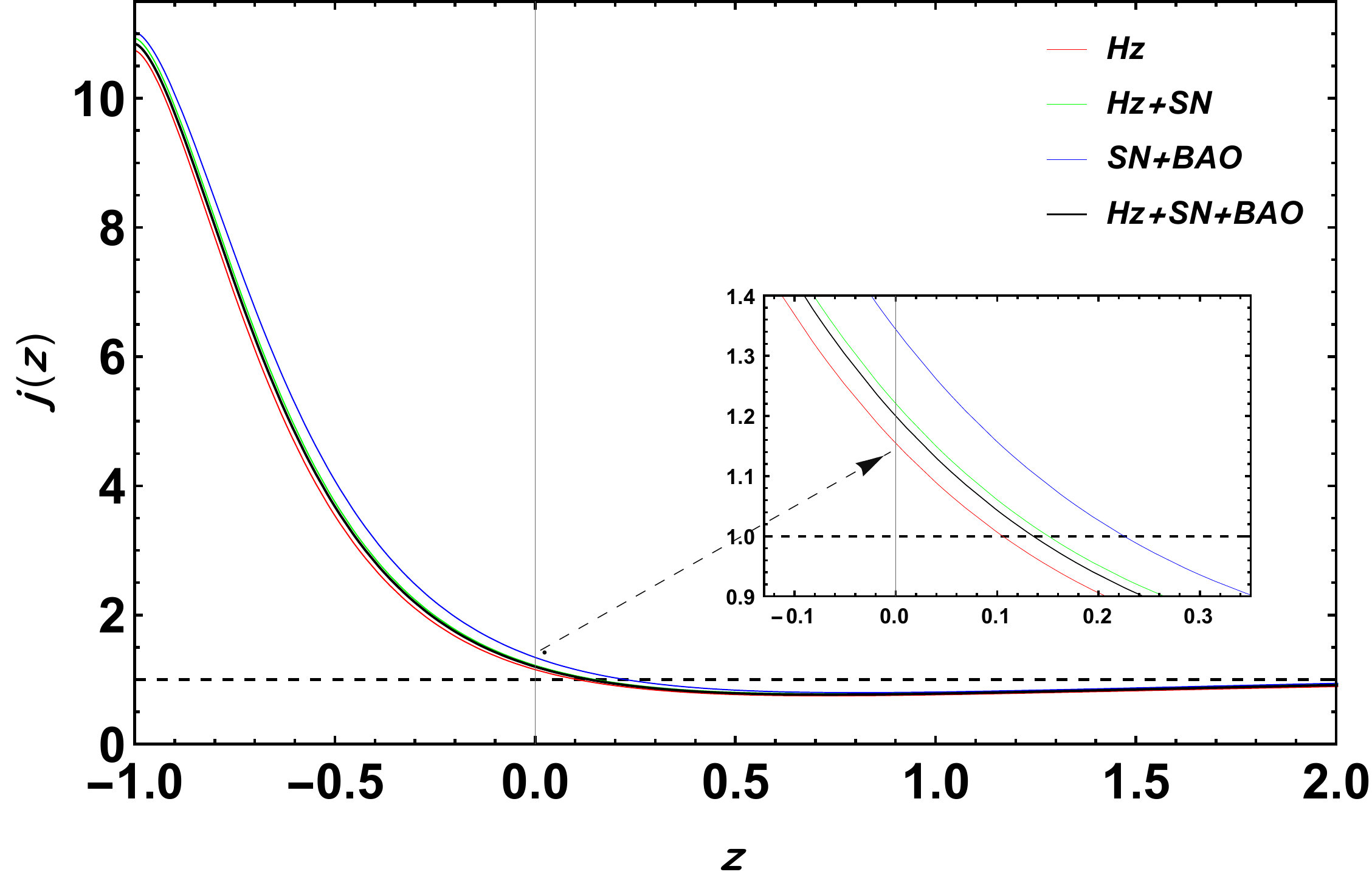} & %
\includegraphics[width=2.7 in, height=2.2 in]{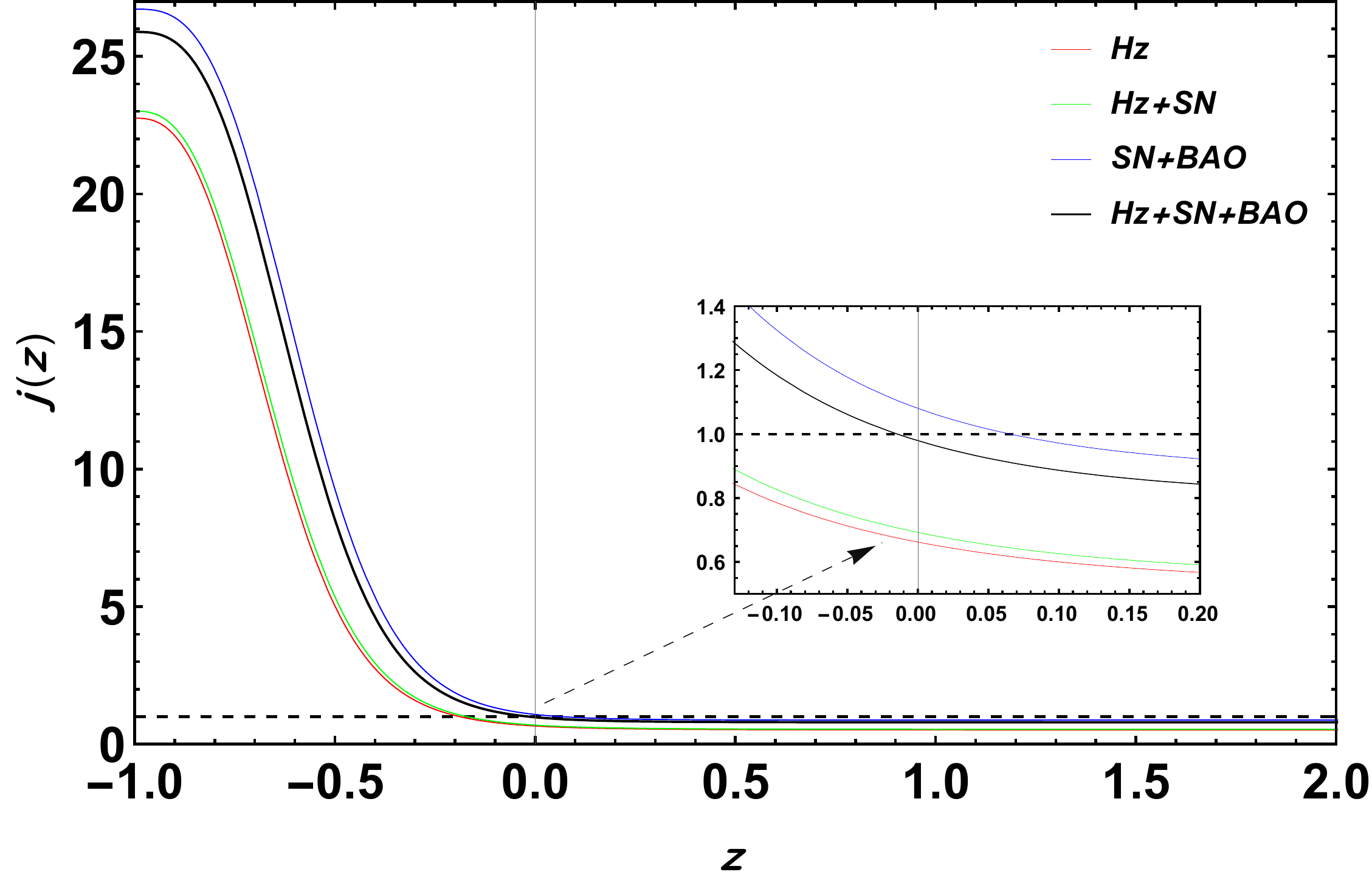} \\ 
\mbox (a) & \mbox (b)%
\end{array}
$%
\end{center}
\caption{ Figures (a) and (b) show the plots for jerk parameter $j(z)$ vs. $%
z $ for models M1 and M2 respectively.}
\label{jerk}
\end{figure}

\begin{figure}[tbp]
\label{snap-M2.pdf}
\par
\begin{center}
$%
\begin{array}{c@{\hspace{.1in}}c}
\includegraphics[width=2.7 in, height=2.2 in]{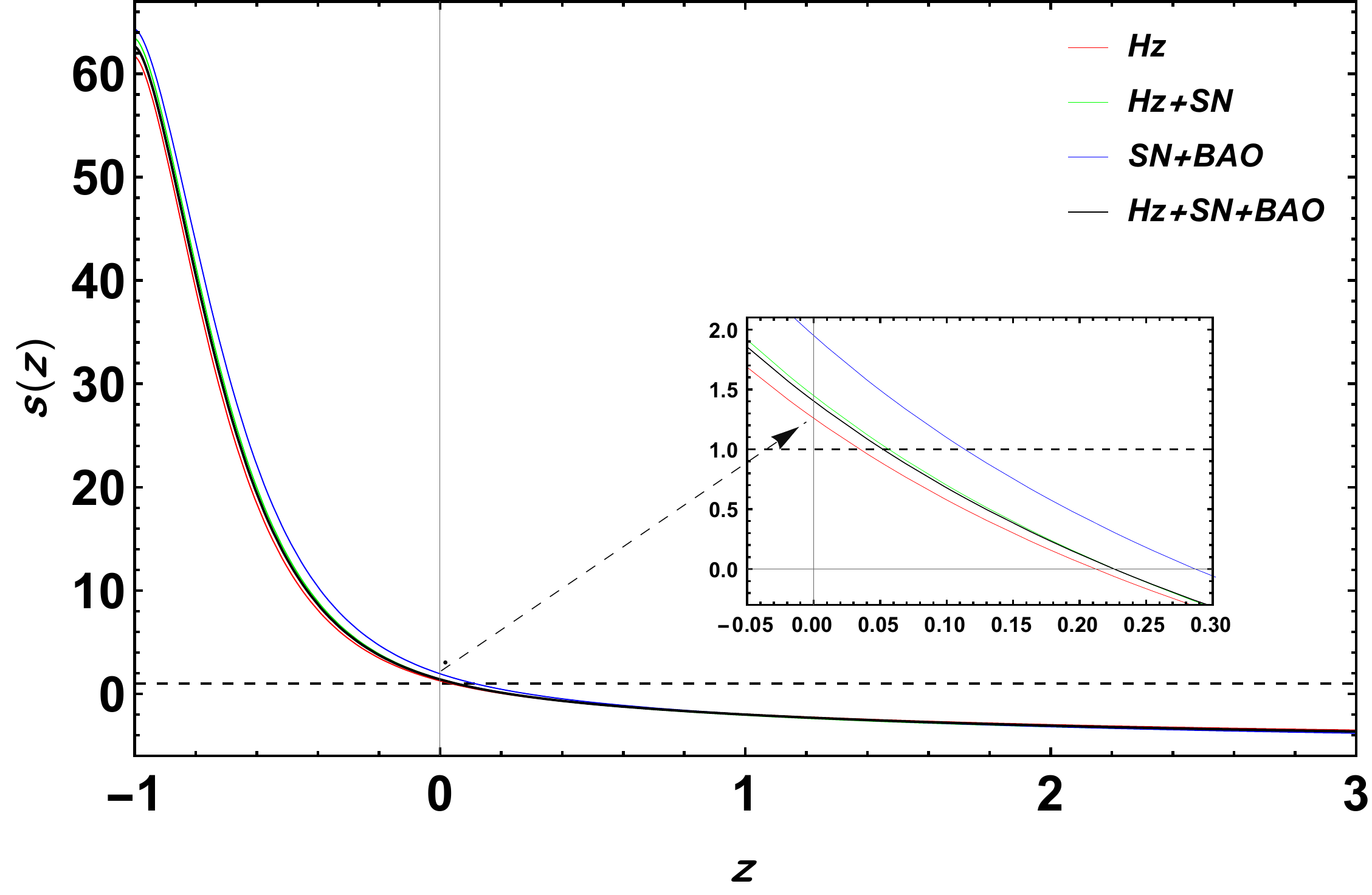} & %
\includegraphics[width=2.7 in, height=2.2 in]{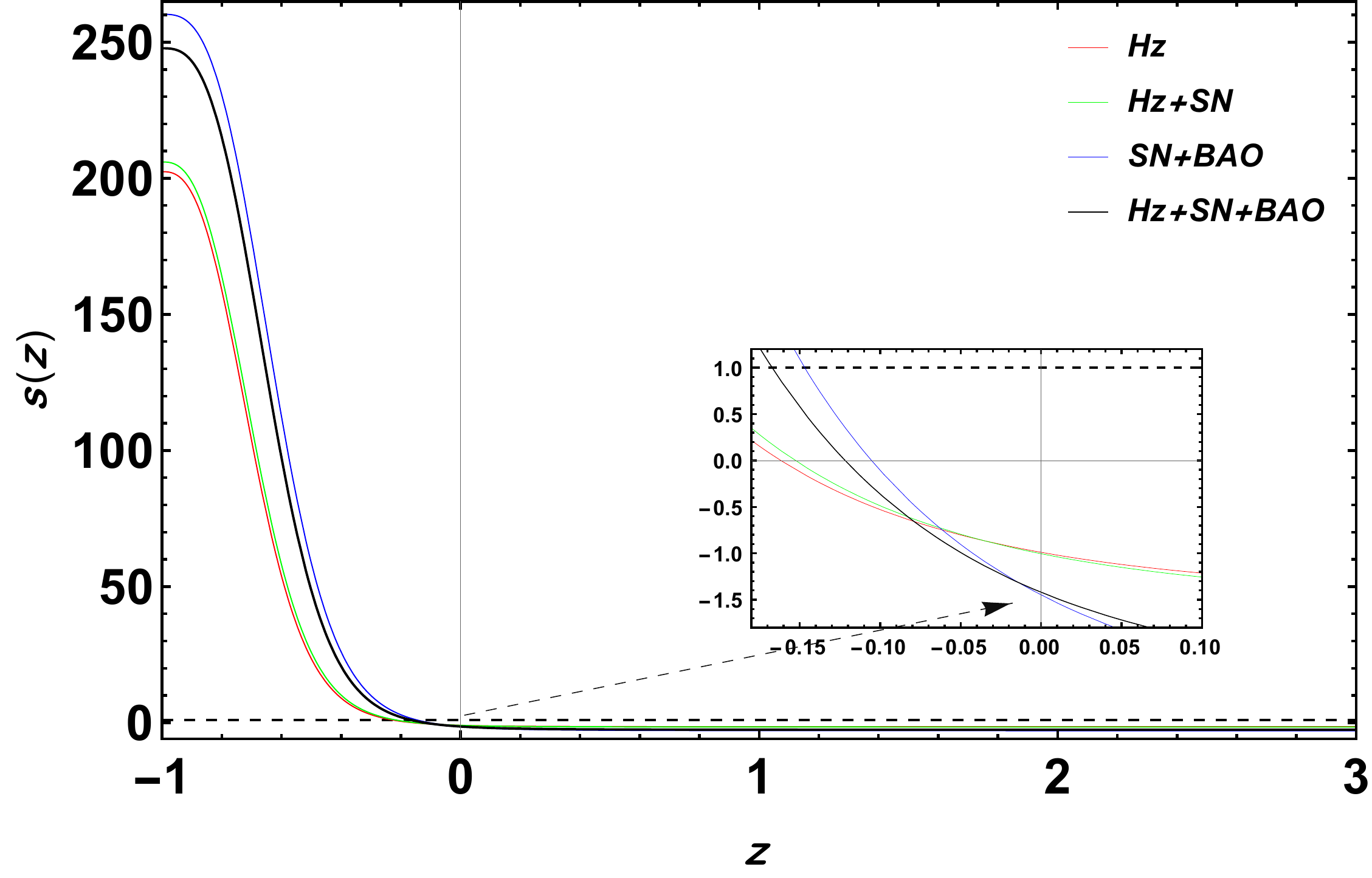} \\ 
\mbox (a) & \mbox (b)%
\end{array}
$%
\end{center}
\caption{ Figures (a) and (b) show the plots for snap parameter $s(z)$ vs. $%
z $ for models M1 and M2 respectively.}
\label{snap}
\end{figure}

\begin{figure}[tbp]
\label{lerk-M2.pdf}
\par
\begin{center}
$%
\begin{array}{c@{\hspace{.1in}}c}
\includegraphics[width=2.7 in, height=2.2 in]{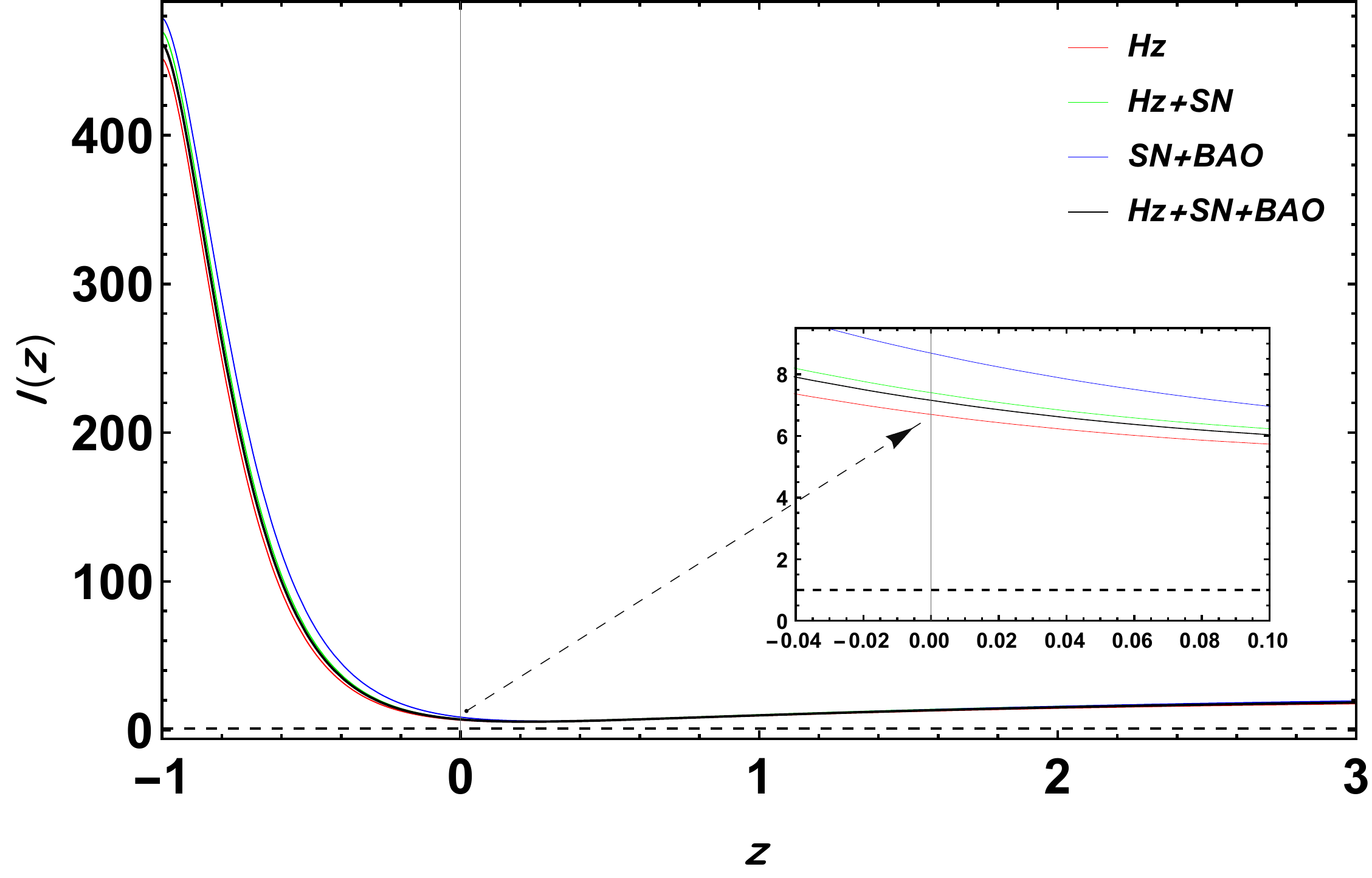} & %
\includegraphics[width=2.7 in, height=2.2 in]{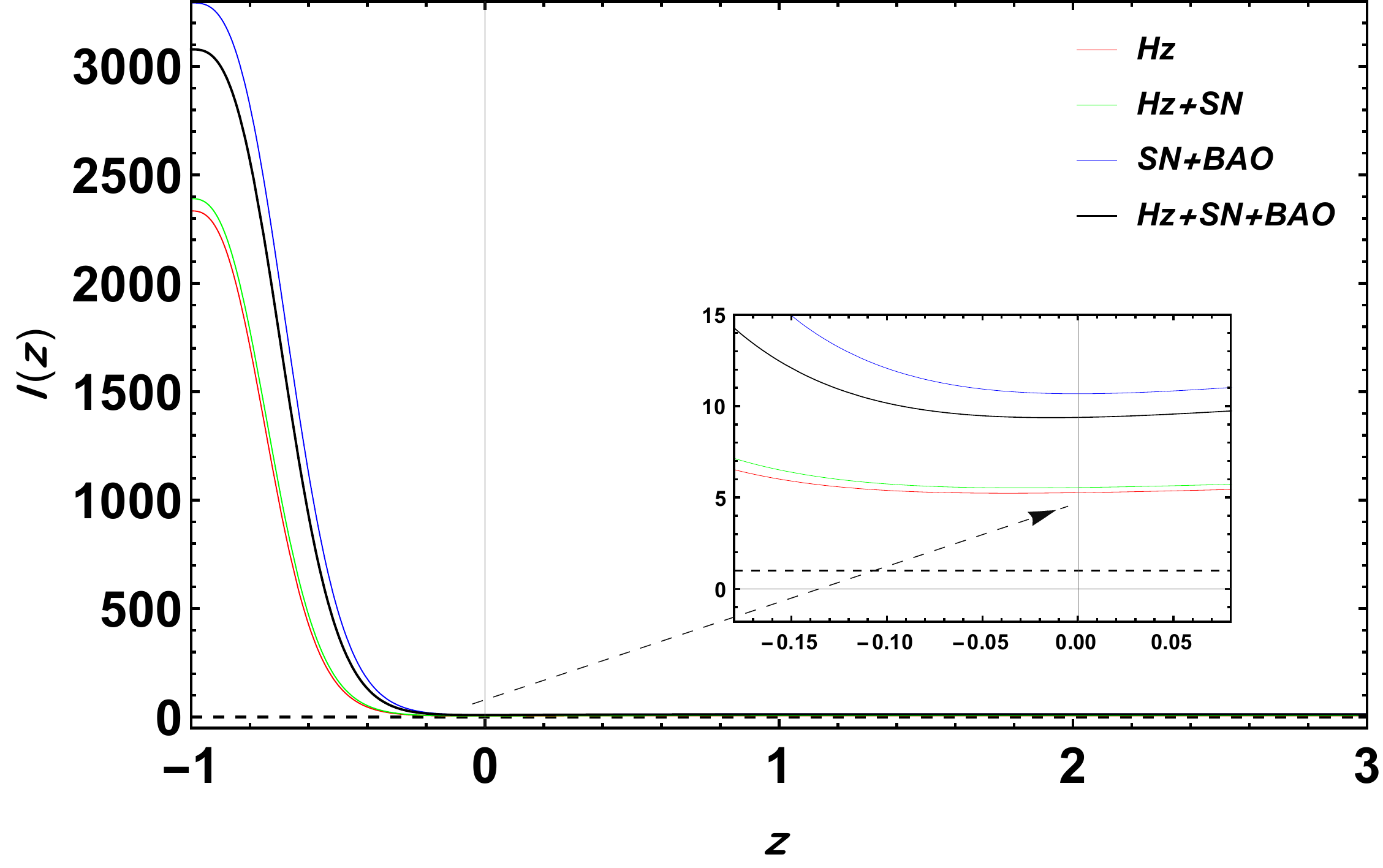} \\ 
\mbox (a) & \mbox (b)%
\end{array}
$%
\end{center}
\caption{ Figures (a) and (b) show the plots for lerk parameter $l(z)$ vs. $%
z $ for models M1 and M2 respectively.}
\label{lerk}
\end{figure}

\qquad Likewise, the Hubble and the deceleration parameters, the other
cosmographic parameters, jerk, snap and lerk parameters also play
significant roles in analyzing a cosmological model. The cosmic jerk $%
j(z=0)\simeq 1$ signify a cosmic acceleration. The evolution of jerk
parameter for the models M1 and M2 are shown in FIG. \ref{jerk} showing that
for all numerical constrained values of model parameters $\alpha $ \& $\beta 
$, $j_{0}\in (1.1,1.4)$ for model M1 and $j_{0}\in (0.6,1.2)$ for model M2.
The increasing values of jerk, snap and lerk parameters in the future ($z<0$%
) showing the deviation from the $\Lambda $CDM model which can also be
interpreted from the statefinder diagrams FIG. 9 and FIG. 10. Similary, the
evolution of snap and lerk parameters are shown in FIG. 14 and FIG. 15
respectively for both the models M1 and M2. From the figures, it can be seen
that for all values of model parameters $\alpha $ \& $\beta $, $s_{0}\in
(1.2,2.2)$ for model M1 and $s_{0}\in (-1.5,0-0.7)$ for model M2 and $%
l_{0}\in (6,9)$ for model M1 and $l_{0}\in (5,13)$ for model M2. These
values of $j_{0}$, $s_{0}$, $l_{0}$ are in good agreement with the expected
values. One can also interpret that the model M2 has better fit to the
observational datasets as compared to model M1 which can also be seen from
FIG. \ref{snap} and FIG. \ref{lerk}.

\section{Physical Dynamics of the models}

\qquad The geometrical part of the Einstein field equations is discussed
elaborately and now the physical interpretations can be discussed for the
obtained models once the matter content of the Universe is specified. In the
introduction, it is mentioned that the candidate of dark energy is still
unknown and it is a matter of speculation only to choose any candidate
described in literature. However, the most discussed candidate and having
best fit with some observations is the Einstein's cosmological constant. So,
in the following, the cosmological constant will be considered as a
candidate of dark energy for further analysis.

So, let us consider the two fluid Universe, cold dark matter and dark energy
only, since the radiation contribution at present is negligible. The matter
pressure is $p=p_{m}=0$ for cold dark matter and for dark energy the
equation of state is $p_{DE}=\omega _{DE}\rho _{DE}$. In the following, the
physical behavior of the matter and dark energy densities and pressures are
found out and their evolutions are shown graphically.

\subsection{Cosmological constant}

\begin{figure}[tbp]
\label{CC-rho-M2.pdf}
\par
\begin{center}
$%
\begin{array}{c@{\hspace{.1in}}c}
\includegraphics[width=2.7 in, height=2.2 in]{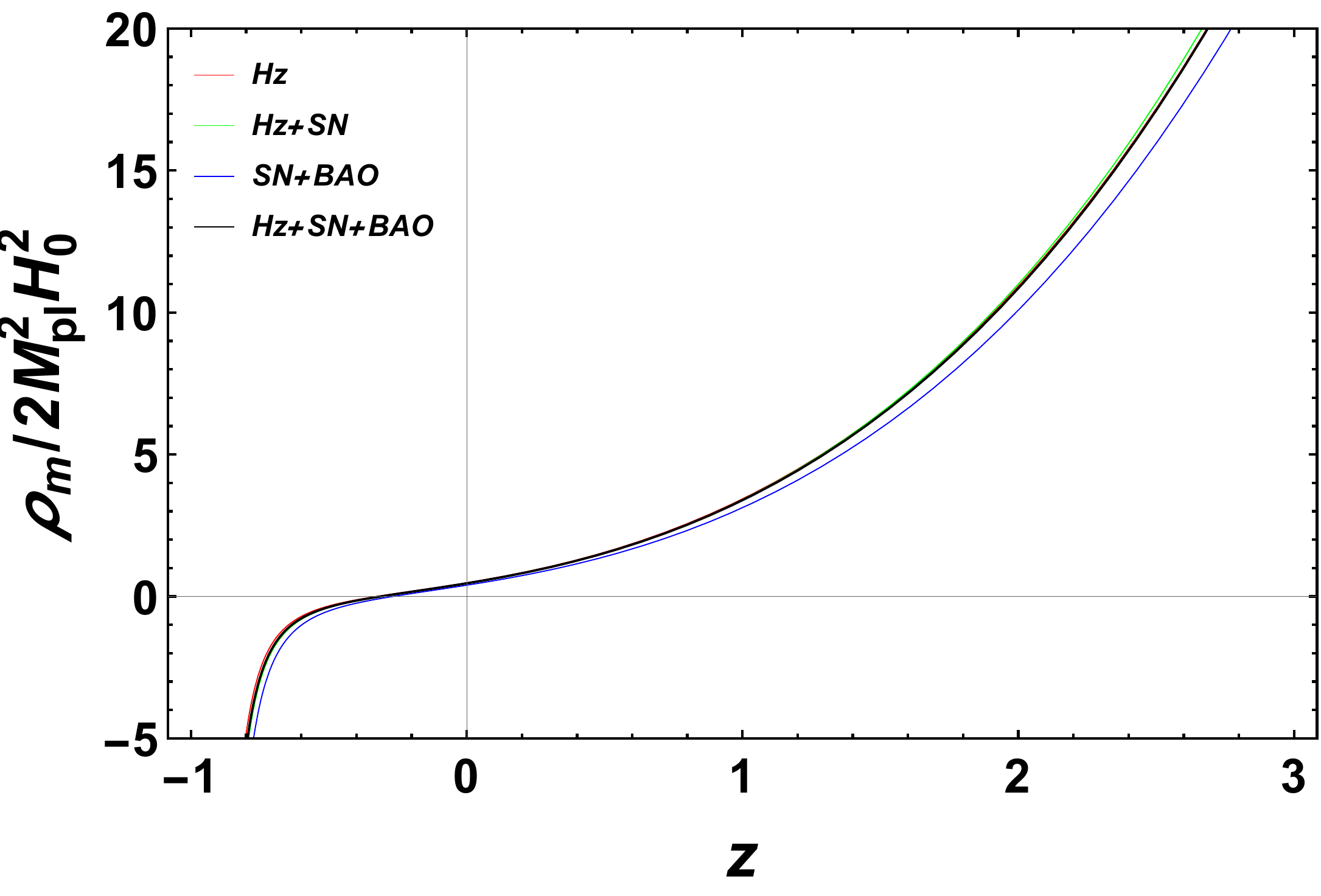} & %
\includegraphics[width=2.7 in, height=2.2 in]{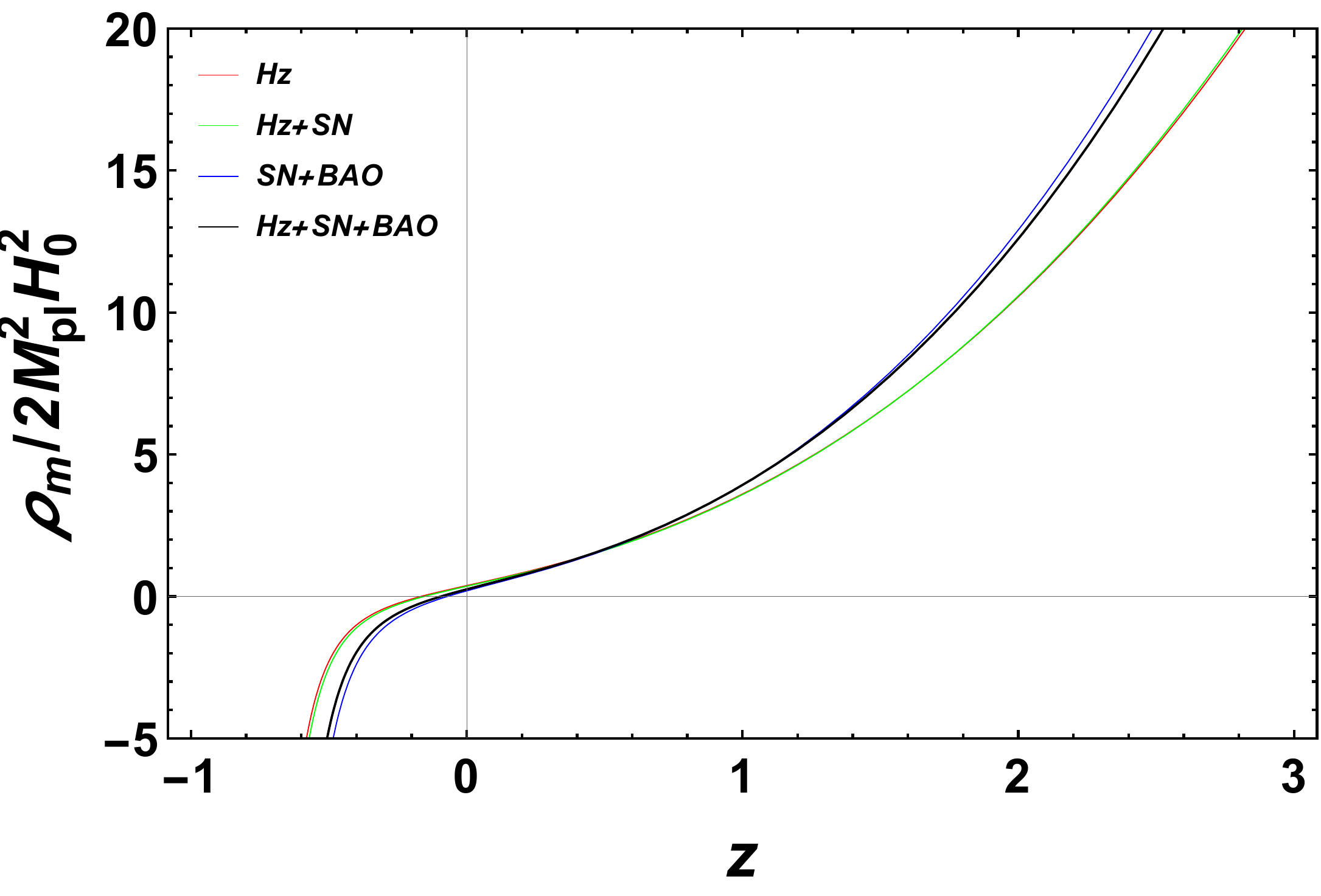} \\ 
\mbox (a) & \mbox (b)%
\end{array}
$%
\end{center}
\caption{ Figures (a) and (b) show the evolution of the matter energy
densities ($\protect\rho _{m}$) for models M1 and M2 respectively.}
\label{rho-m-CC}
\end{figure}

\begin{figure}[tbp]
\label{CC-lam-M2.pdf}
\par
\begin{center}
$%
\begin{array}{c@{\hspace{.1in}}c}
\includegraphics[width=2.7 in, height=2.2 in]{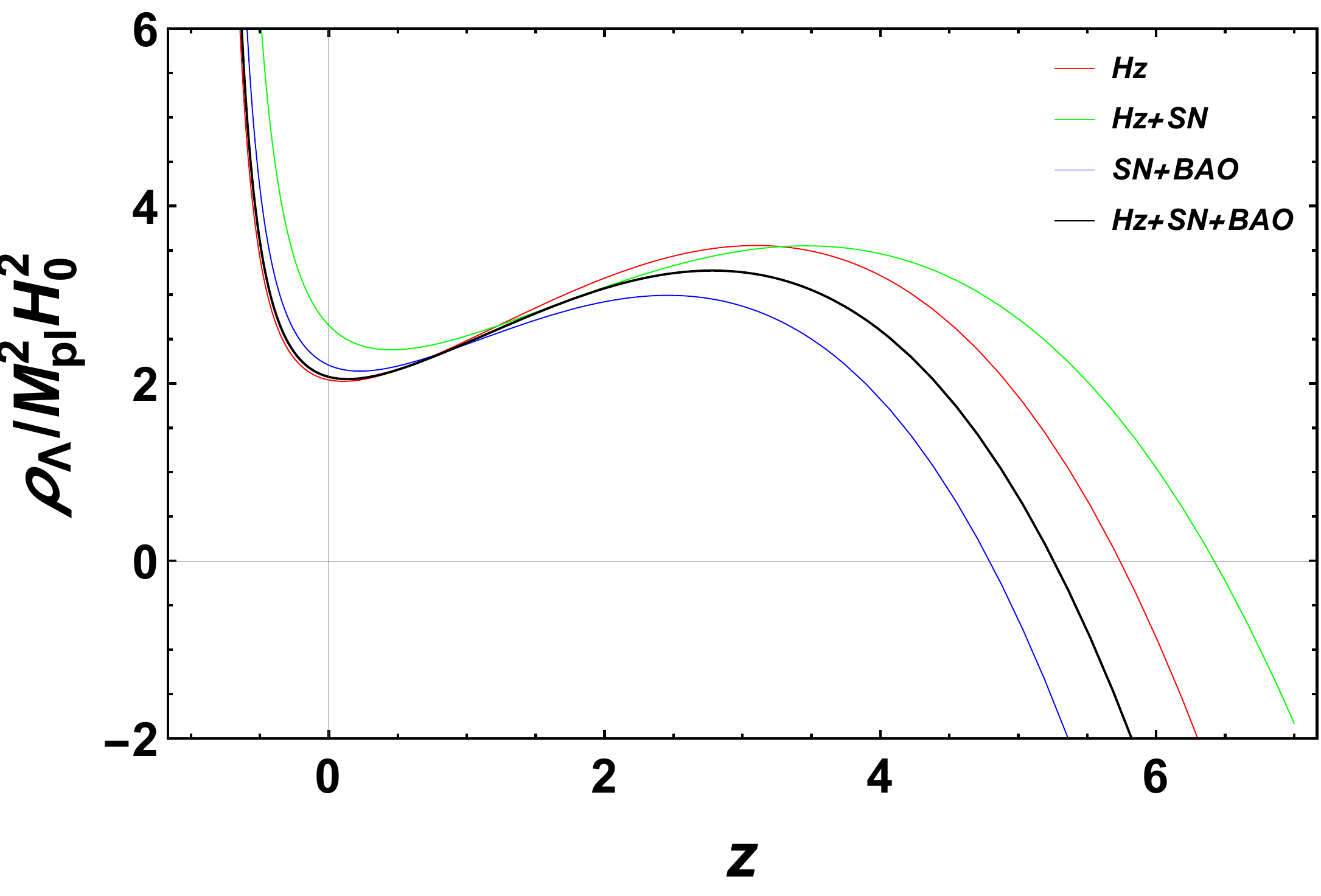} & %
\includegraphics[width=2.7 in, height=2.2 in]{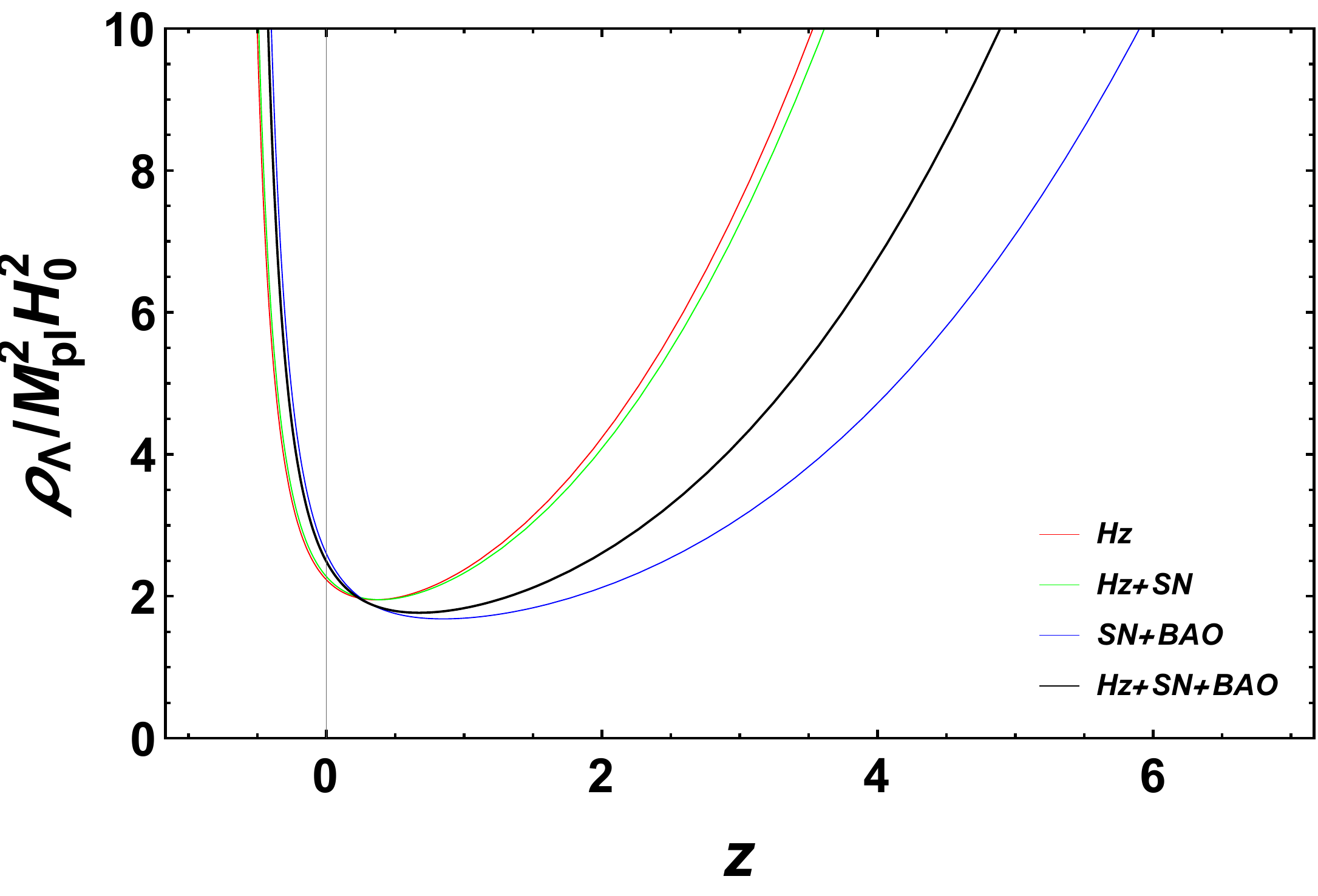} \\ 
\mbox (a) & \mbox (b)%
\end{array}
$%
\end{center}
\caption{ Figures (a) and (b) show the evolution of the energy densities of
the cosmological constant ($\protect\rho _{\Lambda }$) for models M1 and M2
respectively.}
\label{rho-lambda-CC}
\end{figure}

\qquad When the candidate of dark energy is the cosmological constant
implying $\rho _{DE}=\rho _{\Lambda }=M_{pl}^{-2}\Lambda $ and for which the
equation of state parameter $\omega _{DE}$ reduces to $-1$. Solving
equations (\ref{03}) and (\ref{04}), it is easy to obtain the explicit
expressions for the matter energy density and the energy density of
cosmological constant as,%
\begin{eqnarray}
\frac{\rho _{m}}{M_{pl}^{2}H_{0}^{2}} &=&\frac{2\alpha \left[ 1+\left\{
\beta (1+z)\right\} ^{\alpha }\right] ^{4}-4\alpha \left[ 1+\left\{ \beta
(1+z)\right\} ^{\alpha }\right] ^{3}}{\left( 1+\beta ^{\alpha }\right)
^{4}\left( 1+z\right) ^{2\alpha }}\text{,}  \label{ccM1rho} \\
\frac{\rho _{\Lambda }}{M_{pl}^{2}H_{0}^{2}} &=&\frac{\Lambda }{H_{0}^{2}}=%
\frac{(3-2\alpha )\left[ 1+\left\{ \beta (1+z)\right\} ^{\alpha }\right]
^{4}+4\alpha \left[ 1+\left\{ \beta (1+z)\right\} ^{\alpha }\right] ^{3}}{%
\left( 1+\beta ^{\alpha }\right) ^{4}\left( 1+z\right) ^{2\alpha }}\text{.}
\label{ccM1L}
\end{eqnarray}%
for model M1 and

\begin{eqnarray}
\frac{\rho _{m}}{M_{pl}^{2}H_{0}^{2}} &=&\frac{2\alpha \left[ 1+\left\{
\beta (1+z)\right\} ^{2\alpha }\right] ^{3}-6\alpha \left[ 1+\left\{ \beta
(1+z)\right\} ^{2\alpha }\right] ^{2}}{\left( 1+\beta ^{2\alpha }\right)
^{3}\left( 1+z\right) ^{4\alpha }}\text{,}  \label{ccM2rho} \\
\frac{\rho _{\Lambda }}{M_{pl}^{2}H_{0}^{2}} &=&\frac{\Lambda }{H_{0}^{2}}=%
\frac{(3-2\alpha )\left[ 1+\left\{ \beta (1+z)\right\} ^{2\alpha }\right]
^{3}+6\alpha \left[ 1+\left\{ \beta (1+z)\right\} ^{2\alpha }\right] ^{2}}{%
\left( 1+\beta ^{2\alpha }\right) ^{3}\left( 1+z\right) ^{4\alpha }}\text{.}
\label{ccM2L}
\end{eqnarray}%
for model M2. The evolution of these physical parameters are shown in the
FIG. \ref{rho-m-CC} and FIG. \ref{rho-lambda-CC}.

\begin{figure}[tbp]
\label{CC-omega-M2.pdf}
\par
\begin{center}
$%
\begin{array}{c@{\hspace{.1in}}c}
\includegraphics[width=2.7 in, height=2.2 in]{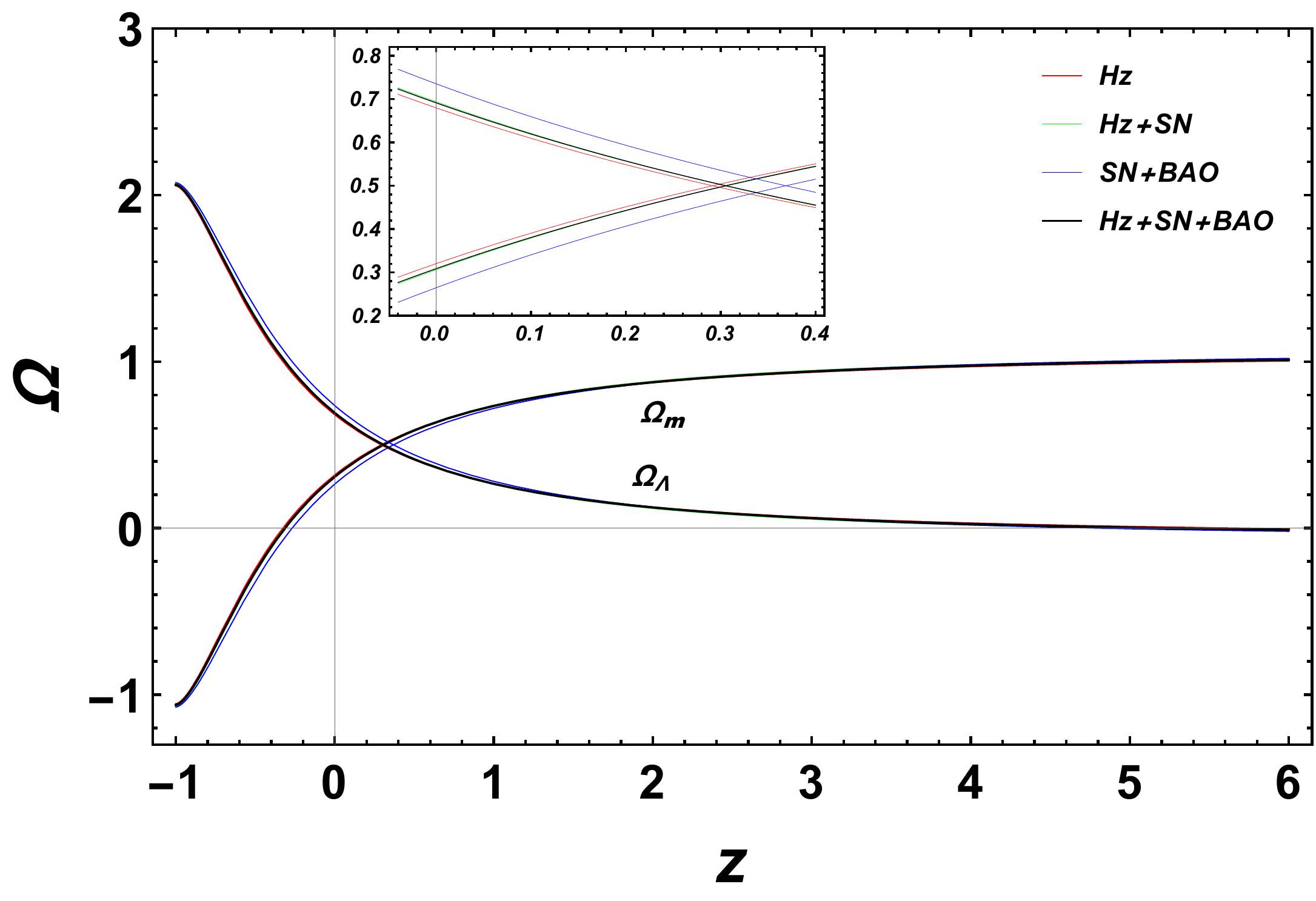} & %
\includegraphics[width=2.7 in, height=2.2 in]{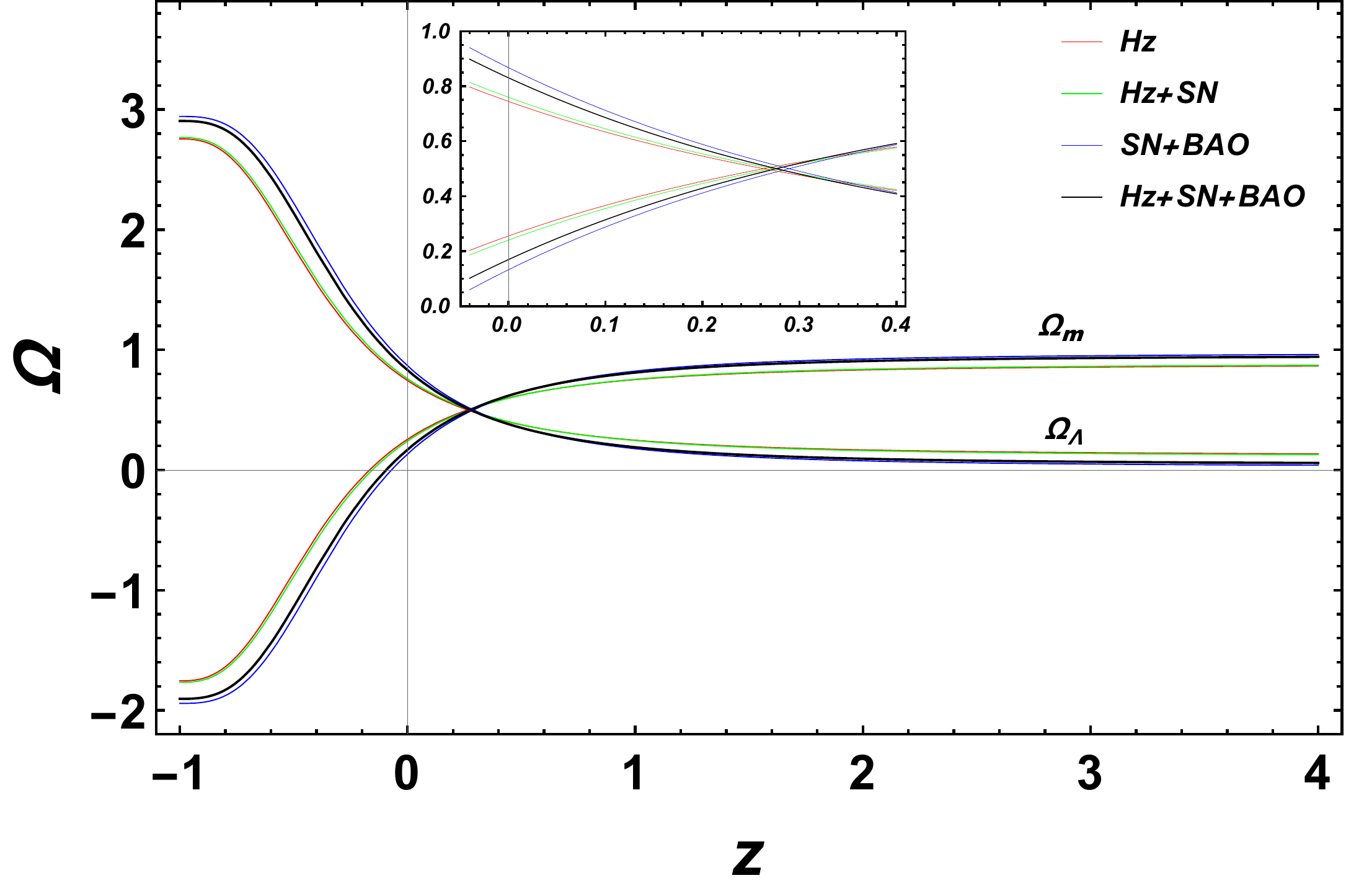} \\ 
\mbox (a) & \mbox (b)%
\end{array}
$%
\end{center}
\caption{ Figures (a) and (b) show the evolution of the density parameters
for matter ($\Omega _{m}$) and cosmological constant ($\Omega _{\Lambda }$)
for models M1 and M2 respectively.}
\label{density-CC}
\end{figure}

The density parameters for matter $\left( \Omega _{m}=\frac{\rho _{m}}{%
3Mpl^{2}H^{2}}\right) $ and density parameter for cosmological constant $%
\left( \Omega _{\Lambda }=\frac{\Lambda }{3H^{2}}\right) $ can also be
computed for both the models M1 and M2 as, 
\begin{equation}
\Omega _{m}=\frac{2\alpha }{3}-\frac{4\alpha }{3\left[ 1+\left\{ \beta
(1+z)\right\} ^{\alpha }\right] }\text{, }\Omega _{\Lambda }=1-\frac{2\alpha 
}{3}+\frac{4\alpha }{3\left[ 1+\left\{ \beta (1+z)\right\} ^{\alpha }\right] 
}  \label{denM1}
\end{equation}%
for model M1 and%
\begin{equation}
\Omega _{m}=\frac{2\alpha }{3}-\frac{2\alpha }{\left[ 1+\left\{ \beta
(1+z)\right\} ^{2\alpha }\right] }\text{, }\Omega _{\Lambda }=1-\frac{%
2\alpha }{3}+\frac{2\alpha }{\left[ 1+\left\{ \beta (1+z)\right\} ^{2\alpha }%
\right] }  \label{denM2}
\end{equation}%
for model M2. One can see from the above expressions that the sum total of
the density parameters with these components is equal to $1$. The evolution
of the density parameters are shown in FIG. \ref{density-CC}.

\subsection{Scalar field}

\qquad Since the equation of state for cosmological constant is non
dynamical and observations reveal it's dynamical characteristics, other
candidates such as a general scalar field came into picture for a suitable
candidate of dark energy. For an ordinary scalar field $\phi $ for the
action can be represented as,

\begin{equation}
S=\int d^{4}x\sqrt{-g}\left\{ \frac{M_{p}^{2}}{2}R-\frac{1}{2}\partial _{\mu
}\phi \partial ^{\mu }\phi -V(\phi )+L_{Matter}\right\} \text{.}  \label{A}
\end{equation}%
The term $V(\phi )$ is the potential function for the scalar field $\phi $.
In the considered FLRW background the energy density $\rho _{\phi }$ will
take the form $\rho _{\phi }=\frac{1}{2}\dot{\phi}^{2}+V\left( \phi \right) $
and pressure $p_{\phi }$ will take the form $p_{\phi }=\frac{1}{2}\dot{\phi}%
^{2}-V\left( \phi \right) $. For a two component Universe, scalar field and
cold dark matter with minimal interaction between them (i.e. they conserve
separately giving $\rho =ca^{-3}=c(1+z)^{3}$, $c$ is a constant of
integration), then the solutions obtained from Eqs. (\ref{03}) and (\ref{04}%
) are,

\begin{equation}
\frac{V(\phi )}{M_{pl}^{2}H_{0}^{2}}=\frac{(3-\alpha )\left[ 1+\left\{ \beta
(1+z)\right\} ^{\alpha }\right] ^{4}+2\alpha \left[ 1+\left\{ \beta
(1+z)\right\} ^{\alpha }\right] ^{3}}{\left( 1+\beta ^{\alpha }\right)
^{4}\left( 1+z\right) ^{2\alpha }}-\frac{c}{2M_{pl}^{2}H_{0}^{2}}(1+z)^{3}
\label{SCPM1}
\end{equation}

\begin{equation}
\frac{\rho _{\phi }}{M_{pl}^{2}H_{0}^{2}}=\frac{3\left[ 1+\left\{ \beta
(1+z)\right\} ^{\alpha }\right] ^{4}}{\left( 1+\beta ^{\alpha }\right)
^{4}\left( 1+z\right) ^{2\alpha }}-\frac{c}{M_{pl}^{2}H_{0}^{2}}(1+z)^{3}
\label{quint-rhoM1}
\end{equation}%
and the expression for the scalar field $\phi (z)$ can be calculated by
integrating,

\begin{equation}
\frac{\phi -\phi _{0}}{\sqrt{2}M_{pl}}=-\int \left[ \frac{\alpha \left[
1+\left\{ \beta (1+z)\right\} ^{\alpha }\right] ^{4}-2\alpha \left[
1+\left\{ \beta (1+z)\right\} ^{\alpha }\right] ^{3}}{\left( 1+\beta
^{\alpha }\right) ^{4}\left( 1+z\right) ^{2\alpha }}-\frac{c}{%
2M_{pl}^{2}H_{0}^{2}}(1+z)^{3}\right] ^{\frac{1}{2}}\frac{(1+\beta ^{\alpha
})^{2}(1+z)^{\alpha -1}}{\left[ 1+\left\{ \beta (1+z)\right\} ^{\alpha }%
\right] ^{2}}dz  \label{quint-fieldM1}
\end{equation}%
for model M1. Here, $\phi _{0}$ is an integrating constant. Similarly, the
potential and energy densities for model M2 are obtained as,

\begin{equation}
\frac{V(\phi )}{M_{pl}^{2}H_{0}^{2}}=\frac{(3-\alpha )\left[ 1+\left\{ \beta
(1+z)\right\} ^{2\alpha }\right] ^{3}+3\alpha \left[ 1+\left\{ \beta
(1+z)\right\} ^{2\alpha }\right] ^{2}}{\left( 1+\beta ^{2\alpha }\right)
^{3}\left( 1+z\right) ^{4\alpha }}-\frac{c}{2M_{pl}^{2}H_{0}^{2}}(1+z)^{3}
\label{SCPM2}
\end{equation}

\begin{equation}
\frac{\rho _{\phi }}{M_{pl}^{2}H_{0}^{2}}=\frac{3\left[ 1+\left\{ \beta
(1+z)\right\} ^{2\alpha }\right] ^{3}}{\left( 1+\beta ^{2\alpha }\right)
^{3}\left( 1+z\right) ^{4\alpha }}-\frac{c}{M_{pl}^{2}H_{0}^{2}}(1+z)^{3}
\label{quint-rhoM2}
\end{equation}%
and the expression for the scalar field $\phi (z)$ can be calculated by
integrating,

\begin{equation}
\frac{\phi -\phi _{0}}{\sqrt{2}M_{pl}}=-\int \left[ \frac{\alpha \left[
1+\left\{ \beta (1+z)\right\} ^{2\alpha }\right] ^{3}-3\alpha \left[
1+\left\{ \beta (1+z)\right\} ^{2\alpha }\right] ^{2}}{\left( 1+\beta
^{2\alpha }\right) ^{3}\left( 1+z\right) ^{4\alpha }}-\frac{c}{%
2M_{pl}^{2}H_{0}^{2}}(1+z)^{3}\right] ^{\frac{1}{2}}\frac{(1+\beta ^{2\alpha
})^{\frac{3}{2}}(1+z)^{2\alpha -1}}{\left[ 1+\left\{ \beta (1+z)\right\}
^{2\alpha }\right] ^{\frac{3}{2}}}dz\text{.}  \label{qunit-phiM2}
\end{equation}

\begin{figure}[tbp]
\label{rho-phi-M2.pdf}
\par
\begin{center}
$%
\begin{array}{c@{\hspace{.1in}}c}
\includegraphics[width=2.7 in, height=2.2 in]{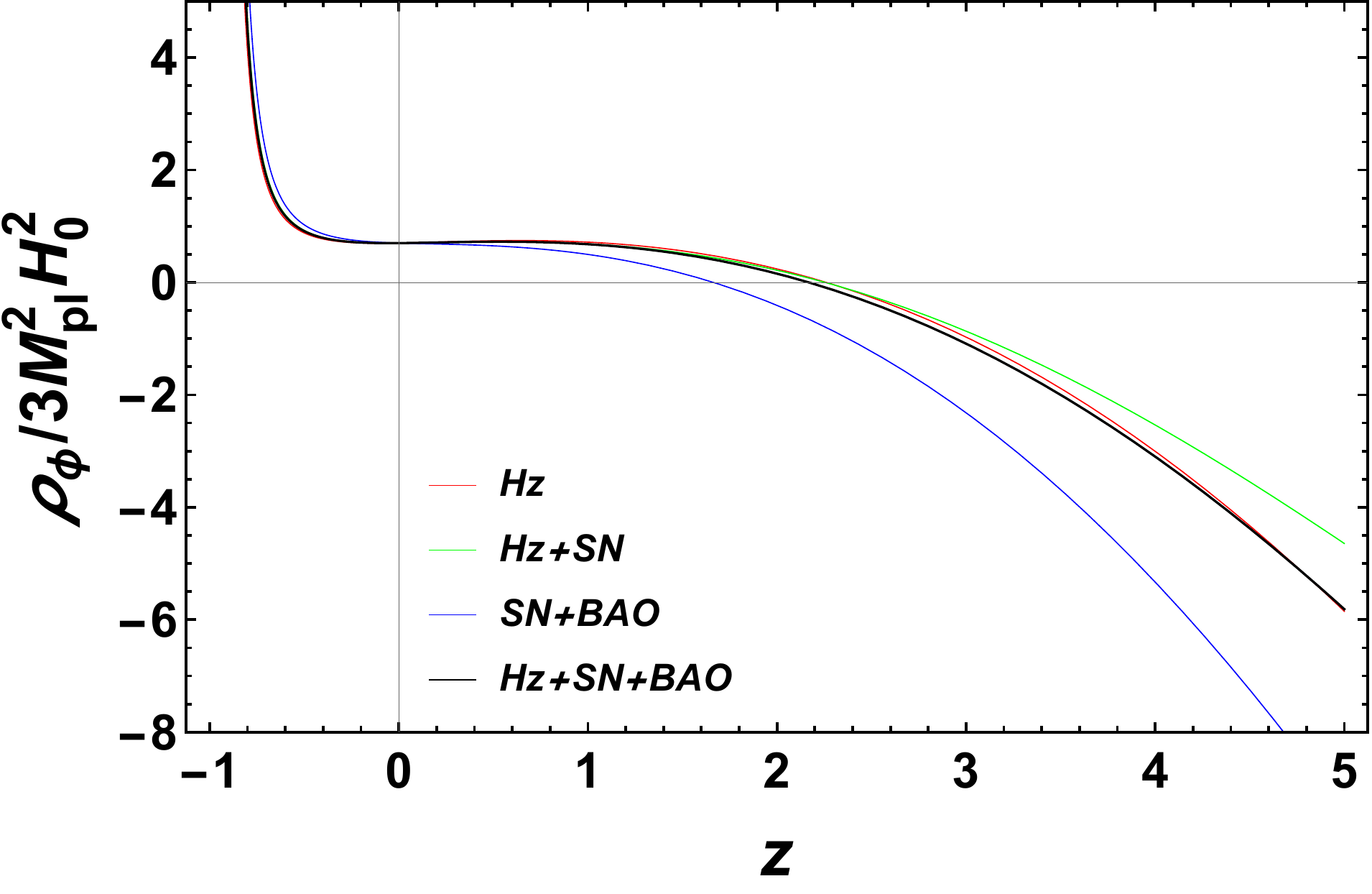} & %
\includegraphics[width=2.7 in, height=2.2 in]{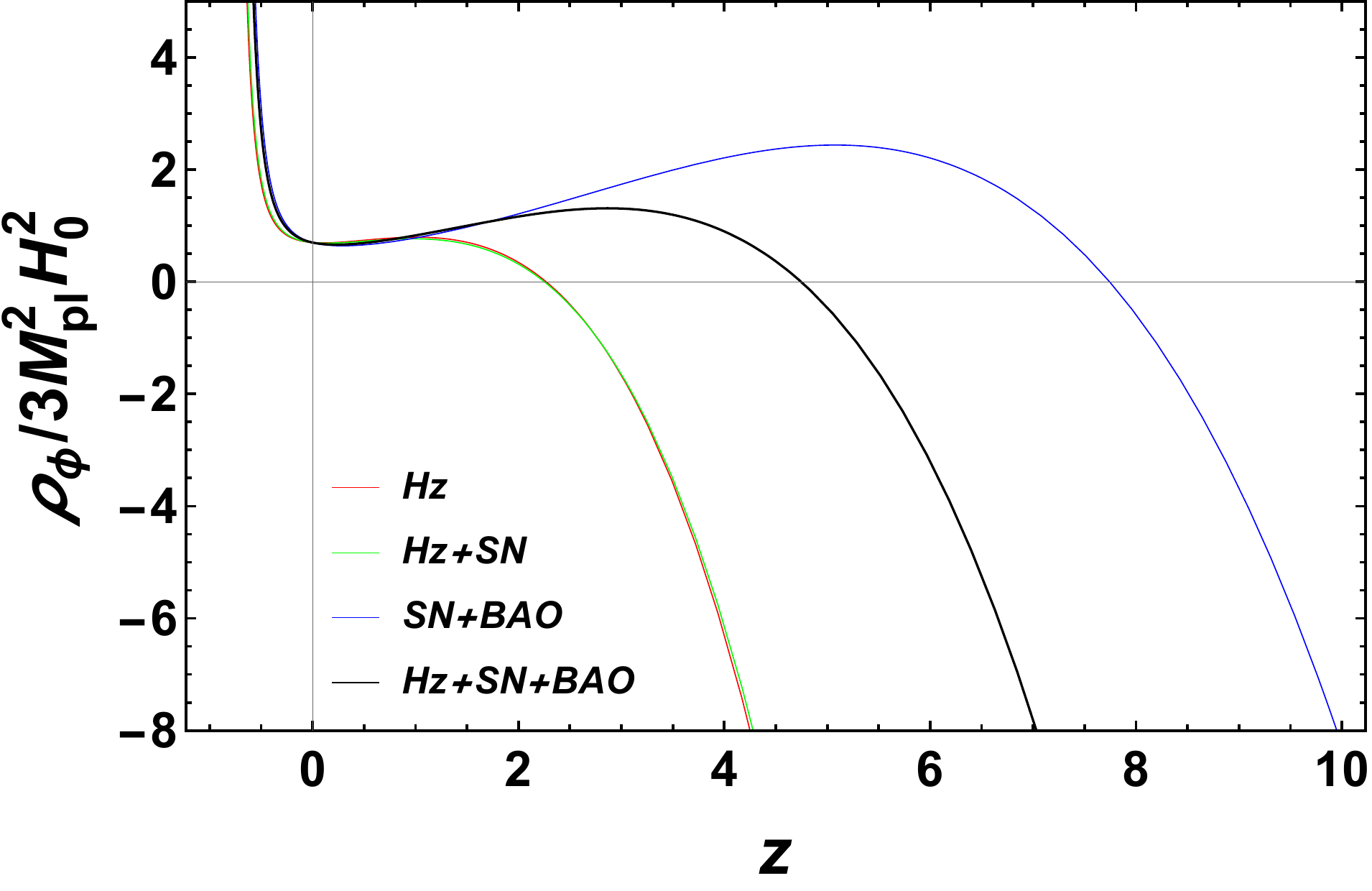} \\ 
\mbox (a) & \mbox (b)%
\end{array}
$%
\end{center}
\caption{ Figures (a) and (b) show the evolution of the scalar field energy
density ($\protect\rho _{\protect\phi }$) for models M1 and M2 respectively.}
\label{rho-phi-SF}
\end{figure}

The density parameters for matter $\left( \Omega _{m}=\frac{\rho _{m}}{%
3M^{2}plH^{2}}\right) $ and density parameter for the scalar field $\left(
\Omega _{\phi }=\frac{\rho _{\phi }}{3M^{2}plH^{2}}=\frac{\frac{1}{2}\dot{%
\phi}^{2}+V\left( \phi \right) }{3M^{2}plH^{2}}\right) $ can be computed for
both the models M1 and M2 as, 
\begin{equation}
\Omega _{\phi }=1-\Omega _{m}\text{\thinspace , }\Omega _{m}=\frac{%
c(1+z)^{3}\left( 1+\beta ^{\alpha }\right) ^{4}\left( 1+z\right) ^{2\alpha }%
}{3M_{pl}^{2}H_{0}^{2}\left[ 1+\left\{ \beta (1+z)\right\} ^{\alpha }\right]
^{4}}  \label{denscM1}
\end{equation}%
for model M1 and%
\begin{equation}
\Omega _{\phi }=1-\Omega _{m}\text{\thinspace , }\Omega _{m}=\frac{%
c(1+z)^{3}\left( 1+\beta ^{2\alpha }\right) ^{3}\left( 1+z\right) ^{4\alpha }%
}{3M_{pl}^{2}H_{0}^{2}\left[ 1+\left\{ \beta (1+z)\right\} ^{2\alpha }\right]
^{3}}  \label{denscM2}
\end{equation}%
for model M2. From equations (\ref{denscM1}) and (\ref{denscM2}), one
obtains $\Omega _{m0}=\frac{c}{3M_{pl}^{2}H_{0}^{2}}\Longrightarrow c=\Omega
_{m0}3M_{pl}^{2}H_{0}^{2}$. The equations of state parameter ($\omega _{\phi
}=\frac{p_{\phi }}{\rho _{\phi }}$) are given by, 
\begin{equation}
\omega _{\phi }^{eff}=\frac{1}{3}\frac{(2\alpha -3)\left[ 1+\left\{ \beta
(1+z)\right\} ^{\alpha }\right] ^{4}-4\alpha \left[ 1+\left\{ \beta
(1+z)\right\} ^{\alpha }\right] ^{3}}{\left[ 1+\left\{ \beta (1+z)\right\}
^{\alpha }\right] ^{4}-\Omega _{m0}\left( 1+\beta ^{\alpha }\right)
^{4}\left( 1+z\right) ^{2\alpha +3}}  \label{eosM1}
\end{equation}%
\begin{equation}
\omega _{\phi }^{eff}=\frac{1}{3}\frac{(2\alpha -3)\left[ 1+\left\{ \beta
(1+z)\right\} ^{2\alpha }\right] ^{3}-6\alpha \left[ 1+\left\{ \beta
(1+z)\right\} ^{2\alpha }\right] ^{2}}{\left[ 1+\left\{ \beta (1+z)\right\}
^{2\alpha }\right] ^{3}-\Omega _{m0}\left( 1+\beta ^{2\alpha }\right)
^{3}\left( 1+z\right) ^{4\alpha +3}}  \label{eosM2}
\end{equation}%
The evolution of the Scalar field energy density, scalar field potential and
the density parameters are shown in FIG. \ref{rho-phi-SF}, FIG. \ref%
{v-phi-SF} and FIG. \ref{density-SF} respectively for models M1 and M2.

\begin{figure}[tbp]
\label{V-phi-M2.pdf}
\par
\begin{center}
$%
\begin{array}{c@{\hspace{.1in}}c}
\includegraphics[width=2.7 in, height=2.2 in]{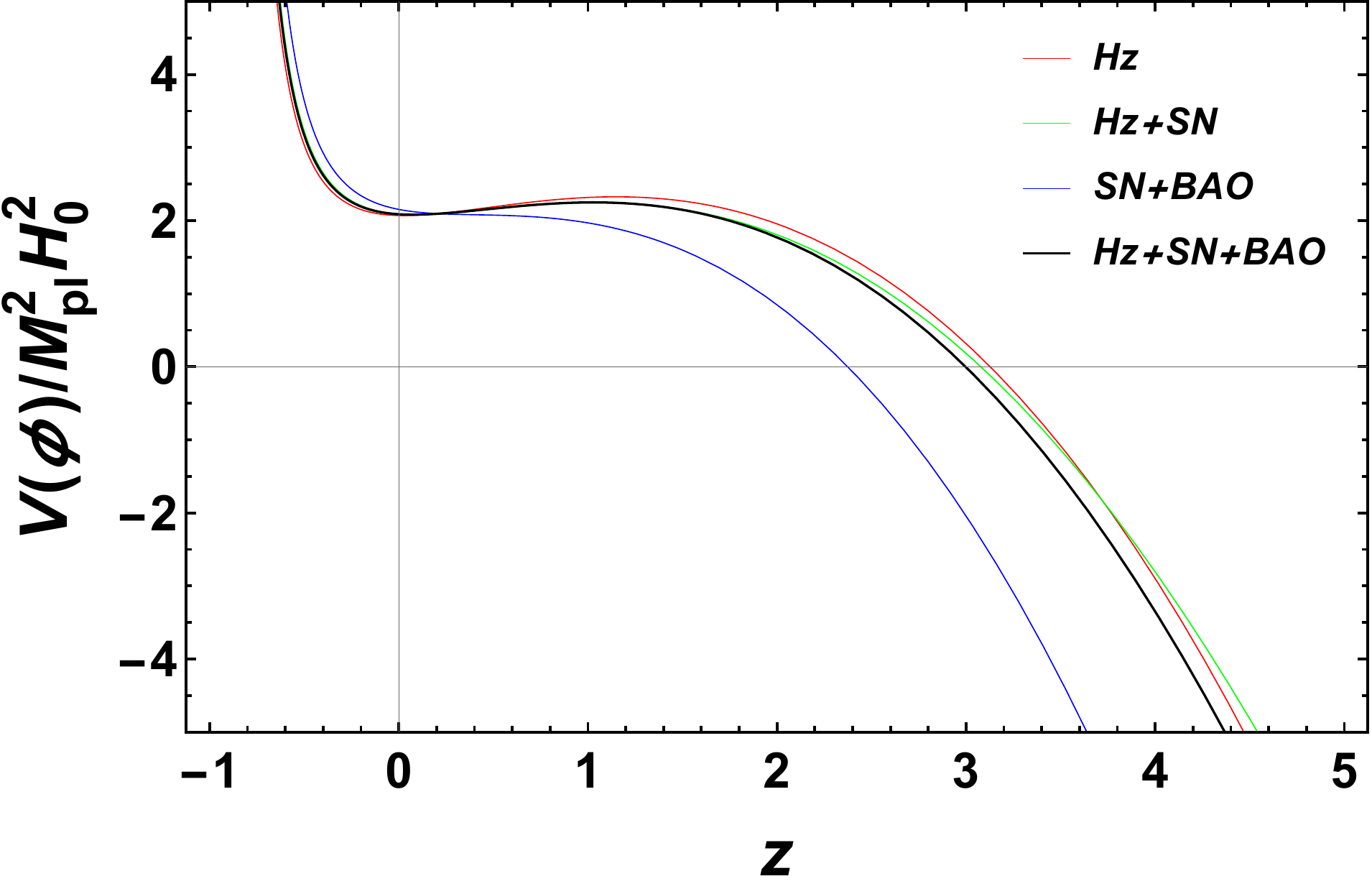} & %
\includegraphics[width=2.7 in, height=2.2 in]{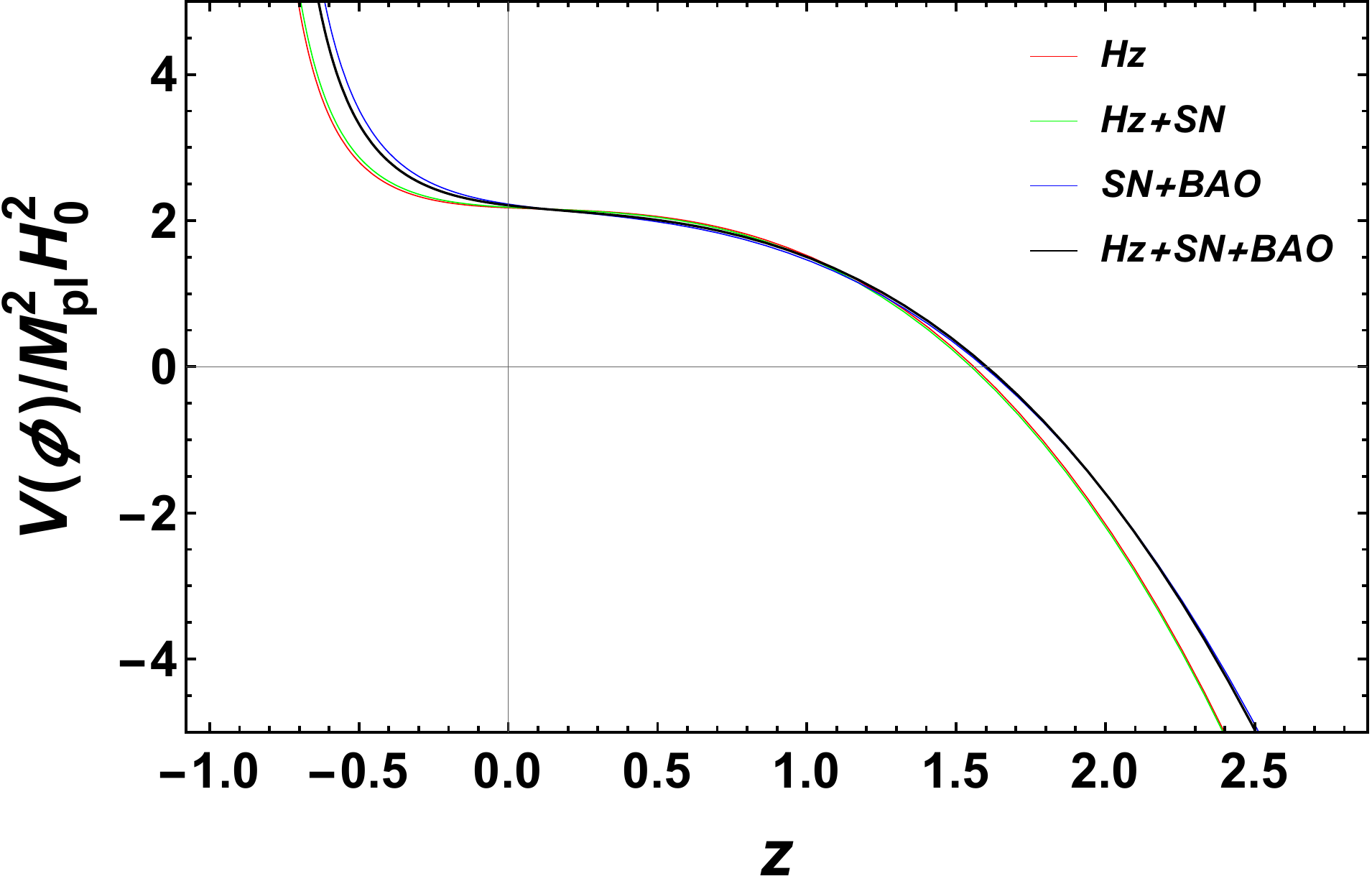} \\ 
\mbox (a) & \mbox (b)%
\end{array}
$%
\end{center}
\caption{ Figures (a) and (b) show the evolution of the scalar field
potential $V({\protect\phi )}\sim z$ for models M1 and M2 respectively.}
\label{v-phi-SF}
\end{figure}

\begin{figure}[tbp]
\label{Density-M2.pdf}
\par
\begin{center}
$%
\begin{array}{c@{\hspace{.1in}}c}
\includegraphics[width=2.7 in, height=2.2 in]{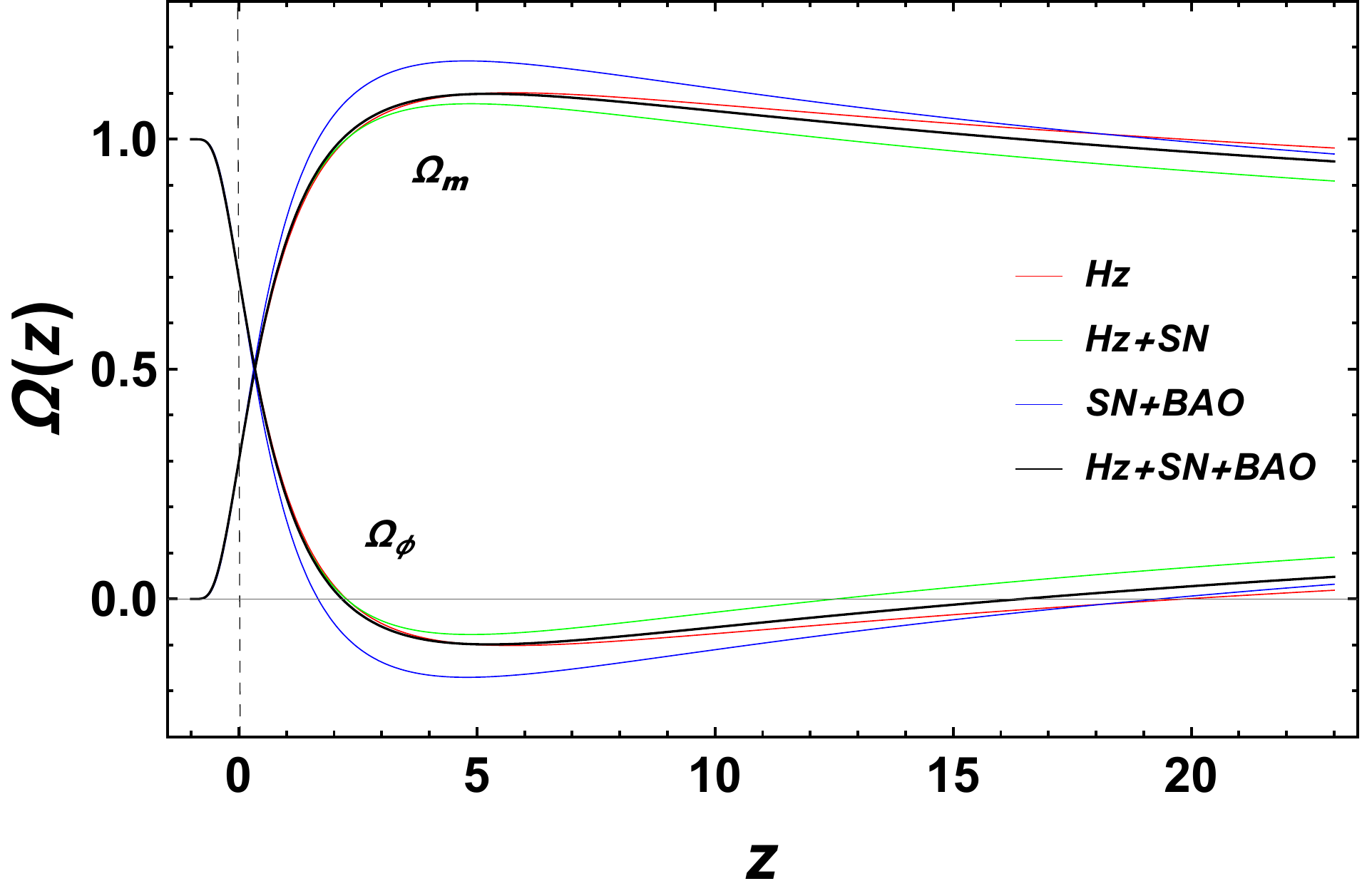} & %
\includegraphics[width=2.7 in, height=2.2 in]{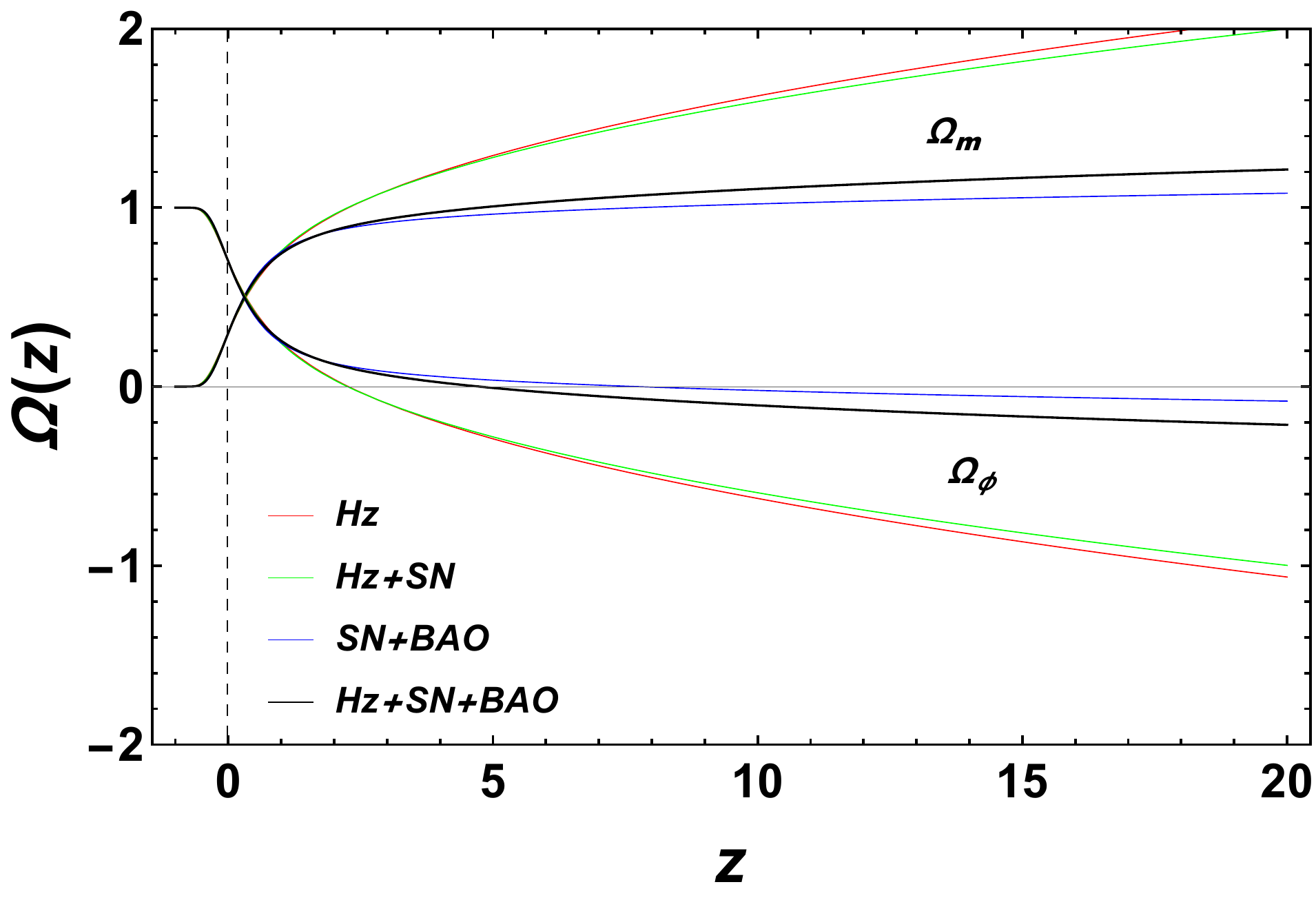} \\ 
\mbox (a) & \mbox (b)%
\end{array}
$%
\end{center}
\caption{ Figures (a) and (b) show the evolution of the density parameters $%
\Omega _{\protect\phi }$ \& $\Omega _{m}$ w.r.t. redshift $z$ for models M1
and M2 respectively.}
\label{density-SF}
\end{figure}

\begin{figure}[tbp]
\label{Eos-Scalar-M2.pdf}
\par
\begin{center}
$%
\begin{array}{c@{\hspace{.1in}}c}
\includegraphics[width=2.7 in, height=2.2 in]{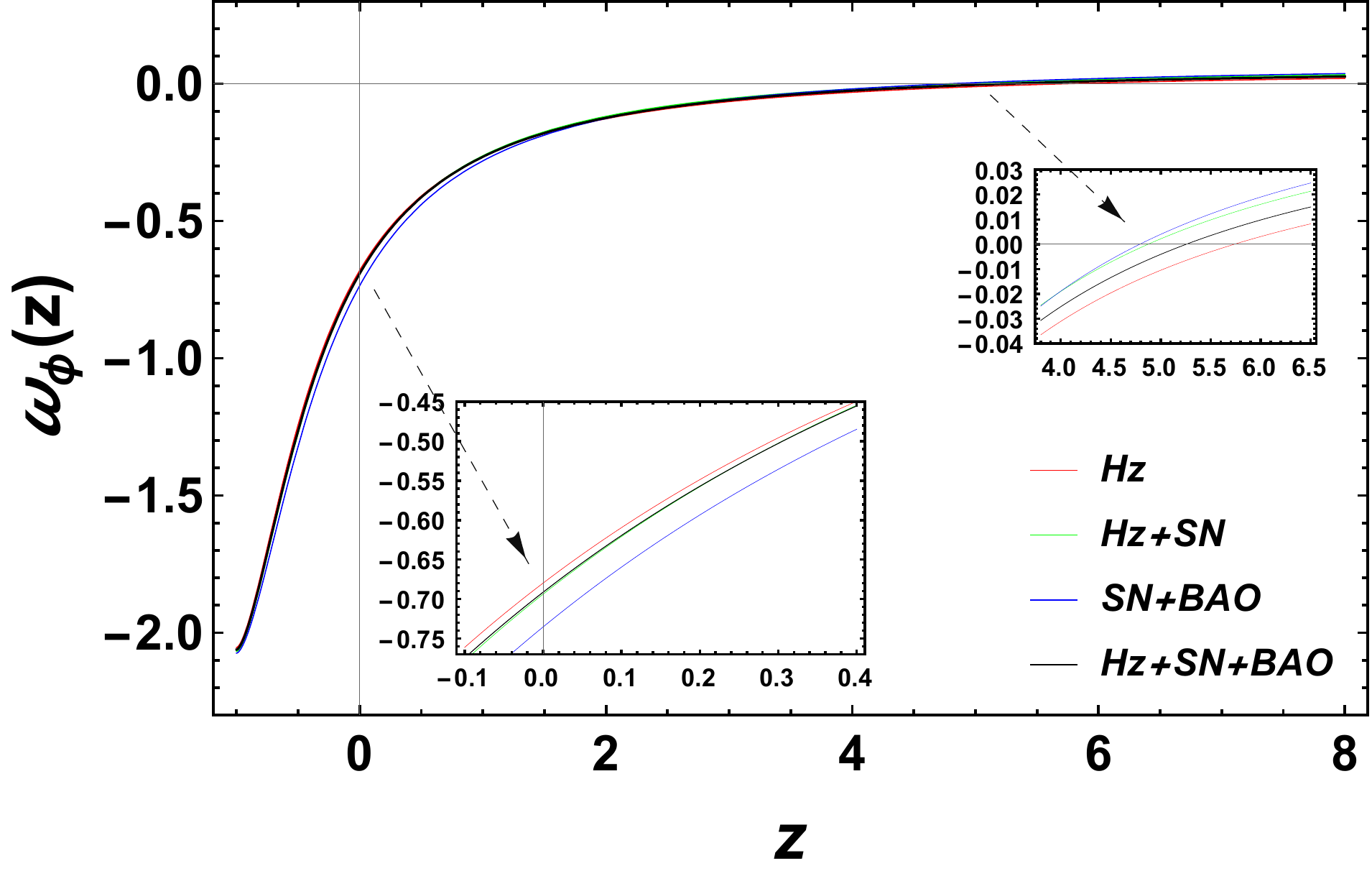} & %
\includegraphics[width=2.7 in, height=2.2 in]{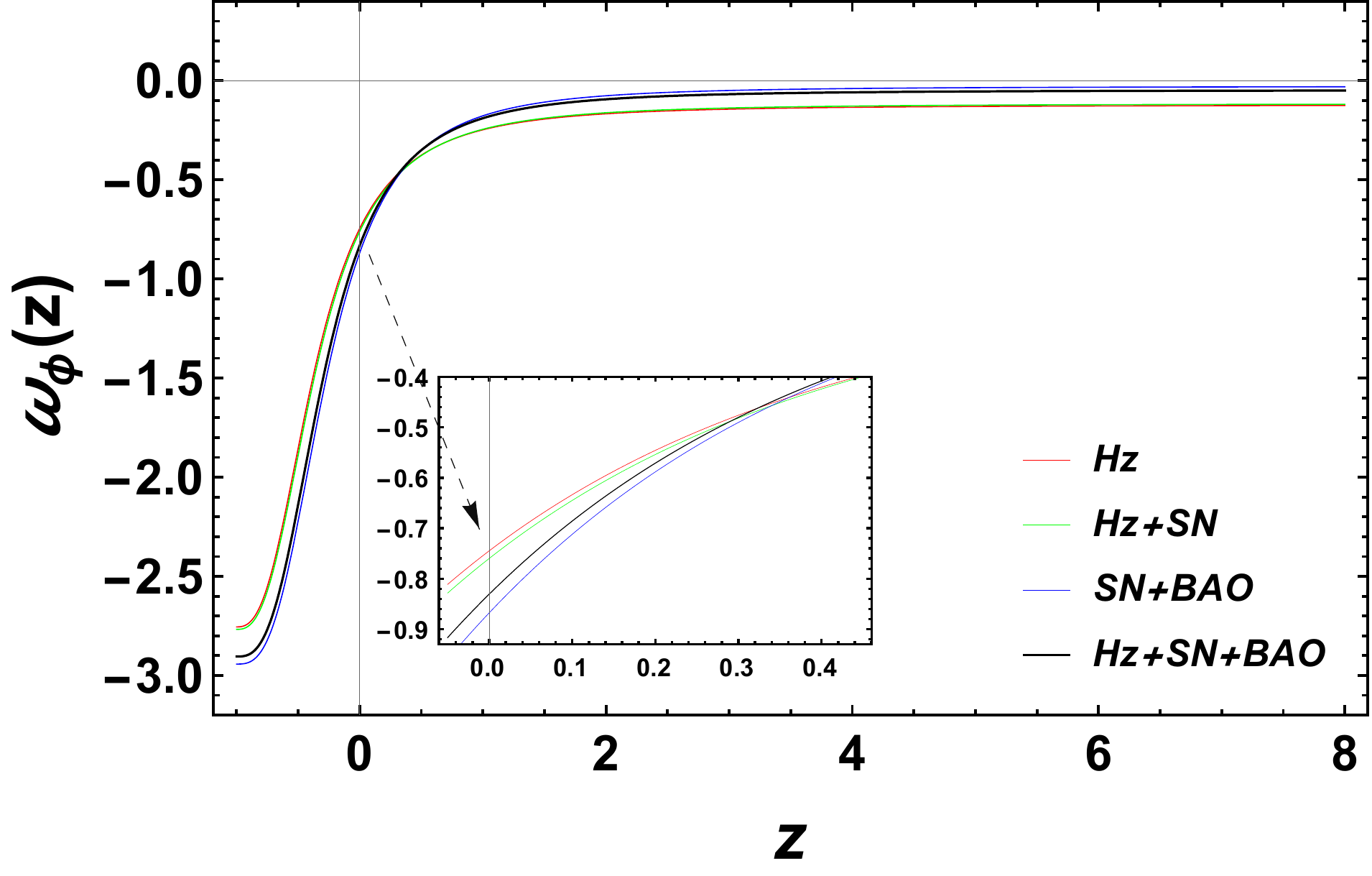} \\ 
\mbox (a) & \mbox (b)%
\end{array}
$%
\end{center}
\caption{ Figures (a) and (b) show the evolution of equation of state
parameter vs. redshift ($\protect\omega _{\protect\phi }(z)\sim z$) for
models M1 and M2 respectively.}
\label{w-phi-SF}
\end{figure}

The evolution of the equation of state parameter ($\omega _{\phi }(z)$) vs.
redshift $z$ is plotted by neglecting the matter contribution and shown in
the FIG. \ref{w-phi-SF} for models M1 and M2.

\section{Age of the Universe}

The calculation of the age of the Universe is associated to the values of
the cosmological parameters, specifically the Hubble parameter. In general,
using the Friedmann equation one can obtain the relation as $t_{0}=\frac{1}{%
H_{0}}F(\Omega _{x})$, $x=$ radiation, matter, dark energy, neutrino etc.
The functional $F$ contributes a fraction and largely the term $1/H_{0}$ in
the age calculation e.g. for $H_{0}=69$ $km/s/Mpc$, one obtains $%
1/H_{0}\approx $ $14.5$ $Gyr$ (Giga years) and the factor $F=0.956$ for $%
\Lambda $CDM model with $\left( \Omega _{m},\Omega _{\Lambda }\right)
=\left( 0.3086,0.6914\right) $ giving pretty good estimate of $t_{0}$ and $%
F=0.666$ for Einstein-de-Sitter model with $\left( \Omega _{m},\Omega
_{\Lambda }\right) =\left( 1,0\right) $ giving much smaller value of $t_{0}$%
. So, the introduction of cosmological constant is significant as
matter-only Universe was not enough to explain the globular clusters in the
Milky Way which appeared to be older than the age of the Universe calculated
then. According to the Planck2015 results age of the universe is estimated
to be $13.799\pm 0.021$ $Gyr$ with $H_{0}=67.74\pm 0.46$ within $68\%$
confidence limits for $\Lambda $CDM model constrained by combined CMB power
spectra, Planck polarization data, CMB lensing reconstruction and external
data of BAO, JLA (Joint light curve analysis) and Hubble datasets.

Here, the present work is a model independent study wherein the geometrical
parameter $H$ is parametrized for which the calculation of the age is
unaffected by the matter content and solely depend on the functional form of
the Hubble parameter $H(t)$. We have already established the $t$-$z$
relationships for models M1 and M2 which can be rewritten as,

\begin{equation*}
t(z)=\frac{\left( 1+\beta ^{\alpha }\right) ^{2}}{\alpha \beta ^{\alpha }%
\left[ 1+\left\{ \beta (1+z)\right\} ^{\alpha }\right] }\frac{1}{H_{0}}\text{
and }t(z)=\frac{\left( 1+\beta ^{2\alpha }\right) ^{\frac{3}{2}}}{\alpha
\beta ^{2\alpha }\left[ 1+\left\{ \beta (1+z)\right\} ^{2\alpha }\right] ^{%
\frac{1}{2}}}\frac{1}{H_{0}}
\end{equation*}%
respectively. By considering the present value of the Hubble parameter, $%
H_{0}=67.8$ $Km/Sec/Mpc$, the terms multiplied to $1/H_{0}$ are calculated
and for both the models are greater than $0.96$ for all constrained values
of $\alpha $ \& $\beta $ and give pretty good estimate for the present age
of the Universe and is larger than the standard model. The age calculation
is tabulated in the following Table-7 for all the constrained numerical
values of $\alpha $ \& $\beta $ for both the models M1 and M2.

\begin{center}
\begin{tabular}{|c|c|c|c||c|c|c|}
\hline
\multicolumn{7}{|c|}{Table-7} \\ \hline
Models & \multicolumn{3}{|c}{M1} & \multicolumn{3}{||c|}{M2} \\ \hline
Datasets & $(\alpha ,\beta )$ & Factor & Age (in $Gyr$) & $(\alpha ,\beta )$
& Factor & Age (in $Gyr$) \\ \hline
$Hz$ & $(1.58064,1.48729)$ & $0.97046$ & $14.0068$ & $(1.31611,1.56124)$ & $%
0.99502$ & $14.3613$ \\ \hline
$Hz+SN$ & $(1.60094,1.44572)$ & $0.97084$ & $14.0123$ & $(1.32551,1.53587)$
& $0.99630$ & $14.3797$ \\ \hline
$SN+BAO$ & $(1.61116,1.36647)$ & $0.99597$ & $14.3751$ & $(1.45677,1.36451)$
& $0.96400$ & $13.9136$ \\ \hline
$Hz+SN+BAO$ & $(1.59173,1.45678)$ & $0.97343$ & $14.0496$ & $%
(1.42829,1.40637)$ & $0.96445$ & $13.9201$ \\ \hline
\end{tabular}
\end{center}

\section{Results and Conclusion}

To summarize the results, the philosophy behind writing this present paper
is to discuss the phenomenology of cosmological parametrization to obtain
exact solutions of Einstein field equations. As an exemplification, a simple
parametrization of Hubble parameter is considered with some model parameters
which reduce to some known models (see \cite{pacif2016}) for some specific
values of the model parameters involved. Two models discussed here in
details and both the models M1 and M2 exhibit a phase transition from
deceleration to acceleration. Also, both the models diverges in finite time
and show big rip singularity. For consistency of the models obtained here,
some observational datasets namely, $H(z)$ datasets with updated $57$ data
points, Supernovae datasets from union 2.1 compilation datasets containing $%
580$ data points and BAO datasets with $6$ data points are considered and
compared with the standard $\Lambda $CDM model. Both the models M1 and M2
contain two model parameters $\alpha $ \& $\beta $ which are constrained
through these datasets and some numerical values are obtained in pairs with
independent $Hz$, combined $Hz+SN$, $SN+BAO$ and $Hz+SN+BAO$ datasets which
are then used for further analysis for geometrical and physical
interpretations of the models. The present values of the deceleration
parameters obtained for these constrained values of model parameters $\alpha 
$ \& $\beta $ are calculated which are tabulated in Table-5 and Table-6
together with the phase transition redshifts and are in certain standard
estimated range. In the future the Universe in both the models enters into
super acceleration phases and diverges in finite times. The other
geometrical parameters such as jerk, snap and lerk parameters are also
discussed and their evolutions are shown graphically. The statefinder
diagnostics and $om$ diagnostics are also presented to compare the obtained
models with the standard $\Lambda $CDM model and the models behavior are
shown in plots compared with the standard $\Lambda $CDM model and SCDM
model. After the brief cosmographic analysis, the physical interpretation of
the models are discussed by considering the cosmological constant and scalar
field as candidates of dark energy. The matter energy density ($\rho _{m}$),
energy density of ($\rho _{\Lambda }$), the density parameters $\Omega
_{\Lambda }$ and $\Omega _{m}$ are also calculated and their dynamical
behavior w.r.t. redshift $z$ are shown graphically for both the models M1
and M2 using the numerical constrained values of the model parameters $%
\alpha $ \& $\beta $. Similarly, the evolution of energy density ($\rho
_{\phi }$), the potential ($V(\phi )$) of the scalar field ($\phi $),
density parameters $\Omega _{\Lambda }$ and $\Omega _{m}$ and also the
equation of state parameter $\omega _{\phi }$ of the scalar field are shown
graphically for both the models M1 and M2. The geometrical and physical
analysis for both the models M1 and M2 interpret that both models M1 and M2
have the quintessence behavior in the past and phantom like behavior in the
future. Finally, the age of the Universe for both the models M1 and M2 are
calculated for the constrained numerical values of model parameters $\alpha $
\& $\beta $. It is found that the age found for both the models are greater
than the standard model and consistent with the age constraints of $\Lambda $%
CDM model.

The conclusion is that the model M2 which is a quadratic varying
deceleration parameter model has better fit to the observational datasets
(see FIG. \ref{error-Hz} and FIG. \ref{error-Sn}) and shows better
approximation to the present cosmological scenario on geometrical as well as
physical grounds as compared to the model M1 which is a linearly varying
deceleration parameter model. The presented study is an example of doing a
comprehensive analysis of any cosmological model that describe a simple
methodology of finding exact solution of the Einstein field equations,
comparing to the observations and estimating model parameters from the
observational datasets. A brief list of various schemes of parametrization
of different geometrical and physical parameters used in the past few
decades to obtain the exact solutions of EFEs are also summarized here which
will help the readers for their studies in cosmological modelling.

\section{Appendix}

A brief list of various parametrization schemes of parametrization of
geometrical and physical parameters used in the past few decades to find
exact solutions of Einstein Field Equations is given below.

\subsection{\textbf{PARAMETRIZATIONS OF GEOMETRICAL PARAMETERS}}

\textbf{Scale factor }$a(t)$

Given below a list of different expansion laws of the scale factor those
have been extensively studied in different contexts.

$a(t)=constant$ \cite{static} (Static model)

$a(t)=ct$ \cite{milne, melia} (Milne model or Linear expansion)

$a(t)\sim \exp (H_{0}t)$ \cite{ellismadsen} ($\Lambda $CDM model or
Exponential expansion)

$a(t)\sim \exp \left[ -\alpha t\ln \left( \frac{t}{t_{0}}\right) +\beta t%
\right] $ \cite{a-zhou} (Inflationary model)

$a(t)\sim \exp \left[ -\alpha t-\beta t^{n}\right] $ \cite{a-zhou}
(Inflationary model)

$a(t)\sim \left[ \exp (\alpha t)-\beta \exp (-\alpha t)\right] ^{n}$ \cite%
{a-zhou} (Inflationary model)

$a(t)\sim \exp \left( \frac{t}{M}\right) \left[ 1+\cos \left( \frac{%
\varsigma (t)}{N}\right) \right] $ \cite{qssc} (quasi steady state
cosmology, Cyclic Universe)

$a(t)\sim t^{\alpha }$ \cite{plc-lohiya} (Power law Cosmology)

$a(t)\sim t^{n}\exp (\alpha t)$ \cite{hsf} (Hybrid expansion)

$a(t)\sim \exp \left[ n(\log t)^{m}\right] $ \cite{logamediate} (Logamediate
expansion)

$a(t)\sim \cosh \alpha t$ \cite{ellismadsen} (Hyperbolic expansion)

$a(t)\sim \left( \sinh \alpha t\right) ^{\frac{1}{n}}$ \cite{ritika}
(Hyperbolic expansion)

$a(t)\sim \left( \frac{t}{t_{s}-t}\right) ^{n}$ \cite{odintsov1} (Singular
model)

$a(t)\sim t^{n}\exp \left[ \alpha (t_{s}-t)\right] $ \cite{odintsov1}
(Singular model)

$a(t)\sim \exp \left( \alpha \frac{t^{2}}{t_{\ast }^{2}}\right) $ \cite%
{bouncing} (Bouncing Model)

$a(t)\sim \exp \left( \frac{\beta }{\alpha +1}(t-t_{s})^{\alpha +1}\right) $ 
\cite{bouncing} (Bouncing Model)

$a(t)\sim \left( \frac{3}{2}\rho _{cr}t^{2}+1\right) ^{\frac{1}{3}}$ \cite%
{bouncing} (Bouncing Model)

$a(t)\sim \left( \frac{t_{s}-t}{t_{\ast }}\right) $ \cite{bouncing}
(Bouncing Model)

$a(t)\sim \sin ^{2}\left( \alpha \frac{t}{t_{\ast }}\right) $ \cite{bouncing}
(Bouncing Model)

\textbf{Hubble parameter }$H(t)$ or $H(a)$

$H(a)=Da^{-m}$\ \cite{H-berman1983}

$H(a)=e^{\frac{1-\gamma a^{2}}{\alpha a}}$\ \cite{H-banerjee2010}

$H(a)=\alpha (1+a^{-n})$\ \cite{H-JPSINGH}

$H(t)=\frac{m}{\alpha t+\beta }$\ \cite{H-Pacif1}

$H(t)=\frac{16\alpha t}{15\left[ 1+(8\alpha t^{2})/5\right] }$ \cite{H-pks1}

$H(t)=m+\frac{n}{t}$ \cite{H-pks2}

$H(t)=\frac{\alpha t_{R}}{t(t_{R}-t)}$\ \cite{H-Cannata}

$H(t)=\frac{\alpha }{3}\left( t+T_{0}\right) ^{3}-\beta \left(
t+T_{0}\right) +\gamma $\ \cite{H-nojiri2006}

$H(t)=\alpha e^{\lambda t}$\ \cite{H-odintsov2012}

$H(t)=\alpha +\beta (t_{s}-t)^{n}$\ \cite{H-odintsov2012}

$H(t)=\alpha -\beta e^{-nt}$\ \cite{H-odin2015}

$H(t)=f_{1}(t)+f_{2}(t)(t_{s}-t)^{n}$\ \cite{H-odin2014}

$H(t)=\frac{\beta t^{m}}{\left( t^{n}+\alpha \right) ^{p}}$ \cite{pacif2016}

$H(t)=n\alpha \tanh (m-nt)+\beta $ \cite{H-ashutosh}

$H(t)=\alpha \tanh \left( \frac{t}{t_{0}}\right) $ \cite{bamba-de}

$H(z)=\left[ \alpha +\left( 1-\alpha \right) \left( 1+z\right) ^{n}\right] ^{%
\frac{3}{2n}}$ \cite{H-abdulla}

\textbf{Deceleration parameter }$q(t)$\textbf{\ or }$q(a)$\textbf{, }$q(z)$

$q(t)=m-1$ \cite{q-berman1998}

$q(t)=-\alpha t+m-1$ \cite{q-akarsu2012}

$q(t)=\alpha \cos (\beta t)-1$ \cite{q-pks1}

$q(t)=-\frac{\alpha t}{1+t}$ \cite{q-pks2}

$q(t)=-\frac{\alpha (1-t)}{1+t}$ \cite{q-pks2}

$q(t)=-\frac{\alpha }{t^{2}}+\beta -1$ \cite{q-ASSRP}

$q(t)=(8n^{2}-1)-12nt+3t^{2}$ \cite{q-bakry}

$q(a)=-1-\frac{\alpha a^{\alpha }}{1+a^{\alpha }}$ \cite{q-banerjee2005}

$q(z)=q_{0}+q_{1}z$ \cite{q-riess}

$q(z)=q_{0}+q_{1}z(1+z)^{-1}$ \cite{q-santos}

$q(z)=q_{0}+q_{1}z(1+z)(1+z^{2})^{-1}$ \cite{q-sdas}

$q(z)=\frac{1}{2}+q_{1}(1+z)^{-2}$ \cite{q-nair}

$q(z)=q_{0}+q_{1}[1+\ln (1+z)]^{-1}$ \cite{q-Xu2008}

$q(z)=\frac{1}{2}+(q_{1}z+q_{2})(1+z)^{-2}$ \cite{q-gong2006}

$q(z)=-1+\frac{3}{2}\left( \frac{\left( 1+z\right) ^{q_{2}}}{%
q_{1}+(1+z)^{q_{2}}}\right) $ \cite{q-campo}

$q(z)=-\frac{1}{4}\left[ 3q_{1}+1-3(q_{1}+1)\left( \frac{q_{1}e^{q_{2}\left(
1+z\right) }-e^{-q_{2}\left( 1+z\right) }}{q_{1}e^{q_{2}\left( 1+z\right)
}+e^{-q_{2}\left( 1+z\right) }}\right) \right] $ \cite{q-pavon}

$q(z)=-\frac{1}{4}+\frac{3}{4}\left( \frac{q_{1}e^{q_{2}\frac{z}{\sqrt{1+z}}%
}-e^{-q_{2}\frac{z}{\sqrt{1+z}}}}{q_{1}e^{q_{2}\frac{z}{\sqrt{1+z}}%
}+e^{-q_{2}\frac{z}{\sqrt{1+z}}}}\right) $ \cite{q-pavon}

$q(z)=q_{f}+\frac{q_{i}-q_{f}}{1-\frac{q_{i}}{q_{f}}\left( \frac{1+z_{t}}{1+z%
}\right) ^{\frac{1}{\tau }}}$ \cite{q-ishida}

$q(z)=q_{0}-q_{1}\left( \frac{(1+z)^{-\alpha }-1}{\alpha }\right) $ \cite%
{q-abdulla1}

$q(z)=q_{0}+q_{1}\left[ \frac{\ln (\alpha +z)}{1+z}-\beta \right] $ \cite%
{q-abdulla2}

$q(z)=q_{0}-(q_{0}-q_{1})(1+z)\exp \left[ z_{c}^{2}-(z+z_{c})^{2}\right] $ 
\cite{q-garza}

\textbf{Jerk parameter }$j(z)$

$j(z)=-1+j_{1}\frac{f(z)}{E^{2}(z)},$ where $f(z)=z$, $\frac{z}{1+z}$, $%
\frac{z}{(1+z)^{2}}$, $\log (1+z)$ and $E(z)=\frac{H(z)}{H_{0}}$\cite%
{jerk-para}

$j(z)=-1+j_{1}\frac{f(z)}{h^{2}(z)},$ where $f(z)=1$, $1+z$, $(1+z)^{2}$, $%
(1+z)^{-1}$ and $h(z)=\frac{H(z)}{H_{0}}$\cite{jerk-para2}

\subsection{\textbf{PARAMETRIZATIONS OF PHYSICAL PARAMETERS}}

\textbf{Pressure }$p\mathbf{(}\rho \mathbf{)}$, $p(z)$

The matter content in the Universe is not properly known but it can be
categorized with its equations of states $p=p(\rho )$. Following is a list
of some cosmic fluid considerations with their EoS. Also, some dark energy
pressure parametrization are listed.

$p(\rho )=w\rho $ (Perfect fluid EoS)

$p{\small (\rho )}=w\rho -f(H)$ \cite{viscous1} (Viscous fluid EoS)

$p{\small (\rho )}=w\rho +k\rho ^{1+\frac{1}{n}}$ \cite{poly0} (Polytropic
gas EoS)

$p{\small (\rho )}=\frac{8w\rho }{3-\rho }-3\rho ^{2}$ \cite{vanderwall}
(Vanderwaal gas EoS)

$p{\small (\rho )}=-(w+1)\frac{\rho ^{2}}{\rho _{P}}+w\rho +(w+1)\rho
_{\Lambda }$ \cite{quad} (EoS in quadratic form)

$p{\small (\rho )}=-\frac{B}{\rho }$\cite{CG1} (Chaplygin gas EoS)

$p{\small (\rho )}=-\frac{B}{\rho ^{\alpha }}$ \cite{GCG1} (Generalized
Chaplygin gas EoS)

$p{\small (\rho )}=A\rho -\frac{B}{\rho ^{\alpha }}$ \cite{MCG1} (Modified
Chaplygin gas EoS)

$p{\small (\rho )}=A\rho -\frac{B(a)}{\rho ^{\alpha }}$ \cite{VMCG}
(Variable modified Chaplygin gas EoS)

$p{\small (\rho )}=A(a)\rho -\frac{B(a)}{\rho ^{\alpha }}$ \cite{NVMCG} (New
variable modified Chaplygin gas EoS)

$p{\small (\rho )}=-\rho -\rho ^{\alpha }$ \cite{NOJI-EOS} (DE EoS)

$p(z)=\alpha +\beta z$ \cite{eos-zhang} (DE EoS)

$p(z)=\alpha +\beta \frac{z}{1+z}$ \cite{eos-zhang} (DE EoS)

$p(z)=\alpha +\beta \left( z+\frac{z}{1+z}\right) $ \cite{eos-wang} (DE EoS)

$p(z)=\alpha +\beta \ln (1+z)$ \cite{eos-jun} (DE EoS)

\textbf{Equation of state parameter }$w(z)$

$w(z)=w_{0}+w_{1}z$ \cite{w-LIN2} (Linear parametrization)

$w(z)=w_{0}+w_{1}\frac{z}{\left( 1+z\right) ^{2}}$ \cite{w-JBP} (JBP
parametrization)

$w(z)=w_{0}+w_{1}\frac{z}{\left( 1+z\right) ^{n}}$ \cite{w-nCPL}
(Generalized JBP parametrization)

$w(z)=w_{0}+w_{1}\frac{z}{1+z}$ \cite{w-CPL1} (CPL parametrization)

$w(z)=w_{0}+w_{1}\left( \frac{z}{1+z}\right) ^{n}$ \cite{w-nCPL}
(Generalized CPL parametrization)

$w(z)=w_{0}+w_{1}\frac{z}{\sqrt{1+z^{2}}}$ \cite{w-sqrt} (Square-root
parametrization)

$w(z)=w_{0}+w_{1}\sin (z)$ \cite{w-sin} (Sine parametrization)

$w(z)=w_{0}+w_{1}\ln (1+z)$ \cite{w-LOG} (Logarithmic parametrization)

$w(z)=w_{0}+w_{1}\ln \left( {\small 1+\frac{z}{1+z}}\right) $ \cite{w-feng}
(Logarithmic parametrization)

$w(z)=w_{0}+w_{1}\frac{z(1+z)}{1+z^{2}}$ \cite{w-BA} (BA parametrization)

$w(z)=w_{0}+w_{1}\left( \frac{\ln (2+z)}{1+z}-\ln 2\right) $ (MZ
parametrization)

$w(z)=w_{0}+w_{1}\left( \frac{\sin (1+z)}{1+z}-\sin 1\right) $ \cite{w-MZ}
(MZ parametrization)

$w(z)=w_{0}+w_{1}\frac{z}{1+z^{2}}$ (FSLL parametrization)

$w(z)=w_{0}+w_{1}\frac{z^{2}}{1+z^{2}}$ \cite{w-FSSL} (FSLL parametrization)

$w(z)=-1+\frac{1+z}{3}\frac{\alpha +2\beta (1+z)}{\gamma +2\alpha
(1+z)+\beta (1+z)^{2}}$ \cite{w-ASSS} (ASSS parametrization)

$w(z)=\frac{1+\left( \frac{1+z}{1+z_{s}}\right) ^{\alpha }}{%
w_{0}+w_{1}\left( \frac{1+z}{1+z_{s}}\right) ^{\alpha }}$ \cite{w-hansted}
(Hannestad Mortsell parametrization)

$w(z)=-1+\alpha (1+z)+\beta (1+z)^{2}$ \cite{w-weller} (Polynomial
parametrization)

$w(z)=-1+\alpha \left[ 1+f(z)\right] +\beta \left[ 1+f(z)\right] ^{2}$ \cite%
{w-sendra} (Generalized Polynomial parametrization)

$w(z)=w_{0}+z\left( \frac{dw}{dz}\right) _{0}$ \cite{w-cooray}

$w(z)=\frac{-2(1+z)d_{c}^{^{\prime \prime }}-3d_{c}^{^{\prime }}}{3\left[
d_{c}^{^{\prime }}-\Omega _{M}(1+z)^{3}\left( d_{c}^{^{\prime }}\right) ^{3}%
\right] }$ where $d_{c}^{^{\prime }}=\int\limits_{0}^{z}\frac{H_{0}dz}{H(z)}$
\cite{w-hai}

$w_{x}(a)=w_{0}\exp (a-1)$ \cite{w-yang}

$w_{x}(a)=w_{0}a(1-\log a)$ \cite{w-yang}

$w_{x}(a)=w_{0}a\exp (1-a)$ \cite{w-yang}

$w_{x}(a)=w_{0}a(1+\sin (1-a))$ \cite{w-yang}

$w_{x}(a)=w_{0}a(1+\arcsin (1-a))$ \cite{w-yang}

$w_{de}(z)=w_{0}+w_{1}q$ \cite{w-elizalde}

$w_{de}(z)=w_{0}+w_{1}q(1+z)^{\alpha }$ \cite{w-elizalde}

$w_{de}(z)=\frac{w_{0}}{\left[ 1+b\ln (1+z)\right] ^{2}}$ \cite{w-wetterich}

$w_{x}(z)=w_{0}+b\left\{ 1-\cos \left[ \ln (1+z)\right] \right\} $ \cite%
{w-supriya}

$w_{x}(z)=w_{0}+b\sin \left[ \ln (1+z)\right] $ \cite{w-supriya}

$w_{x}(z)=w_{0}+b\left[ \frac{\sin (1+z)}{1+z}-\sin 1\right] $ \cite%
{w-supriya}

$w_{x}(z)=w_{0}+b\left( \frac{z}{1+z}\right) \cos (1+z)$ \cite{w-supriya}

$w(z)=w_{0}+w_{a}\left[ \frac{\ln (2+z)}{1+z}-\ln 2\right] $ \cite{w-mazhang}

$w(z)=w_{0}+w_{a}\left[ \frac{\ln (\alpha +1+z)}{\alpha +z}-\frac{\ln
(\alpha +1)}{\alpha }\right] $ \cite{w-sello}

\textbf{Energy density }$\rho $

$\rho =\rho _{c}$ \cite{OT2}, \cite{ABDEL}

$\rho \sim \theta ^{2}$ \cite{SKJP1}

$\rho =\frac{A}{a^{4}}\sqrt{{\small a}^{2}{\small +b}}$ \cite{SKJP3}

$\left( {\small \rho +3p}\right) a^{3}=A$ \cite{RGVgrg}

$\rho +p=\rho _{c}$ \cite{ASRGVprd}

$\rho _{de}(z)=\rho _{de}(0)\left[ 1+\alpha \left( \frac{z}{1+z}\right) ^{n}%
\right] $ \cite{rho-abdulla}

$\rho _{de}(z)=\frac{1}{\rho _{\phi }}\left( \frac{d\rho _{\phi }}{d\phi }%
\right) =-\frac{\alpha a}{\left( \beta +a\right) ^{2}}$ \cite{rho-sdas}

$\rho _{de}(z)=\alpha H(z)$ \cite{rho-rezaei}

$\rho _{de}(z)=\alpha H(z)+\beta H^{2}(z)$ \cite{rho-rezaei}

$\rho _{de}(z)=\frac{3}{\kappa ^{2}}\left[ \alpha +\beta H^{2}(z)\right] $ 
\cite{rho-rezaei}

$\rho _{de}(z)=\frac{3}{\kappa ^{2}}\left[ \alpha +\frac{2}{3}\beta \dot{H}%
(z)\right] $ \cite{rho-rezaei}

$\rho _{de}(z)=\frac{3}{\kappa ^{2}}\left[ \alpha H^{2}(z)+\frac{2}{3}\beta 
\dot{H}(z)\right] $ \cite{rho-rezaei}

$\rho _{de}(z)=\frac{3}{\kappa ^{2}}\left[ \alpha +\beta H^{2}(z)+\frac{2}{3}%
\gamma \dot{H}(z)\right] $ \cite{rho-rezaei}

$\rho _{de}(z)=\rho _{\phi 0}\left( 1+z\right) ^{\alpha }e^{\beta z}$ \cite%
{rho-abdulla2}

\textbf{Cosmological constant (}$\Lambda $\textbf{)}

In order to resolve the long standing cosmological constant problem, authors
have considered some variation laws for the cosmological constant in the
past forty years, commonly known as \textquotedblleft $\Lambda $-varying
cosmologies\textquotedblright\ or \textquotedblleft Decaying vacuum
cosmologies\textquotedblright . Later the idea was adopted to explain the
accelerated expansion of the Universe considering varying $\Lambda $.
Following is list of such decay laws of $\Lambda $.

$\Lambda \sim a^{-n}$ \cite{RGVcqg1}

$\Lambda \sim H^{n}$ \cite{COOPER}

$\Lambda \sim \rho $ \cite{RGVcqg1}

$\Lambda \sim t^{n}$ \cite{COOPER}

$\Lambda \sim q^{n}$ \cite{COOPER}

$\Lambda \sim e^{-\beta a}$ \cite{SGRAJ}

$\Lambda =\Lambda (T)$\ \cite{LINDE} $T$\ is Temperature

$\Lambda \sim C+e^{-\beta t}$ \cite{var-lam3}

$\Lambda =3\beta H^{2}+\alpha a^{-2}$ \cite{CARVALHO}

$\Lambda =\beta \frac{\ddot{a}}{a}$ \cite{SRAY}

$\Lambda =3\beta H^{2}+\alpha \frac{\ddot{a}}{a}$ \cite{ArbabLAM2}

$\frac{d\Lambda }{dt}\sim \beta \Lambda -\Lambda ^{2}$ \cite{var-lammoffat}

\textbf{Scalar field Potentials }$V(\phi )$

$V(\phi )=V_{0}\phi ^{n}$ \cite{copeland-sami} (Power law)

$V(\phi )=V_{0}\exp \left[ -\frac{\alpha \phi }{M_{pl}}\right] $ \cite%
{copeland-sami} (exponential)

$V(\phi )=\frac{V_{0}}{\cosh \left[ \phi /\phi _{0}\right] }$ \cite%
{copeland-sami}

$V(\phi )=V_{0}\left[ \cosh \left( \alpha \phi /M_{pl}\right) \right]
^{-\beta }$ (hyperbolic) \cite{copeland-sami}

$V(\phi )=\frac{\alpha }{\phi ^{n}}$ (Inverse power law) \cite{v-inverse}

$V(\phi )=\frac{V_{0}}{1+\beta \exp (-\alpha \kappa \phi )}$ (Woods-Saxon
potential) \cite{v-wood}

$V(\phi )=\alpha c^{2}\left[ \tanh \frac{\phi }{\sqrt{6}\alpha }\right] ^{2}$
($\alpha $-attractor) \cite{v-shah}

$V(\phi )=V_{0}(1+\phi ^{\alpha })^{2}$ \cite{v-shah2}

$V(\phi )=V_{0}\exp (\alpha \phi ^{2})$ \cite{v-shah2}

$V(\phi )=\frac{1}{4}(\phi ^{2}-1)^{2}$ \cite{v-arefe}

Note: All the parametrization listed above contain some arbitrary constants
such as $\alpha $, $\beta $, $\gamma $, $m$, $n$, $p$, $q_{0}$, $q_{1}$, $%
q_{2}$, $w_{0}$, $w_{1}$, $A$, $B$ are model parameters which are generally
constrained through observational datasets or through any analytical methods
and also some arbitrary functions $f_{1}(t)$, $f_{2}(t)$. $t_{s}$ denote the
bouncing time or future singularity time and $t_{\ast }$ some arbitrary time.


\begin{thebibliography}{999}
\bibitem{perlmutter} S. Perlmutter et al., Astrophys. J. \textbf{517,} 565
(1999).

\bibitem{riess} A. Riess et al., Astrophys. J. \textbf{117}, 707 (1999).

\bibitem{bernadis} de Bernardis et al., Nature \textbf{404}, 955 (2000).

\bibitem{hanany} S. Hanany et al., Astrophy. J. Lett. \textbf{545}, L5
(2000).

\bibitem{netterfield} C. Netterfield et al., Astrophy. J. Lett. \textbf{474}%
, 47 (1997).

\bibitem{mould} J. R. Mould et al., Astrophy. J. \textbf{529}, 786 (2000).

\bibitem{spergel} D. Spergal et al., Astrophys. J. Suppl. \textbf{148,} 175
(2003).

\bibitem{komastu} E. Komatsu et al., Astrophys. J. Suppl. \textbf{192}, 18
(2011).

\bibitem{essence} W. M. Wood-Vasey, Astrophys. J. \textbf{666}, 694 (2007).

\bibitem{turner} M. S. Turner and D. Huterer, J. Phys. Soc. Jap. \textbf{76}%
, 111015 (2007).

\bibitem{globular} B. Chaboyer, Phys. Rept. \textbf{307}, 23 (1998).

\bibitem{capozilo} S. Capozziello and L. Z. Fang, Int. J. Mod. Phys. D 
\textbf{11}, 483 (2002).

\bibitem{dvali} G. R. Dvali et al., Phys. Lett. B \textbf{485}, 208 (2000).

\bibitem{copeland-sami} E. J. Copeland, M. Sami, S. Tsujikawa, Int. J. Mod.
Phys. D \textbf{15}, 1753 (2006).

\bibitem{bamba-de} K. Bamba et al., Astrophys. Space Sci. \textbf{342}, 155
(2012).

\bibitem{sahni} V. Sahni and A. A. Starobinsky, Int. J. Mod. Phys. D \textbf{%
9} (2000) 373.

\bibitem{zlatev} I. Zlatev, L.M. Wang and P. J. Steinhardt, Phys. Rev. Lett. 
\textbf{82} (1999) 896.

\bibitem{brax} P. Brax and J. Martin, Phys. Rev. D \textbf{61} (2000) 103502.

\bibitem{barreiro} T. Barreiro, E.J. Copeland and N.J. Nunes, Phys. Rev. D 
\textbf{61} (2000) 127301.

\bibitem{mukhanov1} C. Armendariz-Picon, T. Damour, and V. Mukhanov, Phys.
Lett. B \textbf{458} (1999) 219.

\bibitem{chiba} T. Chiba, T. Okabe and M. Yamaguchi, Phys. Rev. D \textbf{62}
(2000) 023511.

\bibitem{steinhardt1} C. Armendariz-Picon, V. Mukhanov and P.J. Steinhardt,
Phys. Rev. Lett. \textbf{85} (2000) 4438.

\bibitem{caldwell} R.R. Caldwell, Phys. Lett. B \textbf{545}, 23 (2003).

\bibitem{sen1} A. Sen, J. High Energy Phys., \textbf{0204} 048 (2002).

\bibitem{garousi1} M.R. Garousi, Nucl. Phys. B \textbf{584} 284 (2000).

\bibitem{bergshoeff} E.A. Bergshoeff et al., J. High Energy Phys. \textbf{5}
009 (2000).

\bibitem{gorini} V. Gorini et al., Phys. Rev. D \textbf{67}, 063509 (2003).

\bibitem{chavanis} P. H. Chavanis, Eur. Phys. J. Plus \textbf{129}, 38
(2012).

\bibitem{schwar} K. Schwarzschild, Sitzungsb. der K\"{o}nig. Preuss. Akad.
der Wissen. \textbf{7}, 189 (1916).

\bibitem{static} H. A. Lorentz, A. Einstein, H. Minkowski, H. Weyl, The
Principle of Relativity (New York: Metheun \& Co.) \textbf{175} (1923).

\bibitem{de-Sitter} W. de Sitter, Mon. Not. R. Astron. Soc. \textbf{76}, 699
(1916); \textbf{77}, 155 (1916).

\bibitem{Tolman} R. C. Tolman, Phys. Rev. \textbf{55}, 364 (1939).

\bibitem{Adlar} R. J. Adlar, J. Math. Phys. \textbf{15}, 727 (1974).

\bibitem{Buchd} H. A. Buchdahl, Astrophys. J. \textbf{147}, 310 (1967).

\bibitem{Vaidya} P. C. Vaidya and R. Tikekar, J. Astrophys. \& Astron. 
\textbf{3}, 325 (1982).

\bibitem{Durgapal} M. C. Durgapal, J. Phys. A. Math. Gen. \textbf{15}, 2637
(1982).

\bibitem{Knutsen} H. Knutsen, Astrophys. \& Space Science \textbf{140}, 385
(1988).

\bibitem{Kramer} D. Kramer et al., \textquotedblleft Exact solutions of
Einstein's equations\textquotedblright , Cambridge: CUP, (1980).

\bibitem{Negi} P. S. Negi, Int. J. Theor. Phys. \textbf{45}, 1684 (2006).

\bibitem{milne} Edward A. Milne, Relativity, Gravitation and World
Structure, (Oxford University Press) (1935).

\bibitem{Steady} H. Bondi and T. Gold, Mon. Not. R. Astron. Soc. \textbf{108}%
, 252 (1948); F. Hoyle, Mon. Not. R. Astron. Soc. \textbf{108}, 372 (1948).

\bibitem{knop} R. A. Knop et al., Astrophys. J. \textbf{598},102 (2003).

\bibitem{tegmark} M. Tegmark et al., Astrophys. J. \textbf{606} 702 (2004).

\bibitem{w-de1} T. M. C. Abbott et al., Astrophys. J. Lett., \textbf{872},
L30 (2019).

\bibitem{w-de2} Jing-Fei Zhang et al., Physics Letters B \textbf{799},
135064 (2019).

\bibitem{zimdahl} W. Zimdahl and D. Pavon, D., Gen. Relativ. Gravit. \textbf{%
36}, 1483 (2004).

\bibitem{bertolemi} O. Bertolami and D. Pavon, D., Phys. Rev. D \textbf{61},
064007 (2000).

\bibitem{banerjee} N. Banerjee and D. Pavon, Phys. Rev. D \textbf{63},
043504 (2001).

\bibitem{ellis} G. F. R. Ellis and M. Madsen, M., Class. Quantum Gravity 
\textbf{8}, 667 (1991).

\bibitem{sahni-2} 43. Sahni, V.: Lect. Notes Phys. 653, 141 (2004).

\bibitem{saini} T. D. Saini et al., Phys. Rev. Lett. \textbf{85}, 1162
(2000).

\bibitem{simon} J. Simon et al., Phys. Rev. D \textbf{71}, 123001 (2005).

\bibitem{chuna} J. V. Cunha J. V. J. A. S. Lima, Mon. Not. R. Astr. Soc. 
\textbf{390}, 210 (2008).

\bibitem{edvard} E. \"{M}ortsell and C. Clarkson, J. Cosm. Astropar. Phys., 
\textbf{2009}, 044 (2009).

\bibitem{pacif2016} S. K. J. Pacif et al., Int. J. Geom. Meth. Mod. Phys., 
\textbf{14(7)}, 1750111 (2017).

\bibitem{capozilo-lakoz} S. Capozziello, R. Lazkoz, and V. Salzano, Phys.
Rev. D \textbf{84,} 124061 (2011).

\bibitem{H1} D. Stern et al., J. Cosmol. Astropart. Phys., \textbf{02}, 008
(2010).

\bibitem{H2} J. Simon, L. Verde, R. Jimenez, Phys. Rev. D, \textbf{71},
123001 (2005).

\bibitem{H3} M. Moresco et al., J. Cosmol. Astropart. Phys., \textbf{08},
006 (2012).

\bibitem{H4} C. Zhang et al., Research in Astron. and Astrop., \textbf{14},
1221 (2014).

\bibitem{H5} M. Moresco et al., J. Cosmol. Astropart. Phys., \textbf{05},
014 (2016).

\bibitem{H6} A.L. Ratsimbazafy et al., Mon. Not. Roy. Astron. Soc., \textbf{%
467}, 3239 (2017).

\bibitem{H7} M. Moresco, Mon. Not. Roy. Astron. Soc.: Letters. , \textbf{450}%
, L16 (2015).

\bibitem{H8} E. Gazta\~{n}aga, A. Cabre, L. Hui, Mon. Not. Roy. Astron.
Soc., \textbf{399}, 1663 (2009).

\bibitem{H9} A. Oka et al., Mon. Not. Roy. Astron. Soc., \textbf{439}, 2515
(2014).

\bibitem{H10} Y. Wang et al., Mon. Not. Roy. Astron. Soc. \textbf{469}, 3762
(2017).

\bibitem{H11} C. H. Chuang, Y. Wang, Mon. Not. Roy. Astron. Soc., \textbf{435%
}, 255 (2013).

\bibitem{H12} S. Alam et al., Mon. Not. Roy. Astron. Soc., \textbf{470},
2617 (2017).

\bibitem{H13} C. Blake et al., Mon. Not. Roy. Astron. Soc., \textbf{425},
405 (2012).

\bibitem{H14} C. H. Chuang et al. , Mon. Not. Roy. Astron. Soc., \textbf{433}%
, 3559 (2013).

\bibitem{H15} L. Anderson et al., Mon. Not. Roy. Astron. Soc. , \textbf{441}%
, 24 (2014).

\bibitem{H16} N. G. Busca et al., Astron. Astrop., \textbf{552}, A96 (2013).

\bibitem{H17} J. E. Bautista et al. Astron. Astrophys., \textbf{603}, A12
(2017).

\bibitem{H18} T. Delubac et al., Astron. Astrophys. , \textbf{574}, A59
(2015).

\bibitem{H19} A. Font-Ribera et al., J. Cosmol. Astropart. Phys., \textbf{05}%
, 027 (2014).

\bibitem{sharov} G. S. Sharov, V.O. Vasiliev, Mathematical Modelling and
Geometry \textbf{6}, 1 (2018).

\bibitem{H0-value} E. Macaulay et al., Mon. Not. R. Astron. Soc. \textbf{%
486(2)}, 2184 (2019).

\bibitem{SNeIa} N. Suzuki et al., Astrophys. J. \textbf{746}, 85 (2012).

\bibitem{perivolar} S. Nesseris and L. Perivolaropoulos, Phys. Rev. D 
\textbf{70}, 043531 (2004).

\bibitem{BAO1} C. Blake et al., Mon. Not. Roy. Astron. Soc. \textbf{418},
1707 (2011).

\bibitem{BAO2} W. J. Percival et al., Mon. Not. Roy. Astron. Soc. \textbf{401%
}, 2148 (2010).

\bibitem{BAO3} F. Beutler et al., Mon. Not. Roy. Astron. Soc. \textbf{416},
3017 (2011).

\bibitem{BAO4} N. Jarosik et al., Astrophys. J. Suppl. \textbf{192}, 14
(2011).

\bibitem{BAO5} D. J. Eisenstein et al., Astrophys. J. \textbf{633}, 560
(2005).

\bibitem{BAO6} R. Giostri et al., J. Cosm. Astropart. Phys. \textbf{1203},
027 (2012).

\bibitem{stfnd1} V. Sahni, et al. JETP Lett. \textbf{77}, 201 (2003).

\bibitem{stfnd2} U. Alam, et al., Mon. Not. R. Astron. Soc. \textbf{344},
1057 (2003).

\bibitem{stfnd3} M.~Sami et al., Phys.\ Rev.\ D \textbf{86}, 103532 (2012)

\bibitem{stfnd4} R.~Myrzakulov,et al., J. Cosm. Astrop. Phys. \textbf{1310},
047 (2013).

\bibitem{omdig1} V. Sahni, A. Shaifeloo, A. A. Starobinsky, Phys. Rev. D 
\textbf{78}, 103502 (2008).

\bibitem{omdig2} C.~Zunckel, etal., Phys.\ Rev.\ Lett.\ \textbf{101},181301
(2008).

\bibitem{omdig3} M.~Shahalam, et al., Mon.\ Not.\ Roy.\ Astron.\ Soc.\ 
\textbf{448}, 2948 (2015)

\bibitem{omdig4} A. Agarwal et al., Int. J. Mod. Phys. D \textbf{28},
1950083 (2019).

\bibitem{melia} F. Melia, Astronom. J., \textbf{144}, 110 (2012).

\bibitem{ellismadsen} G. F. R. Ellis and M. S. Madsen, Class. Quantum Grav. 
\textbf{8}, 667 (1991).

\bibitem{a-zhou} X. Zhou, Chinese Phys. B \textbf{18}, 3115 (2009).

\bibitem{qssc} F. Hoyle, G. Burbidge, \& J. V. Narlikar, Astrophys. J., 
\textbf{410(2)}, 437 (1993).

\bibitem{plc-lohiya} D. Lohiya and M. Sethi, Class. Quantum Grav. \textbf{16}%
, 1545 (1999).

\bibitem{hsf} Ozgur Akarsu et al., J. Cosmol. Astropart. Phys. \textbf{01},
22 (2014).

\bibitem{logamediate} J. D. Barrow, N. J. Nunes, Phys. Rev. D \textbf{76},
043501 (2007).

\bibitem{ritika} Ritika Nagpal et al., Annals of Phys. \textbf{405}, 234
(2019).

\bibitem{odintsov1} S. D. Odintsov and V. K. Oikonomou, Phys.Rev. D \textbf{%
92(2)}, 024016 (2015).

\bibitem{bouncing} Alvaro de la Cruz-Dombriz et al., Phys. Rev. D \textbf{97}%
, 104040 (2018).

\bibitem{H-berman1983} M. S. Berman, Nuovo Cimento B, Series \textbf{74B},
182 (1983)

\bibitem{H-banerjee2010} N Banerjee, S. Das and K. Ganguly, Pramana \textbf{%
74(3)}, 481 (2010).

\bibitem{H-JPSINGH} J. P. Singh, Astrophys. Space Sci. \textbf{318}, 103
(2008).

\bibitem{H-Pacif1} S. K. J. Pacif and B. Mishra, Res. Astron. Astrophys. 
\textbf{15(12)}, 2141 (2015).

\bibitem{H-pks1} S. Bhattacharjee and P. K. Sahoo, Phys. Dark Univ. \textbf{%
28}, 100537 (2020).

\bibitem{H-pks2} P. H. R. S. Moraes et al., Adv. Astron. \textbf{2019},
8574798 (2019).

\bibitem{H-Cannata} F. Cannata and A. Y. Kamenshchik, Int. J. Mod. Phys. D 
\textbf{20}, 121 (2010).

\bibitem{H-nojiri2006} S. Noniri and S. D. Odintsov, Gen. Relativ. Grav. 
\textbf{38(8)}, 1285 (2006).

\bibitem{H-odintsov2012} Paul H. Frampton et al., Phys. Lett. B \textbf{708}%
, 204 (2012).

\bibitem{H-odin2015} S. Nojiri et al., J. Cosmol. Astropart. Phys. \textbf{%
1509}, 044 (2015).

\bibitem{H-odin2014} K. Bamba et al., Phys. Lett. B \textbf{732}, 349 (2014).

\bibitem{H-ashutosh} A. Singh, Astrophys. Space Sci. \textbf{365}, 54 (2020).

\bibitem{H-abdulla} Abdulla Al Mamon, Int. J. Mod. Phys. D \textbf{26(11)},
1750136 (2017).

\bibitem{q-berman1998} M. S. Berman and F. M. Gomide, Gen Relativ. Grav. 
\textbf{20}, 191 (1998).

\bibitem{q-akarsu2012} O. Akarsu and T. Dereli, Int. J. Theoret. Phys. 
\textbf{51}, 612 (2012).

\bibitem{q-pks1} P. K. Sahoo et al., Mod. Phys. Lett. A, \textbf{33(33)},
1850193 (2018).

\bibitem{q-pks2} P. K. Sahoo et al., New Astron. \textbf{60}, 80 (2018).

\bibitem{q-ASSRP} Abdussattar and S. R. Prajapati, Astrophys. Space Sci. 
\textbf{331}, 657 (2011).

\bibitem{q-bakry} M. A. Bakry, A. T. Shafeek, Astrophys. Space Sci. \textbf{%
364},135 (2019).

\bibitem{q-banerjee2005} N. Banerjee and S. Das, Gen. Relativ. Grav. \textbf{%
37}, 1695 (2005).

\bibitem{q-riess} A. G. Riess et al., Astrophy. J. \textbf{607}, 665 (2004).

\bibitem{q-santos} B. Santos, J. C. Carvalho, J. S. Alcaniz, Astropart.
Phys. \textbf{35}, 17 (2011).

\bibitem{q-sdas} Abdulla Al Mamon and Sudipta Das, Int. J. Mod. Phys. D 
\textbf{25(03)}, 1650032 (2016).

\bibitem{q-nair} R. Nair et al., J. Cosmol. Astropart. Phys. \textbf{01},
018 (2012).

\bibitem{q-Xu2008} L. Xu and H. Liu, Mod. Phys. Lett. A \textbf{23}, 1939
(2008).

\bibitem{q-gong2006} Y. G. Gong and A. Wang, Phys. Rev. D \textbf{73},
083506 (2006).

\bibitem{q-chunalima} J. V. Cunha and J. A. S. Lima, Mon. Not. R. Astron.
Soc. \textbf{390}, 210 (2008).

\bibitem{q-campo} S. del Campo et al., Phys. Rev. D \textbf{86}, 083509
(2012).

\bibitem{q-pavon} D. Pavon et al., Proc. MG13 Meeting Gen. Relativ.,
Stockholm Univ., Sweden (2012).

\bibitem{q-ishida} E. E. O. Ishida et al., Astroparticle Phys. \textbf{28},
7 (2007).

\bibitem{q-abdulla1} Abdulla Al Mamon, Mod. Phys. Lett. A \textbf{33(10)},
1850056 (2018).

\bibitem{q-abdulla2} Abdulla Al Mamon and S. Das, Eur. Phys. J. C \textbf{%
77(7)}, 495 (2017).

\bibitem{q-garza} Jaime Roman-Garza et al., Eur. Phys. J. C \textbf{79}, 890
(2019).

\bibitem{jerk-para} Z. X. Zhai et al., Physics Letters B \textbf{727}, 8
(2013).

\bibitem{jerk-para2} A. Mukherjee and N. Banerjee, Phys. Rev. D \textbf{93},
043002 (2016).

\bibitem{viscous1} C. Ekart, Phys. Rev. \textbf{58}, 919 (1940).

\bibitem{poly0} K. Karami, S. Ghaari, and J. Fehri, Eur. Phys. J. C \textbf{%
64(1)}, 85 (2009).

\bibitem{vanderwall} G. M. Kremer, Phys. Rev. D \textbf{68} 123507 (2003).

\bibitem{quad} P. H. Chavanis, Universe \textbf{1}, 357 (2015).

\bibitem{CG1} A. Yu. Kamenshchik, U. Moschella and V. Pasquier, Phys. Lett.
B \textbf{511}, 265 (2001).

\bibitem{GCG1} M. C. Bento, O. Bertolami and A. A. Sen, Phys. Rev. D \textbf{%
66}, 043507 (2002).

\bibitem{MCG1} H. B. Benaoum, arxiv:hep-th/0205140v1 (2002).

\bibitem{VMCG} U. Debnath, Astrophys. Space Sci. \textbf{312}, 295 (2007).

\bibitem{NVMCG} W. Chakraborty, U. Debnath, Gravit. and Cosmol. \textbf{16},
223 (2010).

\bibitem{NOJI-EOS} S. Nojiri and S. D. Odintsov, Phys. Rev. D \textbf{70},
103522 (2004).

\bibitem{eos-zhang} Q. Zhang et al., Eur. Phys. J. C \textbf{75(7)}, 300
(2015).

\bibitem{eos-wang} D. Wang et al., Eur. Phys. J. C \textbf{77(4)}, 263
(2017).

\bibitem{eos-jun} Jun-Chao Wang, Xin-He Meng, Comm. Theor. Phys., \textbf{%
70(12)}, 67 (2018).

\bibitem{w-LIN2} J. Weller and A. Albrecht, Phys. Rev. D \textbf{65}, 103512
(2002).

\bibitem{w-JBP} H. K. Jassal, J. S. Bagla, and T. Padmanabhan, Mon. Not. R.
Astron. Soc. Letters \textbf{356(1)}, L11 (2005).

\bibitem{w-nCPL} Dao-Jun Liu et al., Mon. Not. R. Astron. Soc. \textbf{388},
275 (2008).

\bibitem{w-CPL1} M. Chevallier and D. Polarski, Int. J. Mod. Phys. D \textbf{%
10}, 213 (2001).

\bibitem{w-sqrt} G. Pantazis, S. Nesseris and L. Perivolaropoulos, Phys.
Rev. D \textbf{93}, 103503 (2016).

\bibitem{w-sin} R. Lazkoz, V. Salzano, and I. Sendra, Phys. Lett. B \textbf{%
694}, 198 (2010).

\bibitem{w-LOG} G. Efstathiou, Mon. Not. R. Astron. Soc. \textbf{310,} 842
(1999).

\bibitem{w-feng} L. Feng and T. Lu, J. Cosmol. Astropart. Phys. \textbf{11},
034 (2011).

\bibitem{w-BA} E. M. Barboza, Jr. and J. S. Alcaniz, J. Cosmol. Astropart.
Phys. \textbf{02}, 042 (2012).

\bibitem{w-MZ} J. -Z. Ma and X. Zhang, Phys. Lett. B \textbf{699}, 233
(2011).

\bibitem{w-FSSL} C. -J. Feng et al., J. Cosmol. Astropart. Phys. \textbf{09}%
, 023 (2012).

\bibitem{w-ASSS} U. Alam et al., Mon. Not. R. Astron. Soc. \textbf{354}, 275
(2004).

\bibitem{w-hansted} S. Hannestad, E. Mrtsell, J. Cosm. Astropart. Phys., 
\textbf{0409}, 001 (2004).

\bibitem{w-weller} J. Weller, A. Albrecht, Phys.Rev. D \textbf{65}, 103512
(2002).

\bibitem{w-sendra} I. Sendra and R. Lazkoz, Mon. Not. R. Astron. Soc., 
\textbf{422(1)}, 776 (2012).

\bibitem{w-cooray} A. R. Cooray, D. Huterer, Astrophys. J. \textbf{513}, L95
(1999).

\bibitem{w-hai} Hai-Nan Lin, Xin Li, Li Tang, Chinese Physics C \textbf{43},
075101 (2019).

\bibitem{w-yang} W. Yang et al., Phys. Rev. D \textbf{99}, 043543 (2019).

\bibitem{w-elizalde} E. Elizalde et al., Int. J. Mod. Phys. D \textbf{28(01)}%
, 1950019 (2019).

\bibitem{w-wetterich} C. Wetterich, Physics Letters B 594(1-2), 17 (2004).

\bibitem{w-supriya} S. Pan et al., Phys. Rev. D \textbf{98}, 063510 (2018).

\bibitem{w-mazhang} J. Z. Ma and X. Zhang, Phys. Lett. B \textbf{699}, 233
(2011).

\bibitem{w-sello} S. Sello, arXiv:1308.0449 [astro-ph.CO]

\bibitem{OT2} M. Ozer and M.O. Taha, Nucl. Phys. B \textbf{287}, 776 (1987).

\bibitem{ABDEL} A-M. M. Abdel-Rahman, Gen. Relativ. Gravit. \textbf{22}, 655
(1990).

\bibitem{SKJP1} S. K. J. Pacif and Abdussattar, Eur. Phys. J. Plus \textbf{%
129}, 244 (2014).

\bibitem{SKJP3} Abdussattar and S. R. Prajapati, Chin. Phys. Lett. Vol. 
\textbf{28(2)}, 029803 (2011).

\bibitem{RGVgrg} R. G. Vishwakarma, Gen. Relativ. Grav. \textbf{33(11)},
1973 (2001).

\bibitem{ASRGVprd} R. G. Vishwakarma, Abdussattar and A. Beesham, Phys. Rev.
D \textbf{60}, 063507 (1999).

\bibitem{rho-abdulla} Abdulla Al Mamon, Mod. Phys. Lett. A \textbf{33(20)},
1850113 (2018).

\bibitem{rho-sdas} S. Das et al., Res. Astron. Astrophys. \textbf{18(11)},
131 (2018).

\bibitem{rho-rezaei} M. Rezaei et al., Phys. Rev. D \textbf{100}, 023539
(2019).

\bibitem{rho-abdulla2} Abdulla Al Mamon, K. Bamba and S. Das, Eur. Phys. J.
C \textbf{77(1)}, 29 (2017).

\bibitem{RGVcqg1} R. G. Vishwakarma, Class. Quantum Grav. \textbf{18}, 1159
(2001).

\bibitem{COOPER} J. M. Overduin and F. I. Cooperstock, Phys. Rev. D \textbf{%
58}, 043506 (1998).

\bibitem{SGRAJ} S. G. Rajeev, Phys. Lett. B \textbf{125}, 144 (1983).

\bibitem{LINDE} A. D. Linde, JETP Lett. \textbf{19}, 183 (1974).

\bibitem{var-lam3} A. Beesham, Phys. Rev. D \textbf{48}, 3539 (1993).

\bibitem{CARVALHO} J. C. Carvalho, J A S Lima and I Waga, Phys. Rev. D 
\textbf{46}, 2404 (1992).

\bibitem{SRAY} S. Ray et al., Int. J. Theor. Phys. \textbf{48(9)}, 2499
(2009).

\bibitem{ArbabLAM2} Arbab I Arbab, Grav. Cosmol. \textbf{8}, 227 (2002).

\bibitem{var-lammoffat} J. W. Moffat, Los Alamos report astro-ph/9608202
(1996).

\bibitem{v-inverse} P.J.E. Peebles, B. Ratra, Astrophys. Phys. J. \textbf{325%
}, L17 (1988).

\bibitem{v-wood} V.K. Oikonomou, N.Th. Chatzarakis, Annals of Phys. \textbf{%
411}, 167999 (2019).

\bibitem{v-shah} M. Shahalam, M. Sami, Anzhong Wang, Phys. Rev. D \textbf{98}%
, 043524 (2018).

\bibitem{v-shah2} M. Shahalam et al., Gen. Relativ. Gravit. \textbf{51}, 125
(2019).

\bibitem{v-arefe} I. Ya. Aref'eva, AIP Conf. Proc. \textbf{826}, 301 (2006).
\end{thebibliography}
\end{document}